\UseRawInputEncoding

\documentclass{aastex63}

\usepackage[ppl]{mathcomp}
\usepackage{comment}
\usepackage{multirow}
\usepackage{color}

\newcommand{\mjy}{$\mathrm{mJy}\,\mathrm{beam^{-1}}$}
\newcommand{\jy}{$\mathrm{Jy}\,\mathrm{beam^{-1}}$}
\newcommand{\kms}{$\mathrm{km}\,\mathrm{s^{-1}}$}

\newcommand{\jykms}{$\mathrm{Jy}\,\mathrm{beam^{-1}}\,\mathrm{km}\,\mathrm{s^{-1}}$}
\newcommand{\msun}{$\mathrm{M_{\odot}}$}
\newcommand{\lsun}{$\mathrm{L_{\odot}}$}

\newcommand{\vlsr}{$v_\mathrm{LSR}$}
\newcommand{\vsys}{$v_\mathrm{sys}$}

\newcommand{\two}{I\hspace{-.1em}I}

\accepted{November 22, 2022}
\submitjournal{ApJ}

\shorttitle{ALMA Fragmented Source Catalogue} \shortauthors{Sato et al.}
\graphicspath{{./}{figures/}}

\begin{document}

\title{ALMA Fragmented Source Catalogue in Orion (FraSCO)\\
I. Outflow interaction within an embedded cluster in OMC-2/FIR\,3, FIR\,4, and FIR\,5}

\correspondingauthor{Asako Sato}
\email{sato.asako.322@s.kyushu-u.ac.jp}

\author[0000-0001-5817-6250]{Asako Sato}
\affiliation{Department of Earth and Planetary Sciences, Graduate School of Science, Kyushu University, Fukuoka 819-0395, Japan}
\nocollaboration{1}

\author[0000-0002-7287-4343]{Satoko Takahashi}
\affiliation{National Astronomical Observatory of Japan, 2-21-1 Osawa, Mitaka, Tokyo 181-8588, Japan}
\affiliation{Department of Astronomical Science, School of Physical Sciences, The Graduate University for Advanced Studies (SOKENDAI), 2-21-1 Osawa, Mitaka, Tokyo 181-8588, Japan}
\nocollaboration{1}

\author[0000-0001-8337-4961]{Shun Ishii}
\affiliation{National Astronomical Observatory of Japan, 2-21-1 Osawa, Mitaka, Tokyo 181-8588, Japan}
\affiliation{Department of Astronomical Science, School of Physical Sciences, The Graduate University for Advanced Studies (SOKENDAI), 2-21-1 Osawa, Mitaka, Tokyo 181-8588, Japan}
\nocollaboration{1}

\author[0000-0002-3412-4306]{Paul T.P. Ho}
\affiliation{Academia Sinica Institute of Astronomy and Astrophysics, 
11F of AS/NTU Astronomy-Mathematics Building, No.1, Sec. 4, Roosevelt Rd, Taipei 10617, Taiwan, R.O.C.}
\affiliation{East Asian Observatory, Hilo 96720, HI, USA}
\nocollaboration{2}

\author[0000-0002-0963-0872]{Masahiro N. Machida}
\affiliation{Department of Earth and Planetary Sciences, Faculty of Science, Kyushu University, Fukuoka 819-0395, Japan}
\nocollaboration{1}

\author[0000-0003-2251-0602]{John Carpenter}
\affiliation{Joint ALMA Observatory, Alonso de C\'{o}́rdova 3107, Vitacura, Santiago 763-0355, Chile}
\nocollaboration{2}

\author[0000-0003-2343-7937]{Luis A.Zapata}
\affiliation{Instituto de Radioastronomía y Astrofísica, Universidad Nacional Autónoma de México, Antigua Carretera a Pátzcuaro \#8701, Ex-Hda. San José de la Huerta, Morelia, Michoacán, México C.P. 58089}
\nocollaboration{2}

\author[0000-0002-3665-5784]{Paula Stella Teixeira}
\affiliation{SUPA, School of Physics \& Astronomy, University of St Andrews, North Haugh, St Andrews KY16 9SS, UK}
\nocollaboration{2}

\author{Sümeyye Suri}
\affiliation{University of Vienna, Department of Astrophysics, Türkenschanzstrasse 17, 1180 Vienna, Austria}
\nocollaboration{2}



\begin{abstract}
We present a high angular resolution ($\sim1''$) and wide-field ($2'.9 \times 1'.9$) image of the 1.3-mm continuum, CO\,($J$ = 2--1) line, and SiO\,($J$ = 5--4) line emissions toward an embedded protocluster, FIR\,3, FIR\,4, and FIR\,5, in the Orion Molecular Cloud 2 obtained from the Atacama Large Millimeter/submillimeter Array (ALMA). We identify 51 continuum sources, 36 of which are newly identified in this study. Their dust masses,  projected sizes, and $\mathrm{H_2}$ gas number densities are estimated to be $3.8 \times 10^{-5}$--$ 1.1 \times 10^{-2}$\,\msun, 290--2000\,au, and $6.4 \times 10^{6}$--$3.3 \times 10^{8}\,\mathrm{cm^{-3}}$, respectively. The results of a Jeans analysis show that $\sim80\,\%$ of the protostellar sources and $\sim15\,\%$ of the prestellar sources are gravitationally bound. We identify 12 molecular outflows traced in the CO\,($J$ = 2--1) emission, six of which are newly detected. We spatially resolve shocked gas structures traced by the SiO\,($J$ = 5--4) emission in this region for the first time. We identify shocked gas originating from outflows and other shocked regions. These results provide direct evidence of an interaction between a dust condensation, FIR\,4, and an energetic outflow driven by HOPS-370 located within FIR\,3. A comparison of the outflow dynamical timescales, fragmentation timescales, and protostellar ages shows that the previously proposed triggered star-formation scenario in FIR\,4 is not strongly supported. We also discuss the spatial distribution of filaments identified in our continuum image by comparing it with a previously identified hub-fiber system in the $\mathrm{N_2H^+}$ line.
\end{abstract}


\section{introduction}
\label{intro}
Most stars are formed within cluster environments \citep{ladalada2003} and thus, to understand the formation of stars, it is vital to study young protoclusters.
Protoclusters are formed within filamentary molecular clouds extending to parsec scale sizes \citep{schneider1979, evans1999, Motte2010A, andre2014}. 
Recent observational studies suggest that filamentary molecular clouds have rather complex structures such as hub-filament systems where the filaments intersect around the hub \citep{myers2009}. In a hub-filament system, a mass accumulation process (i.e., gas inflow) through the filaments onto the hub plays an essential role in determining the final mass of stars within the protocluster \citep{McLaughlin1996ApJ...469..194M, bonnell2001, mckee2003, wang2010, Smith2011, krumholz2012, cheng2018, Ohashi2016ApJ...833..209O, Sanhueza2019}. Filamentary molecular clouds are also resolved into fibers defined as velocity coherent structures by \cite{hacar2013}. These fibers also exhibit complex structures such as hub-fiber systems, and dense cores and clumps seem to form at the hub (e.g., \citealt{zhang2020, clarke2020}). In this paper, we use the term “filaments” based on the filamentary morphological structures, while we use the term “fibers” when the filamentary structures show velocity coherence.
The feedback within protoclusters, such as through outflows and stellar radiation, is another important factor in the star-forming cluster environment \citep{wang2010, nakamura2011, hansen2012, Offner2017}.
To understand the star formation processes within a protocluster, it is important to observe nearby embedded clusters (within 1 kpc), identify individual sources and associated outflows, and reveal how each source interacts within the protocluster.

The closest known giant molecular cloud is the Integral Shaped Filament (ISF; \citealt{bally1987}) located within the northern part of the Orion A giant molecular cloud (at a distance $d$ = 400 pc; \citealt{groschedl2018}) \citep{Maddalena1985, bally1987, Tatematsu1993ApJ...404..643T, sakamoto1994, Nagahama1998, wilson2005, ikeda2007, ODell2008, buckle2012, kong2018, ishii2019}.
The ISF extends $\sim7$ pc in length and consists of remarkable filamentary molecular clouds hosting hundreds of protostars \citep{chini1997, lis1998, johnstone1999, nielbock2003, nutter2007, takahashi2013, teixeira2016, sadavoy2016}.
Due to its proximity, the ISF is a well-studied region across a wide range of wavelengths \citep{bally1987, chini1997, lis1998, johnstone1999, megeath2012, stutz2013}.

The Orion Molecular Cloud 2 region (the OMC-2 region) is located within the ISF and is classified as an embedded protocluster containing a large number of infrared sources ($\sim$400 pc$^{-2}$; \citealt{ladalada2003}). 
A dozen bright millimeter sources were identified in the OMC-2 region by 1.3\,mm single-dish observations \citep{chini1997}. 
In this study, we focus on the brightest regions identified by the 1.3\,mm single dish observations, named FIR\,3, FIR\,4, and FIR\,5.  
Fifteen individual sources have been identified within FIR\,3, 4, and 5 by multi-wavelength higher angular resolution observations, consisting of one Class 0, three Class I, three Class II, three flat spectrum, and five non-classified mm/sub-mm sources \citep{reipurth1999, nielbock2003,allen2007, megeath2012, stutz2013,furlan2016, osorio2017, vanterwisga2019}. 
FIR\,4 has the largest luminosity of $L_{\mathrm{bol}} \sim 1000$\,\lsun\,in total \citep{crimier2009} among FIR 3, 4, and 5. FIR\,4 was identified as a single peaked source first from the 1.3\,mm continuum observations \citep{chini1997}, then the interferometric observations have spatially resolved the internal structure consisting of several millimeter sources \citep{shimajiri2008}. \cite{furlan2016} reported that two protostellar candidates, HOPS-108 (Class 0) and HOPS-64 (Class I), are associated with FIR\,4. Due to the large luminosity and complexity of the structure, the origin of the star formation environment in FIR\,4 has been discussed in previous studies.
\cite{shimajiri2008} proposed that an energetic outflow driven by HOPS-370, which is a Class I protostellar source located at the peak position of FIR\,3 \citep{furlan2016, tobin2020b}, has collided with the dust condensation, FIR\,4. HOPS-370 has a bolometric luminosity of 360\,\lsun, a stellar mass of 2.5\,\msun, and a disk radius of 94\,au  \citep{furlan2016, tobin2020b}. The energetic outflow has been observed at various wavelengths \citep{yu1997, Aso2000, stanke2002, williams2003, takahashi2008, shimajiri2008, stutz2013, gonzalez2016, osorio2017, nakamura2019, tanabe2019, feddersen2020, tobin2020b}. \cite{shimajiri2008} suggested that the outflow interaction has impacted the star formation process within FIR\,4. \cite{osorio2017} and \cite{nakamura2019} also supported the outflow interaction scenario with their independent observing data sets.
On the other hand, \cite{lopez2013} suggested a presence of a B-type star within FIR\,4 in order to explain the large bolometric luminosity.
In contrast, \cite{fontani2017} and \cite{favre2018} proposed another scenario using foreground radiation to explain the large bolometric luminosity. 
It is crucial to obtain high angular resolution and wide field imaging to disentangle the proposed scenarios.
In addition, other groups have suggested the presence of strong UV radiation, possibly emitted from the interior of FIR\,4 \citep{lopez2013} or radiated by the foreground region \citep{fontani2017, favre2018}. 
This radiation may be important for explaining the origin and environment of FIR\,4. 

We report ALMA 12-m-array and ACA 7-m-array observations toward FIR\,3, 4, and 5 within the OMC-2 region. 
In this paper, we define FIR\,3, FIR\,4, and FIR\,5 as the FIR\,3 region, FIR\,4 region, and FIR\,5 region, respectively, as we discuss substructures spatially resolved both within and around the sources.
Our primary goals are to identify protocluster members and to study their physical properties. 
We observed the 1.3-mm continuum emission and CO\,($J$ = 2--1) emission to identify dust sources and molecular outflows, respectively. We simultaneously observed SiO\,($J$ = 5--4) emission tracing gas originating from outflows and other shocked regions.

We describe the observations, data reduction, and imaging methods in Section\,\ref{obs}.
In Section\,\ref{result}, we present synthesized images of the 1.3\,mm continuum, CO\,($J$ = 2--1), and SiO\,($J$ = 5--4) emissions to identify individual fragmented sources, molecular outflows, and shocked regions, respectively.
In Section\,\ref{dis}, we discuss the possibility of outflow interactions within the protocluster and how this possibly affects star formation within the embedded protocluster. Based on the continuum data, we also discuss a hub-filament system whose center is located within the FIR\,4 region. Finally, we summarize this study in Section\,\ref{summary}.

\section{observations}
\label{obs}
\begin{figure}
\centering
    \includegraphics[width=15cm]{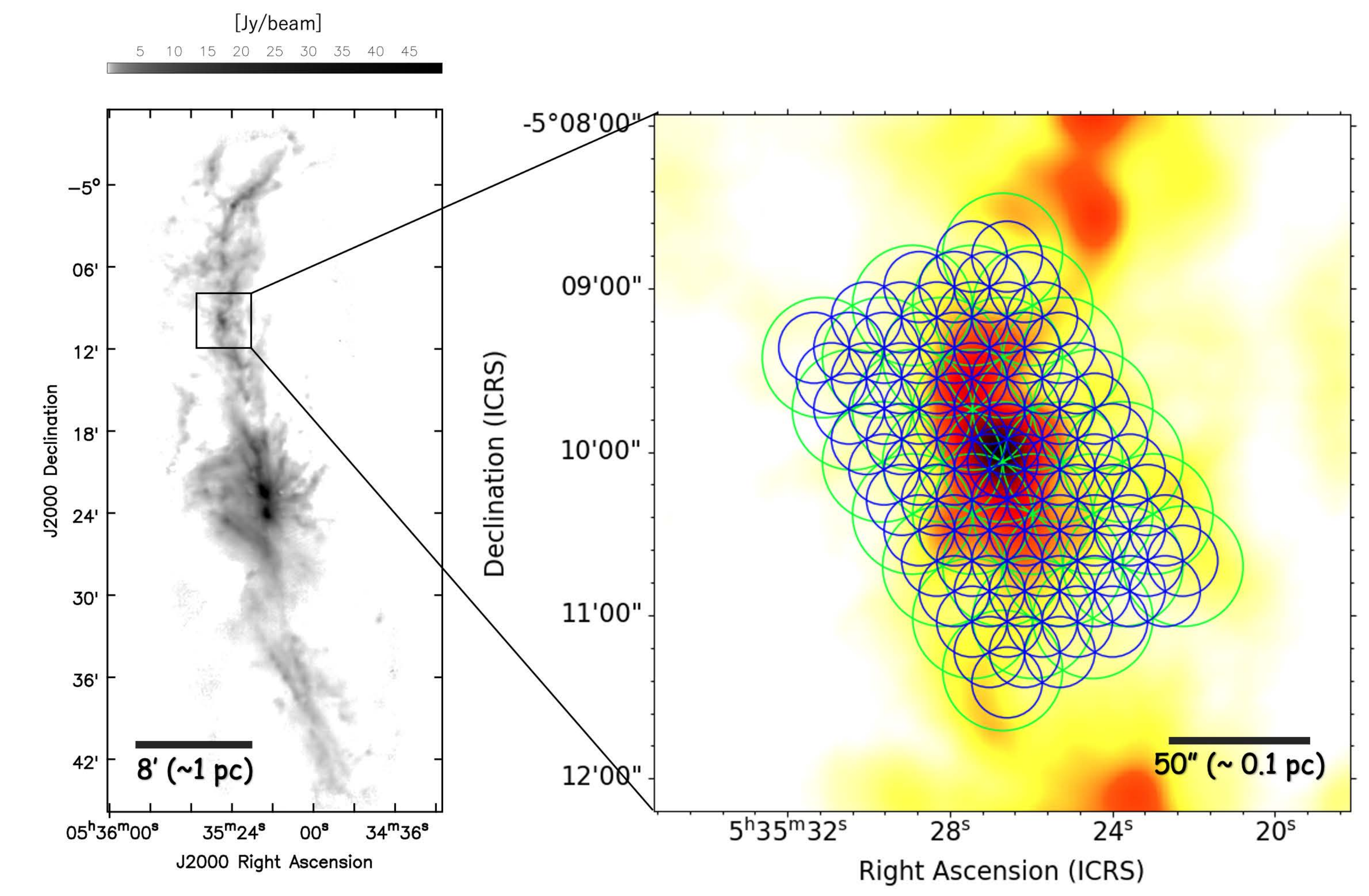}
    \caption{Left: 850\,\micron\,continuum image of the northern end of the Integral Shaped Filament taken with the JCMT/SCUBA \citep{johnstone1999}. Right: Observing fields of the ALMA 12-m array (blue circles: 92 pointings) and the ACA 7-m array (green circles: 32 pointings).}
     \label{field}
\end{figure}

ALMA observations of the 1.3-mm continuum, CO\,($J$ = 2--1; 230.53797\,GHz) emission, and SiO\,($J$ = 5--4; 217.10498\,GHz) emission were carried out with the ALMA 12-m array on 2018 April 19 and with the ACA 7-m array (Morita array) on 2018 April 19, 20, and 23 (project code: 2017.1.01353.S; PI: S. Takahashi). 
In order to cover the OMC-2/FIR\,3, 4, and 5 regions, we mapped a $ 3'.08 \times 2'.23 $ area (32 fields) with the ACA 7-m array and a $ 2'.92 \times 1'.92 $ (92 fields) area with the 12-m array with Nyquist sampling (see Figure\,\ref{field}). The field center of the map was set to  R.A. = $05^{h}35^{m}26^{s}.7140$, Dec. = $-05 \tcdegree 10'03''.800$. 
The on-source times per field of the ACA 7-m array and 12-m array were 540\,seconds and 40\,seconds, respectively. 
A 937.5 MHz wide dual polarization spectral window (244.14 kHz resolution), centered at the frequency of the CO\,($J$ = 2--1) and SiO\,($J$ = 5--4) lines, was placed in two of the four basebands. 
The velocity resolution in these windows is 0.32\,$\mathrm{km}\,\mathrm{s^{-1}}$ for CO\,($J$ = 2--1) and 0.34\,$\mathrm{km}\,\mathrm{s^{-1}}$ for SiO\,($J$ = 5--4).
The other two basebands (two spectral windows), 1875 MHz wide, were allocated to the continuum observations. 
In addition to the two basebands allocated for the continuum emission, the channels that detected no molecular line emissions from the basebands allocated for CO and SiO were used to produce the continuum image. 
After subtracting the line emissions, the line-free emission channels provided a total effective continuum bandwidth of 3.4 GHz for the datasets of both arrays.
The datasets cover the projected baseline ranges between 7\,m and 47\,m for the ACA 7-m array and between 12\,m and 423\,m for the ALMA 12-m array, and are hence insensitive to structures extending to more than $23''.0$  for the ACA 7-m array and $13''.4$ for the ALMA 12-m array at the 10\,\% level of the total flux density. Here, $\theta_{\mathrm{MRS}} \approx {0.6 \lambda}/{L_{\mathrm{min}}}$ was used for the estimation (ALMA Technical handbook), where $ \theta_{\mathrm{MRS}} $ is the maximum recoverable size in radians, $\lambda$ is the observing wavelength in meters, and $L_{\mathrm{min}}$ is the shortest baseline in meters.
Details of the observation parameters including further information such as the calibrators and observing conditions are summarized in Table \ref{para-1}. 

The data were calibrated using the Common Astronomy Software Application (CASA: \citealt{CASA2022}; \url{https://arxiv.org/abs/2210.02276}) version 5.4.0 with the ALMA pipeline. The data were imaged with CASA version 5.6.1 and 5.6.2.
The final 1.3\,mm continuum image and the data cubes for the molecular line observations were produced using a CASA task ``tclean''. 
Robust weighting with a Briggs parameter of 0.5 was used for both the continuum and molecular line images.
The synthesized beams and noise levels ($1\sigma $) obtained for the continuum images were $1''.13 \times 0''.65$ (P.A. = $-$68 deg.) and $\sim 0.22$\,\mjy\,for the ALMA 12-m array, and $9''.63 \times 4''.19$ (P.A. = $-$76 deg.) and $ \sim 0.78$\,\mjy\,for the ACA 7-m array.
The CO and SiO data cubes were produced using a velocity resolution of 
5.0\,$\mathrm{km}\,\mathrm{s^{-1}}$ and 1.0\,$\mathrm{km}\,\mathrm{s^{-1}}$ for the ALMA 12-m array. 
The synthesized beams obtained for the CO and SiO data cubes for the ALMA 12-m array were $1''.19 \times 0''.74$ (P.A. = $-$68 deg.) and $1''.26 \times 0''.79$ (P.A. = $-$69 deg.), respectively.
Table \ref{para-2} summarizes the achieved angular resolutions and noise levels, with the figure numbers indicating the image with the corresponding data set.
After some trials, we decided to present the ALMA 12-m array and the ACA 7-m array images separately, rather than combine the both data in order to demonstrate the most compact and the most extended structures using individual images.

\begin{table}
	\centering
	\caption{Observing parameters}
	\label{para-1}
	\small
	\begin{tabular}{l|ll} 
		\hline
		Parameters & ACA 7m-array & ALMA 12m-array\\
		\hline
		\hline
		Observation date (YYYY-MM-DD) & 2018-04-19, 2018-04-20, 2018-04-23& 2018-04-19 \\
		Number of antennas & 12 & 44 \\
		Mapping center (ICRS) & \multicolumn{2}{c}{ $05^{h}35^{m}26^{s}.7140$, $-05 \tcdegree 10'03''.800$ } \\
		Mapping area & $ 3'.1 \times 2'.2 $ & $ 2'.9 \times 1'.9 $ \\
		Number of pointings & 32 & 92 \\
		Primary beam size at 230.53 GHz (arcsec) & 44 & 26 \\
		Continuum representative frequency (GHz) & \multicolumn{2}{c}{231} \\
		Continuum effective bandwidth (GHz) & \multicolumn{2}{c}{3.4} \\
		PWV (mm)& 1.4 -- 3.0 & 1.9 -- 2.3 \\
		Phase rms$^{a}$ (deg.) & 8.4 -- 17.5 & 8.6 -- 21.6 \\
		Projected baseline coverage (m) & 7 -- 47 &
		12 -- 493 \\
		Maximum recoverable size (arcsec) & 23 & 13 \\
		On-source time per field (second) & 540 & 40 \\
        System temperature (K) & 80 -- 120 & 70 --170  \\
		Flux calibrator & J0522-3627 & J0522-3627 \\
		Bandpass calibrator & J0522-3627 & J0522-3627 \\
		Phase calibrator & J0542-0913, J0607-0834 & J0541-0541 \\
		\hline
	\end{tabular}
	\tablenotetext{a}{Antenna-based phase differences on phase calibrator at the baseline of Q4 (phase rms values measured in the fourth quartile of the baseline length) }
\end{table}

\begin{table}
	\centering
	\caption{Summary of the imaging parameters}
	\label{para-2}
	\small
	\begin{tabular}{lcccccc} 
		\hline
		\hline
		Dataset & Array & Briggs& \multicolumn{1}{c}{Synthesized beam, P.A.}& Noise level & Velocity resolution & Figures\\
		& &Weight &\multicolumn{1}{c}{[arcsec $\times$ arcsec, deg.]} & [\mjy] & [\kms]\\
		\hline
		1.3\,mm continuum & 12\,m& 0.5 & $1.13 \times 0.65$, -68  & 0.22 & - & \ref{cont-3}, \ref{cont-2}, \ref{dendro-1}, \ref{dis-1}, \ref{dis-4} \ref{appendix-comp}\\	
		1.3\,mm continuum & 7\,m & 0.5 & $9.63 \times 4.19$, -76 & 0.78 & - & \ref{cont-1}, \ref{dis-4}\\
		$^{12}$CO\,($J$=2--1) & 12\,m& 0.5 & $1.19 \times 0.74$, -68 & 5.3 & 5.0 &\ref{LINE-1} -- \ref{flow1-3}, \ref{flow-11}, \ref{flow-12}, \ref{dis-1}, \ref{app-ch-5kms}\\
		SiO\,($J$=5--4) & 12\,m& 0.5 & $1.26 \times 0.79$, -69 & 4.8  & 5.0 & \ref{app-ch-5kms}\\
		$^{12}$CO\,($J$=2--1) & 12\,m& 0.5 & $ 1.19 \times 0.74$, -68 &  10.0& 1.0 & \ref{flow4-7} -- \ref{flow-10}, \ref{interact-1}, \ref{app-ch-1kms}\\
		SiO\,($J$=5--4) & 12\,m& 0.5 &$ 1.26 \times 0.79$, -69 & 9.0 & 1.0 & \ref{LINE-1} -- \ref{flow-9}, \ref{flow-11}, \ref{interact-1},  \ref{dis-1}, \ref{app-ch-1kms}\\
		\hline
	\end{tabular}
\end{table}

\section{results}
\label{result}
\subsection{1.3\,mm Continuum Emission}
\label{result-cont}
\begin{figure}
\centering
    \includegraphics[width=13cm]{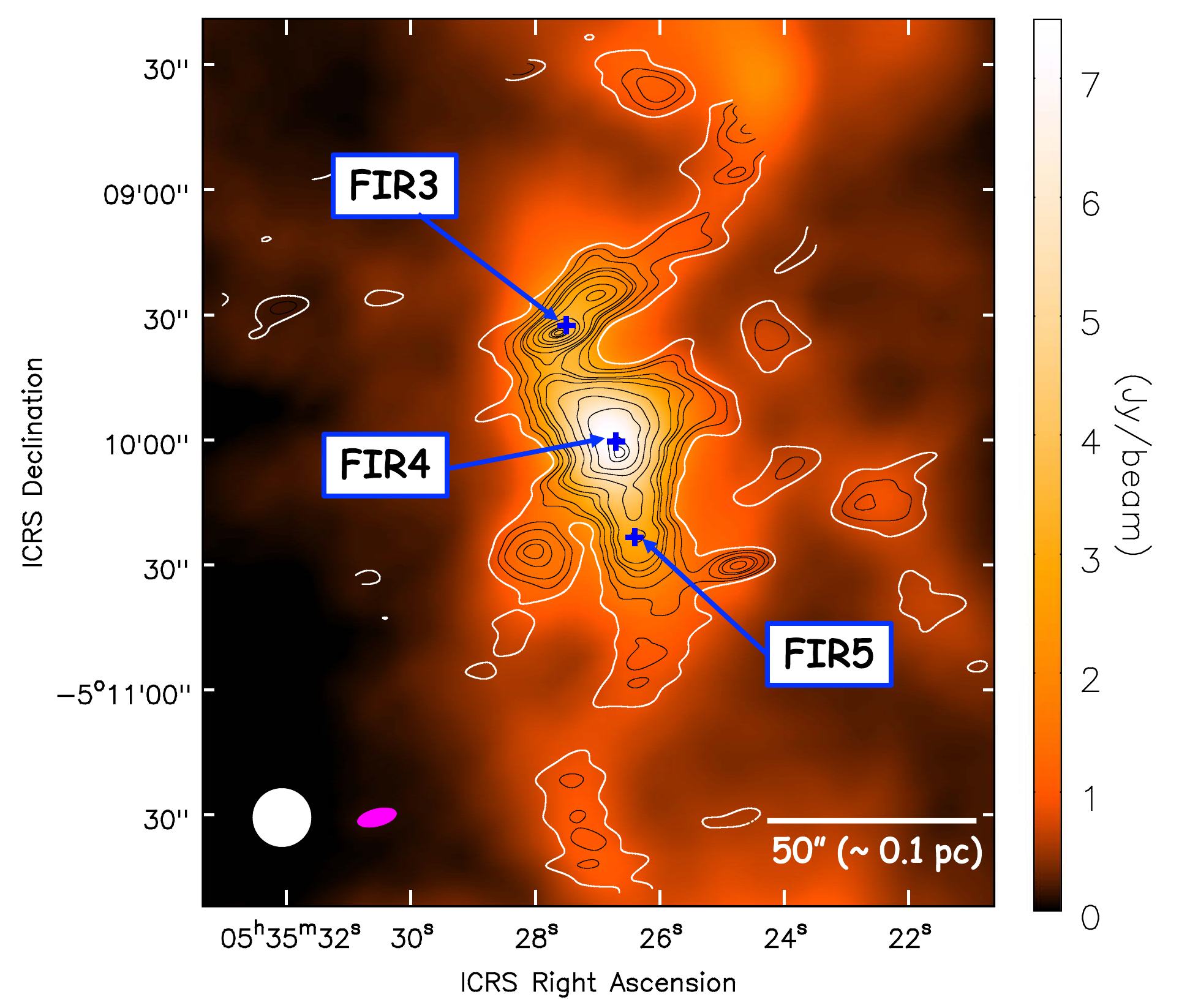}
    \caption{ACA 7-m array 1.3\,mm continuum image (white and black contours, magenta beam ellipse) overlaid with JCMT/SCUBA 850\,\micron\,continuum image (color scale and white beam ellipse, \citealt{johnstone1999}). The white contour level is 10$\sigma$ and the black contour levels are [20, 30, 50, 60, 80, 130, 180, 230, 300, 320, 340] $\times 1\sigma$ ($1\sigma = 0.78$\,\mjy). The crosses denote locations of FIR\,3, 4, and 5 \citep{chini1997}.}
     \label{cont-1}
\end{figure}

Figure\,\ref{cont-1} shows that the spatial distribution of the 1.3\,mm continuum image taken with the ACA 7-m array (white and black contours) is consistent with the 850\,\micron\,continuum image taken with the JCMT/SCUBA (color scale, \citealt{johnstone1999}).
The ACA 7-m array 1.3\,mm continuum image shows several substructures within the 1.3\,mm sources, FIR\,3, 4, and 5, previously identified by \cite{chini1997}.
In addition, we detected two single peaked bright components with the ACA 7-m array, with peaks at R.A. = $05^{h}35^{m}28^{s}.047$, Dec. = $-05 \tcdegree 10'26''.474$ and R.A. = $05^{h}35^{m}24^{s}.729$, Dec. = $-05 \tcdegree 10'30''.145$.

\begin{figure}
    \centering
    \includegraphics[width=15cm]{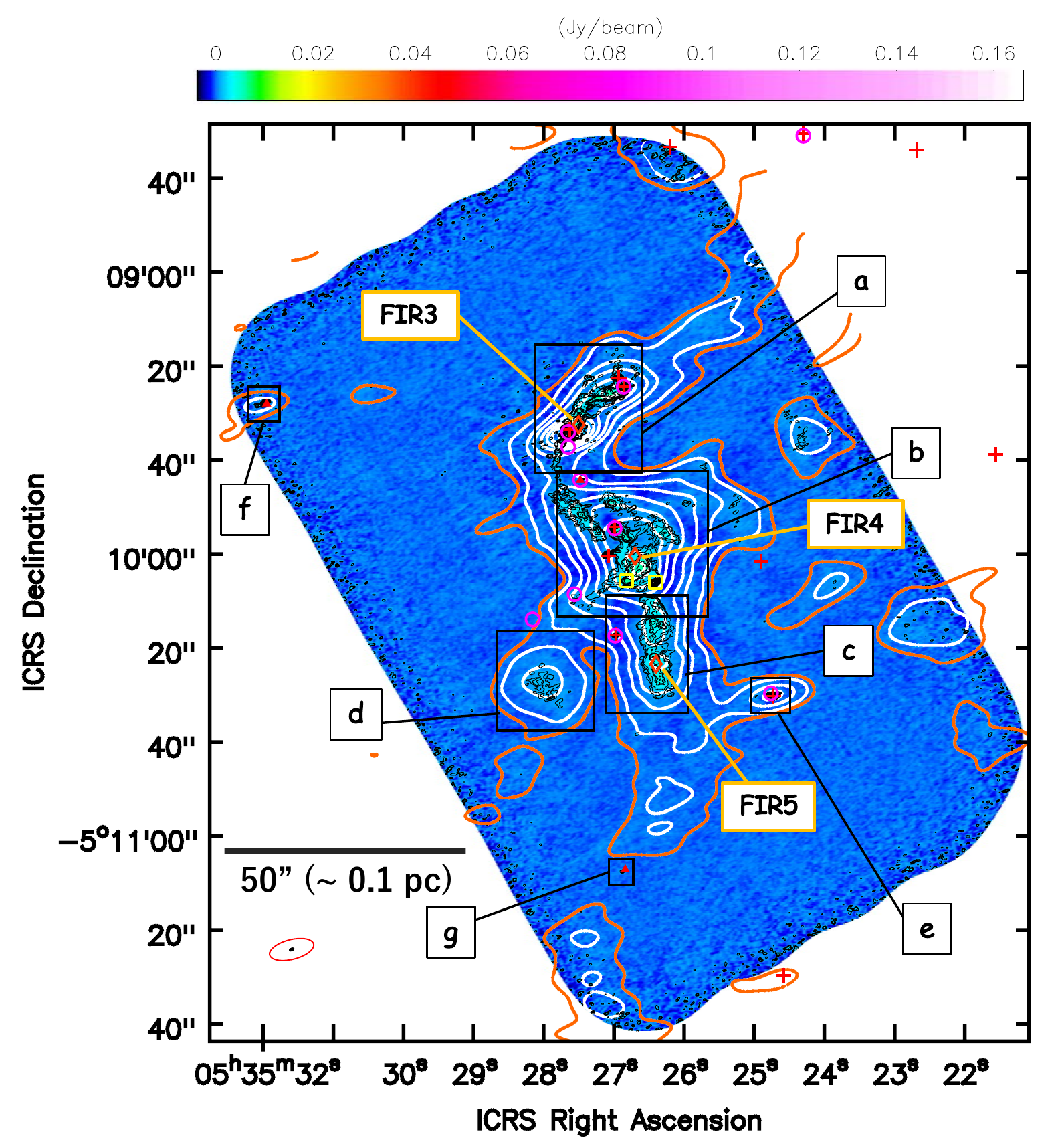}
    \caption{Continuum image by the ALMA 12-m array (color and black contours) and ACA 7-m array (orange and white contours).  The noise levels ($1\sigma$) for the 12-m array and 7-m array are 0.22\,\mjy and 0.78\,\mjy, respectively. The black contour levels are [5, 7, 10, 15, 17, 20, 25, 32, 40, 50, 60, 80, 100, 150,170, 220, 300, 500, 700] $\times 1\sigma$ of the ALMA 12-m array. The orange contour level is 10$\sigma$ and the white contour levels are [20, 24, 50, 80, 130, 180, 230, 300, 320, 340] $\times 1\sigma$ of the ACA 7-m array. The synthesized beams of the 12-m array and 7-m array are denoted by black filled and red open ellipses in the bottom-left corner. The symbols represent HOPS sources (red crosses, \citealt{furlan2016}), TIMMI2 mid-infrared sources (open pink circles, \citealt{nielbock2003}), Spitzer disk sources (red triangles, \citealt{megeath2012}), VLA sources (yellow squares, \citealt{osorio2017}), and the central positions of the FIR regions (red diamonds, \citealt{chini1997}).}
    \label{cont-3}
\end{figure}

\begin{figure}
 \begin{minipage}[b]{0.3\linewidth}
  \centering
     \includegraphics[width=17cm]{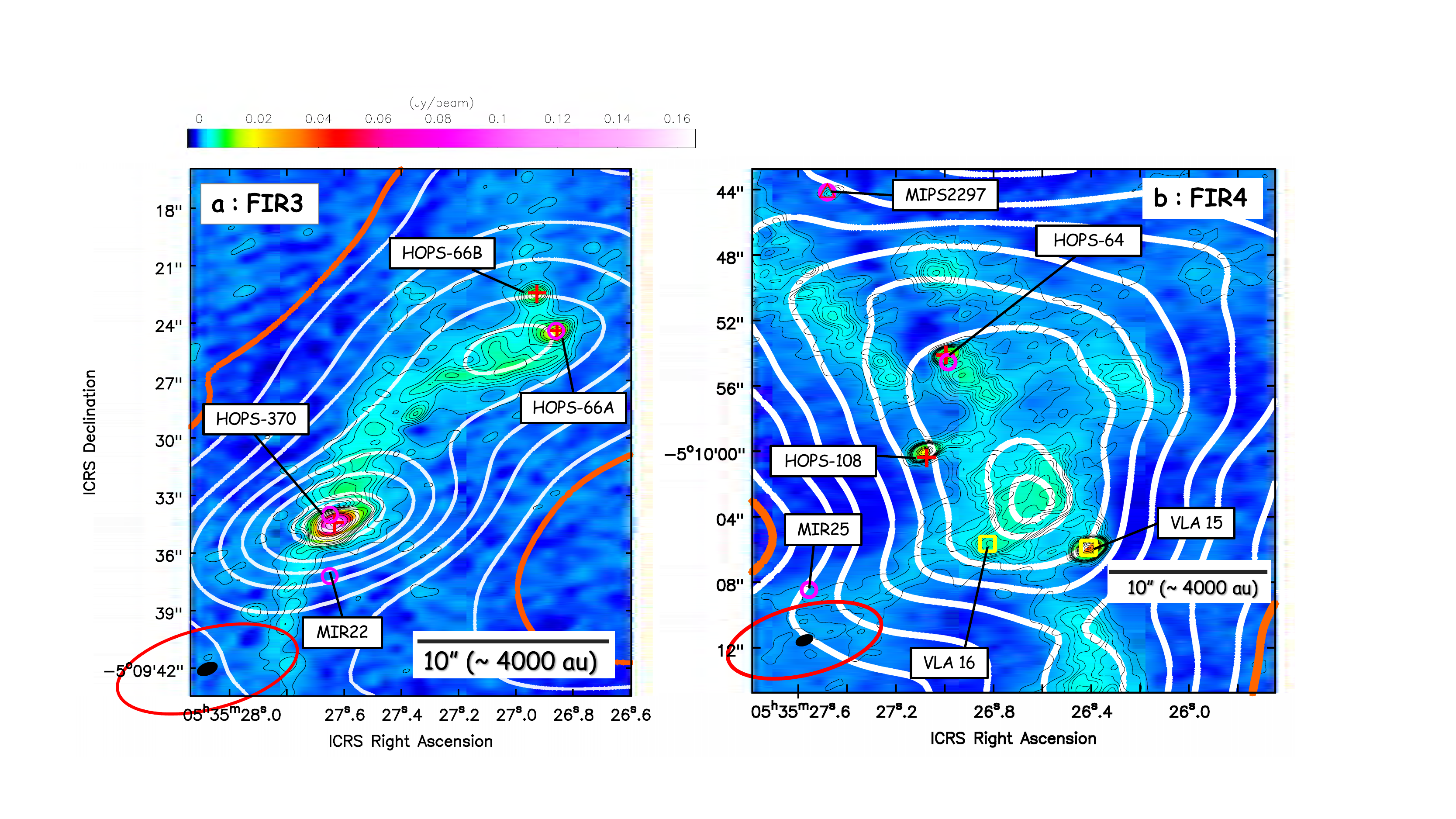}
 \end{minipage}\\
 \begin{minipage}[b]{0.3\linewidth}
  \centering
  \includegraphics[width=17cm]{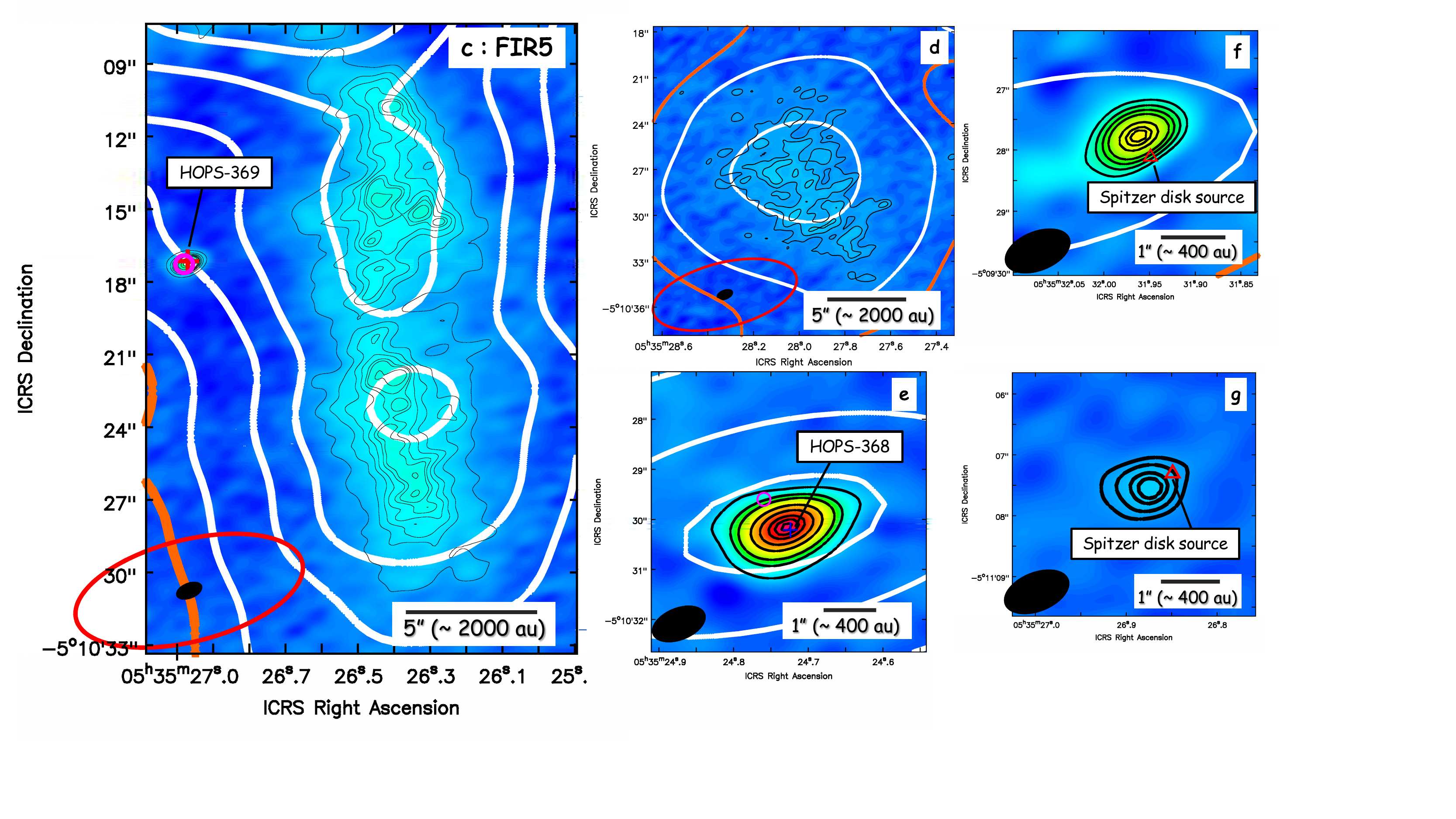}
 \end{minipage}
 \caption{Zoomed-in images of regions (a)--(g) in Figure\,\ref{cont-3}.  The orange and white contour levels and the color scale in all figures are the same as those in Figure\,\ref{cont-3}.  The black contour levels in Figure\,(a) are [5, 10, 15, 20, 25, 30, 40, 50, 70, 100, 150, 200, 300, 500, 700] $\times 1\sigma$ ($1\sigma = 0.22$\,\mjy). The black contour levels in Figure\,(b) are [3, 6, 10, 13, 15, 17, 20, 22, 25, 30, 50, 60, 68, 70, 80, 100, 120, 150, 175] $\times 1\sigma$ ($1\sigma = 0.22$\,\mjy). The black contour levels in Figure\,(c) are [5, 10, 13, 15, 17, 18, 20, 25, 30, 33] $\times 1\sigma$ ($1\sigma = 0.22$\,\mjy). The black contour levels in Figure\,(d) are [3, 5, 7] $\times 1\sigma$ ($1\sigma = 0.22$\,\mjy). The black contour levels in Figure\,(e) are [10, 30, 50, 100, 150, 200, 230, 250] $\times 1\sigma$ ($1\sigma = 0.22$\,\mjy). The black contour levels in Figure\,(f) are [20, 30, 40, 50, 70, 80, 85] $\times 1\sigma$ ($1\sigma = 0.22$\,\mjy). The black contour levels in Figure\,(g) are [3, 5, 7, 8] $\times 1\sigma$ ($1\sigma = 0.22$\,\mjy). The black filled and red open ellipses are the same as those in Figure\,\ref{cont-3}.}
 \label{cont-2}
 \end{figure}

Figures\,\ref{cont-3} and \ref{cont-2} show the 1.3\,mm continuum image obtained from the ALMA 12-m array.
The spatial resolution of the 1.3\,mm continuum for the ALMA 12-m array is about seven times higher than that for the ACA 7-m array. 
The ALMA 12-m array provides further spatial resolution of the continuum emission detected with the ACA 7-m array.
The ALMA 12-m array also detected structures associated with continuum sources detected in previous works such as at 3\,mm continuum sources detected by \citet{kainulainen2017} and \citet{vanterwisga2019} and 0.87\,mm continuum sources detected by \citet{tobin2019}.
The spatial resolution of our 1.3\,mm continuum image obtained from the ALMA 12-m array is approximately three times higher than that of the 3\,mm continuum in previous mosaic mapping observations \citep{kainulainen2017, vanterwisga2019}.
\cite{tobin2019} observed 0.87 mm continuum emissions toward known IR sources with $0''.1$ ($\sim 40$\,au) resolution, which is approximately ten times higher than that of our ALMA 12-m array image, while the observations by \cite{tobin2019} are single pointing toward individual protostellar sources.
With mosaic mapping, we were able to image structures in the region where \cite{tobin2019} did not cover.
For example, relatively luminous sources were detected in our ALMA 12-m array image outside the FIR\,3, 4, and 5 regions in addition to those in the densest part of the regions.
We also detected fainter structures than those previously detected by \cite{kainulainen2017}, \cite{vanterwisga2019} and \cite{tobin2019}, as our observations have approximately two orders of magnitude better dust mass sensitivity compared to their observations (i.e., 3$\sigma$ dust mass sensitivity of $3.4 \times 10^{-5}$\,\msun\,assuming $T$=15 K; see Section\,\ref{result-cont-mass}). 
Here, the dust mass sensitivity was estimated per corresponding an observed beam.
Finally, we successfully resolved substructures such as filamentary and compact structures, including those not previously identified. 
In the following, 
we describe more details of the individual regions presented in Figure \ref{cont-2}a--\ref{cont-2}g.
Identification of the described sources below are made with Astronomical Dendrograms \citep{rosolowsky2008} as introduced in Section\,\ref{result-cont-id}

\textbf{FIR\,3 region} (Figure\,\ref{cont-2}a): Three continuum sources were detected with the ALMA 12-m array. 
One of the detected sources is associated with a Class I source with a peak flux of 156\,\mjy, HOPS-370.
In previous works (e.g., \citealt{nielbock2003}), HOPS-370 was identified as a member of a binary system (the two pink circles around HOPS-370 in Figure\,\ref{cont-2}a). 
However, we only detected the northern source in the 1.3\,mm continuum emission.
This result is consistent with a recent ALMA study using 0.87\,mm continuum observations (e.g., \citealt{tobin2019}).
The other two sources are for a binary system, HOPS-66A/B, which has a flat spectrum (Class flat categorized in \citealt{furlan2016}) corresponding to the Class I/\two \,boundary.
HOPS-66 were identified as a single source by the infrared observations \citep{furlan2016}.
The following up 0.87\,mm and 9\,mm continuum observations, however, resolved out the single source and newly identified is as a binary system, HOPS-66A/B \citep{tobin2019}.

\textbf{FIR\,4 region} (Figure\,\ref{cont-2}b): Within a single peaked structure imaged with the ACA 7-m array (white contours), we detected substructures with the ALMA 12-m array (black contours).
They are part of filamentary structures within which several compact sources are embedded. 
Although the 1.3\,mm continuum emission obtained from the ACA 7-m array shows a single peak, there is no corresponding compact peaked component in the higher-resolution 12-m array image.
Instead of a single peak, an extended fluffy structure was detected with the ALMA 12-m array (black contours).
The four most compact 1.3\,mm continuum sources are associated with previously identified sources: two IR sources (HOPS-64 and HOPS-108; \citealt{furlan2016}) and two centimeter sources (VLA15 and VLA16; \citealt{osorio2017}).
These four continuum sources were identified in previous ALMA continuum observations by \cite{kainulainen2017}, \cite{vanterwisga2019}, and \cite{tobin2019}.

\textbf{FIR\,5 region} (Figure\,\ref{cont-2}c): Two extended filamentary structures were detected with the ALMA 12-m array. No compact sources were detected within the filamentary structures.
This result is different from those for both the FIR\,3 and FIR\,4 regions described above. A compact 1.3\,mm continuum source associated with HOPS-369 \citep{furlan2016} was detected outside the filamentary structure.

\textbf{Figure\,\ref{cont-2}d}: 
With the ACA 7-m array, we detected a single-peak structure ($\sim$ 4000\,au scale), which has already been reported in previous ALMA observations (e.g., \citealt{kainulainen2017, hacar2018, zhang2020}).
Our ALMA continuum observations spatially resolved patchy substructures within the single-peak structure corresponding to the 3$\sigma$ to 8$\sigma$ emission levels, which is the first time these substructures have been resolved. 

\textbf{Figure\,\ref{cont-2}e--\ref{cont-2}g}: We detected three compact sources associated with previously detected IR sources (HOPS-368 and two Spitzer disk sources) outside  the FIR\,3, 4, and 5 regions.

\vspace{0.5cm}

To evaluate the concentration of the structures, we defined $C_{\mathrm{(ACA/JCMT)}}$ as the ratio between the flux measured by the ACA 7-m array and that measured by the JCMT after taking account of the beam size and wavelength differences, and $C_{\mathrm{(12m/ACA)}}$ as the ratio between the flux measured by the ALMA 12-m array and that measured by the ACA 7-m array after taking account of the beam size differences.
 The structure concentration factor $C$ indicates the flux recovery rate for comparing two different experiments. 
For $C_{\mathrm{(ACA/JCMT)}}$ presented in Table\,\ref{concent-1}, we estimated ratio ranges of between 31\,\% and 50\,\%. The number did not vary much across the regions.
Here, we assumed spectral index $\beta = $ 1.5--2.1 listed in Table\,\ref{concent-1} to estimate the extrapolated flux at 1.3\,mm.
For $C_{\mathrm{(12m/ACA)}}$ presented in Table\,\ref{concent-2}, the estimated ratio varies depending on the regions (30\,\%--100\,\%). 
$C_{\mathrm{(12m/ACA)}}$ obtained toward HOPS-368 (a relatively isolated Class I source) was $\sim$ 100\,\%, indicating no missing flux. 
This can be interpreted as indicating that the emission is mostly due to compact emission associated with a compact dusty disk and that the extended emission from the envelope or core can be ignored. 
For the same reason, the FIR\,3 region, including one Class I source and a binary system of flat sources, shows a relatively high flux-recovery rate ($\sim 66\,\%$).  
In contrast, the FIR\,4 region, FIR\,5 region, and the region presented in Figure\,\ref{cont-2}d show small flux-recovery rates of $C_{\mathrm{(12m/ACA)}}$ = 31\,\%--44\,\%.
These regions contain Class 0 and prestellar sources embedded within the filamentary structures.
They are considered to be young compared with HOPS-368 and the sources in the FIR\,3 region.
Thus, the extended emission from the envelope and core remains.

\begin{table}[]
    \centering
    \caption{Integrated flux density in the selected area ($F_{\nu}$) and concentration factor ($C_{\mathrm{(ACA/JCMT)}}$). The beam sizes of the data for the ACA 7-m array 1.3\,mm continuum was adjusted to $\sim 14''$ using the CASA task ``imsmooth'' (sixth column). All the cells in the region over $10\sigma$ for the ACA 7-m array continuum were selected (white and orange contours in Figure\,\ref{cont-1} and \ref{cont-2}, respectively), and $F_{\nu}$ was calculated using them. $C_{\mathrm{(ACA/JCMT)}}$ is the ratio of $F_{\nu}$ for the ACA 7-m array 1.3\,mm continuum (beam$\sim 14''$) to $F_{\nu}$ for the extrapolated 850\,\micron\,continuum.}
    
    \begin{tabular}{rr|rrrrr}
    \footnotesize
    Region & Fig. & $F_{\mathrm{850\,\mu m}}$ & extrapolated flux at 1.3\,mm & Sp. index  & $F_{\mathrm{1.3\,mm}}$ for the ACA 7-m array& $C_{\mathrm{(ACA/JCMT)}}$  \\
    \,& \,& [mJy] & [mJy] $^{[1]}$ &  ($\beta$) &[mJy] (JCMT beam)&  [$\%$]\\
    \hline
    \hline
    FIR\,3& \ref{cont-2}a & 9030 & 1875 & 1.7 $^{[2]}$ & 874 & 46.6  \\
    FIR\,4& \ref{cont-2}b &  20930 & 4731 & 1.5 $^{[2]}$ & 2120  & 44.8  \\
    FIR\,5& \ref{cont-2}c & 8770 & 1982 & 1.5 $^{[2]}$ & 691 & 34.9   \\
     -  & \ref{cont-2}d & 2060 & 466 & 1.5 $^{[2]}$ & 143 & 30.7  \\
    HOPS-368  & \ref{cont-2}e & 632 & 111 & 2.1 $^{[3]}$ & 55 & 49.7  \\
    \hline
    \end{tabular}
    \tablenotetext{1}{calculated using $F_{\nu} \propto \nu^{\beta + 2}$}
    \tablenotetext{2}{\cite{sadavoy2016}}
    \tablenotetext{3}{\cite{tobin2019} (index with 0.87 -- 0.9\,mm)}
    \label{concent-1}
\end{table}

\begin{table}[]
    \centering
    \caption{Integrated flux density in the selected area ($F_{\nu}$) and concentration factor ($C_{\mathrm{(12m/ACA)}}$). The beam sizes of the data for the ALMA 12-m array 1.3\,mm continuum was adjusted to $\sim 9''$ using the CASA task ``imsmooth'' (fourth column). All the cells in the region over $10\sigma$ for the ACA 7-m array continuum were selected (white and orange contours in Figure\,\ref{cont-1} and \ref{cont-2}, respectively), and $F_{\nu}$ was calculated using them. $C_{\mathrm{(12m/ACA)}}$ is the ratio of $F_{\nu}$  for the ALMA 12-m array 1.3\,mm continuum (beam$\sim 9''$) to $F_{\nu}$ for the ACA 7-m array 1.3\,mm continuum (original data). }
    
    \begin{tabular}{rr|rrr}
    \footnotesize
    Region & Fig. & $F_{\mathrm{1.3\,mm}}$ for the ACA 7-m array& $F_{\mathrm{1.3\,mm}}$ for the ALMA 12-m array& $C_{\mathrm{(12m/ACA)}}$   \\
    \,& \,&  [mJy] &  [mJy] (ACA beam)  & [$\%$]\\
    \hline
    \hline
    FIR\,3& \ref{cont-2}a &  919 & 610 & 66.4 \\
    FIR\,4& \ref{cont-2}b  & 2220 & 709 & 31.9 \\
    FIR\,5& \ref{cont-2}c &  752 & 318 & 42.3 \\
     -  & \ref{cont-2}d &  166 & 74 & 44.5 \\
    HOPS-368  & \ref{cont-2}e &  76 & 80 & 104.7 \\
    \hline
    \end{tabular}
    \label{concent-2}
\end{table}

\subsubsection{Source Identifications Using the ALMA 12-m Array Continuum Data}
\label{result-cont-id}
\begin{figure}
 \begin{minipage}[b]{0.3\linewidth}
  \centering
\includegraphics[width=15cm]{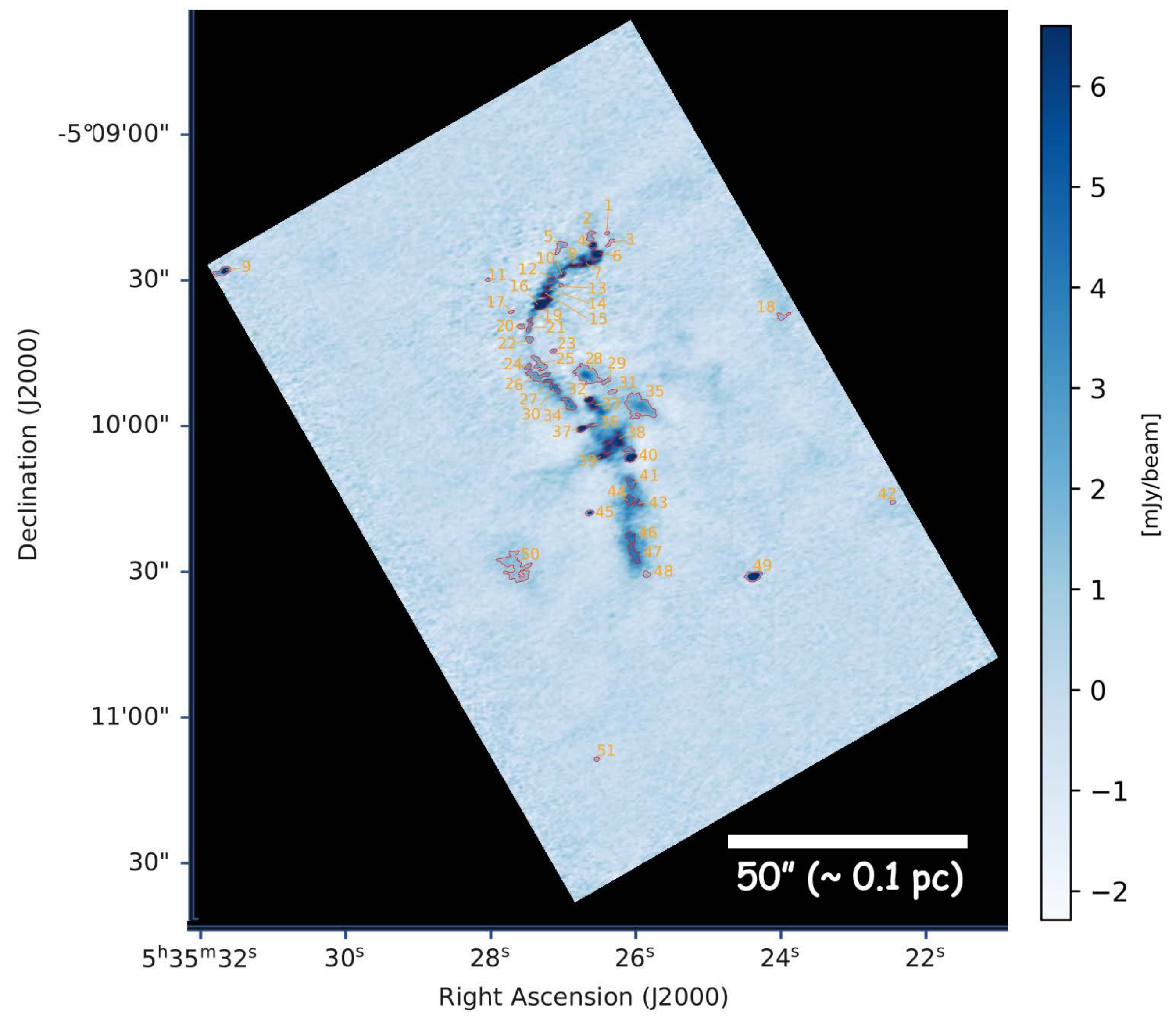}
 \end{minipage}\\
 \begin{minipage}[b]{0.3\linewidth}
  \centering
     \includegraphics[width=15cm]{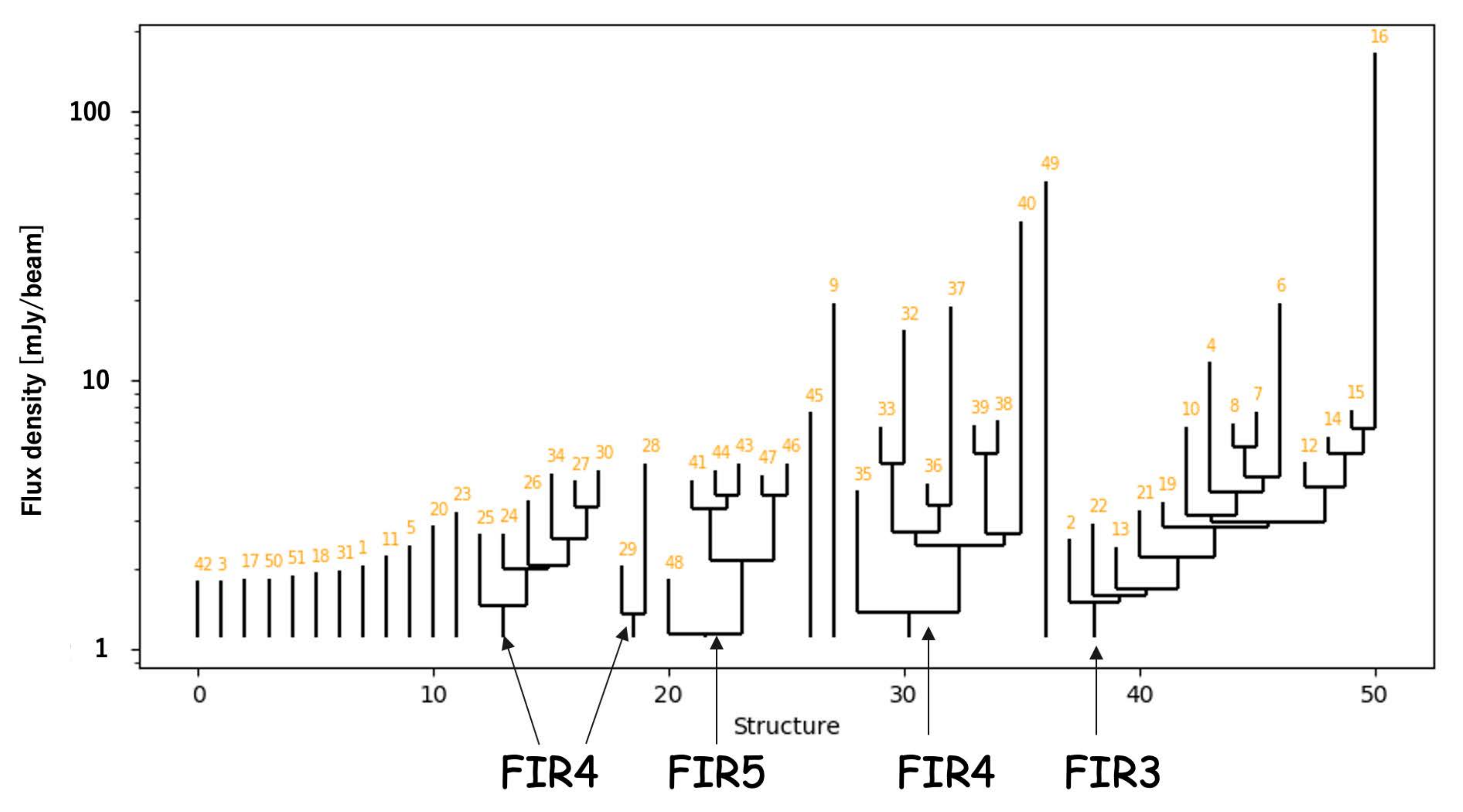}
 \end{minipage}
  \caption{Top: Location of 1.3\,mm continuum sources. 51 sources (ID = 1--51) were identified as leaves applying dendrograms to the ALMA 12-m array image. The red contours represent individual structures of continuum sources. Bottom: Tree structure of continuum sources with ID. The five tree structures belong to the FIR\,3, 4, and 5 regions respectively.}
    \label{dendro-1}
\end{figure}

In order to uniformly identify the continuum sources for our 1.3\,mm continuum image, we used Astronomical Dendrograms (hereafter dendrograms, \citealt{rosolowsky2008, goodman2009}).
dendrograms is a structure identifying algorithm package that abstracts the hierarchical structure of 2D ($p$-$p$) or 3D ($p$-$p$-$v$) data cubes into a visualized representation \citep{goodman2009}. 
dendrograms uses three types of representation: ``leaves'', ``branches'', and ``trunk''.
Trunk is the base of the hierarchical structure.
Branches are structures that split into multiple sub-structures.
Leaves do not include further substructures and are the most compact components.
We applied dendrograms to the 1.3\,mm continuum image obtained by the ALMA 12-m array with the following three criteria:
(1) the structure size is larger than the beam size ({\tt min$\_$npix}=40, 1 pixel = $0.''1$),
(2) the minimum value of a leaf is more than $5\sigma$ ({\tt min$\_$value} = $5\sigma$, $1\sigma = 0.22$\,\mjy), and (3) the peak flux density of a leaf measured from a nearby local minimum value is more than $3\sigma$ ({\tt min$\_$delta} = $3\sigma$).
Figure\,\ref{dendro-1} shows the location of identified leaves (top panel) and the tree structure of the 1.3\,mm continuum emission (bottom panel) by a dendrograms analysis. 
As shown in the top panel of Figure\,\ref{dendro-1}, dendrograms identified compact structures and faint extended structures as leaves.

Figure\,\ref{dendro-1} (bottom) shows five branches.
They are located in the FIR\,3, 4, and 5 regions indicating that the three regions contain hierarchical structures.
In this study, we used the dendrogram ``leaves'' to identify the 1.3\,mm continuum sources.
In a future paper, we will further discuss the hierarchical structures within the molecular cloud.

We identified 51 continuum sources in total. We call the set the ``Fragmented Source Catalogue in Orion'' (FraSCO) and indicate the sources within the set by numbers, ordered from north to south in R.A..
Most of the identified sources are distributed in the main filamentary structure. 
The source name, position, peak flux, total flux, and radius of the identified sources are summarized  in Table\,\ref{cont-source1}.
The source position is measured at the peak flux of the leaf. 
The total flux is measured within the identified leaf.
The radius is defined as $\sqrt{A_{\mathrm{leaf}}/\pi}$, where $A_{\mathrm{leaf}}$ is the projected area of a leaf onto the plane of the sky.
Note that dendrograms also computes the radius based on the geometric mean of the second moments along the major axis and minor axis.
For our dataset, the radius computed from the second moments is a factor of two smaller than that measured from the projected area.
The radius comparisons made in \cite{rosolowsky2008} for large scale clumps (comparisons of $\sim$pc scale structures) imply that the radii computed from the second moments are smaller than those measured in previous studies.
Although our study focuses on smaller size scale structures ($\sim$1000\,au) compared to \cite{rosolowsky2008}, the trend is consistent with what was reported in their study.  

We also compare the identified sources with the previously detected sources in Table\,\ref{cont-source2} (also see Appendix\,\ref{app-comp}).
For 15 out of the identified 51 sources, there are counterparts previously detected at other wavelengths and located within $1''$ from our identified sources.
The 15 sources are composed of three Class \two $\,\,$sources, seven Class 0/I/flat sources, and five sources previously identified with only mm/sub-mm wavelength (no classification). The five mm/sub-mm sources are considered to be still deeply embedded within the molecular core or too cold to detect clearly at infrared wavelengths.
The remaining 36 sources were identified for the first time in this study.
We also confirmed these by eye and compared them with the synthesized beam pattern. 
We made sure that the detected faint sources are not side-lobe components created by a strong continuum source such as FraSCO-16 (i.e., HOPS-370). 

\begin{table}
	\centering
	\caption{Source identifications with dendrograms. The source ID ($1$--$51$) corresponds to the number in Figure\,\ref{dendro-1}. The source names in bold represent newly identified 1.3\,mm continuum sources.}
	\label{cont-source1}
	\footnotesize
	\begin{tabular}{llllrrr} 
		\hline
		\hline
		ID & source name& R.A. (ICRS) & Dec. (ICRS)& Radius [arcsec] & Peak flux [mJy/beam] & Integrated flux [mJy]  \\
		\hline
		1&	\textbf{FraSCO-1}&$	05^{h}35^{m}26^{s}.7325$&	$-05 \tcdegree09'20''.6600$&0.36 &	2.06 	&0.75 \\
		2&	\textbf{FraSCO-2}&$	05^{h}35^{m}26^{s}.9636$&	$-05 \tcdegree09'20''.6044$&0.80 &	2.57 	&4.62 \\
		3&	\textbf{FraSCO-3}&$	05^{h}35^{m}26^{s}.6859$&	$-05 \tcdegree09'21''.9646$&0.57 &	1.79 	&1.66 \\
		4&	FraSCO-4&$	05^{h}35^{m}26^{s}.9328$&	$-05 \tcdegree09'22''.5670$&0.70 &	11.70 &	12.41\\ 
		5&	\textbf{FraSCO-5}&$	05^{h}35^{m}27^{s}.3703$&	$-05 \tcdegree09'22''.7829$&0.97 &	2.45 	&5.93 \\
		6&	FraSCO-6&$	05^{h}35^{m}26^{s}.8765$&	$-05 \tcdegree09'24''.4338$&0.99 &	19.36 &	30.84\\ 
		7&	\textbf{FraSCO-7}&$	05^{h}35^{m}27^{s}.2180$&	$-05 \tcdegree09'26''.1418$&0.89 &	7.69 	&19.11 \\
		8&	\textbf{FraSCO-8}&$	05^{h}35^{m}27^{s}.1984$&	$-05 \tcdegree09'26''.7078$&0.54 &	6.93 	&6.97 \\
		9&	FraSCO-9&$	05^{h}35^{m}31^{s}.9781$&	$-05 \tcdegree09'27''.8434$&	1.12  &	19.40 &	24.52\\ 	
		10&	\textbf{FraSCO-10}&$	05^{h}35^{m}27^{s}.3491$&	$-05 \tcdegree09'28''.6333$&	0.60 &	6.76 	&6.21 \\
		11&	\textbf{FraSCO-11}&$	05^{h}35^{m}28^{s}.3645$&	$-05 \tcdegree09'29''.5757$&	0.37 &	2.22 	&0.81 \\
		12&	\textbf{FraSCO-12}&$	05^{h}35^{m}27^{s}.4959$&	$-05 \tcdegree09'29''.5413$&	0.59 &	4.97 	&5.92 \\
		13&	\textbf{FraSCO-13}&$	05^{h}35^{m}27^{s}.3666$&	$-05 \tcdegree09'30''.7253$&0.38 	&	2.40 	&1.11 \\
		14&	\textbf{FraSCO-14}&$	05^{h}35^{m}27^{s}.5490$&	$-05 \tcdegree09'31''.2365$&0.49 &	6.21 	&5.12 \\
		15&	\textbf{FraSCO-15}&$	05^{h}35^{m}27^{s}.5914$&	$-05 \tcdegree09'32''.7387$&	0.47 &	7.77 	&5.90 \\
		16&	FraSCO-16&$	05^{h}35^{m}27^{s}.6322$&	$-05 \tcdegree09'34''.4991$&1.25 &	164.64 &	223.11\\ 
		17&	\textbf{FraSCO-17}&$	05^{h}35^{m}28^{s}.4580$&	$-05 \tcdegree09'36''.1552$&0.37 &	1.83 &	0.76\\ 
		18&	\textbf{FraSCO-18}&$	05^{h}35^{m}24^{s}.3357$&	$-05 \tcdegree09'37''.6100$&0.77 &	1.92 	&3.13 \\
		19&	\textbf{FraSCO-19}&$	05^{h}35^{m}27^{s}.7866$&	$-05 \tcdegree09'37''.7639$&0.39 &	3.54 	&1.88 \\
		20&	\textbf{FraSCO-20}&$	05^{h}35^{m}27^{s}.9120$&	$-05 \tcdegree09'39''.1008$&	0.60 &	2.88 	&2.51 \\
		21&	\textbf{FraSCO-21}&$	05^{h}35^{m}27^{s}.7985$&	$-05 \tcdegree09'39''.3549$&	0.62 &	3.30 	&3.84 \\
		22&	\textbf{FraSCO-22}&$	05^{h}35^{m}27^{s}.7952$&	$-05 \tcdegree09'41''.8262$&	0.64 &	2.94 	&3.36 \\
		23&	FraSCO-23&$	05^{h}35^{m}27^{s}.4689$&	$-05 \tcdegree09'44''.1767$&0.49 &	3.26 	&1.82 \\
		24&	\textbf{FraSCO-24}&$	05^{h}35^{m}27^{s}.8163$&	$-05 \tcdegree09'47''.3895$&	0.54 &	2.69 	&2.57 \\
		25&	\textbf{FraSCO-25}&$	05^{h}35^{m}27^{s}.6642$&	$-05 \tcdegree09'46''.7290$&	1.09 &	2.68 	&8.30 \\
		26&	\textbf{FraSCO-26}&$	05^{h}35^{m}27^{s}.6948$&	$-05 \tcdegree09'49''.1693$&1.22 &	3.58 	&14.09\\ 
		27&	\textbf{FraSCO-27}&$	05^{h}35^{m}27^{s}.5511$&	$-05 \tcdegree09'50''.3347$&0.52 &	4.23 &	3.79\\ 
		28&	FraSCO-28&$	05^{h}35^{m}27^{s}.9700$&	$-05 \tcdegree09'49''.5700$&	2.00 &	4.92 	&37.32 \\
		29&	\textbf{FraSCO-29}&$	05^{h}35^{m}26^{s}.7462$&	$-05 \tcdegree09'50''.3119$&	0.63 &	2.04 	&2.54 \\
		30&	\textbf{FraSCO-30}&$	05^{h}35^{m}27^{s}.4488$&	$-05 \tcdegree09'51''.7606$&0.85 &	4.66 	&10.68 \\
		31&	\textbf{FraSCO-31}&$	05^{h}35^{m}26^{s}.6624$&	$-05 \tcdegree09'52''.4792$&0.61 &	1.97 	&2.04 \\
		32&	FraSCO-32&$	05^{h}35^{m}26^{s}.9843$&	$-05 \tcdegree09'54''.1045$&0.70 &	15.37 &	15.73\\
		33&	\textbf{FraSCO-33}&$	05^{h}35^{m}26^{s}.9178$&	$-05 \tcdegree09'55''.3142$&	0.77 &	6.72 	&12.68 \\ 
		34&	\textbf{FraSCO-34}&$	05^{h}35^{m}27^{s}.2507$&	$-05 \tcdegree09'54''.9636$&	1.12 &	4.52 	&15.41 \\
		35&	FraSCO-35&$	05^{h}35^{m}26^{s}.2906$&	$-05 \tcdegree09'55''.4723$&	2.45&	3.88 	&51.81 \\
		36&	\textbf{FraSCO-36}&$	05^{h}35^{m}26^{s}.9453$&	$-05 \tcdegree09'59''.3577$&	0.42&	4.15 	&2.52 \\
		37&	FraSCO-37&$	05^{h}35^{m}27^{s}.8060$&	$-05 \tcdegree10'00''.1570$&	0.78&	18.81& 	19.79\\ 
		38&	\textbf{FraSCO-38}&$	05^{h}35^{m}26^{s}.6430$&	$-05 \tcdegree10'03''.1720$&	1.43&	7.08 	&45.36 \\
		39&	FraSCO-39&$	05^{h}35^{m}26^{s}.7936$&	$-05 \tcdegree10'05''.6117$&	0.70&	6.84 	&10.98 \\
		40&	FraSCO-40&$	05^{h}35^{m}26^{s}.4139$&	$-05 \tcdegree10'05''.7640$&	1.28 &	39.13 &	55.45\\ 
		41&	\textbf{FraSCO-41}&$	05^{h}35^{m}26^{s}.4059$&	$-05 \tcdegree10'11''.9170$&	0.89 &	4.26 	&11.12 \\
		42&	\textbf{FraSCO-42}&$	05^{h}35^{m}22^{s}.8239$&	$-05 \tcdegree10'15''.7530$&	0.43 &	1.79 	&0.97 \\
		43&	\textbf{FraSCO-43}&$	05^{h}35^{m}26^{s}.3173$&	$-05 \tcdegree10'15''.1954$&	0.63 &	4.93 	&6.17 \\
		44&	\textbf{FraSCO-44}&$	05^{h}35^{m}26^{s}.4303$&	$-05 \tcdegree10'14''.6234$&	0.84 &	4.64 	&11.12 \\
		45&	FraSCO-45&$	05^{h}35^{m}26^{s}.9738$&	$-05 \tcdegree10'17''.2435$&	0.66 &	7.60 	&5.87 \\
		46&	FraSCO-46&$	05^{h}35^{m}26^{s}.4099$&	$-05 \tcdegree10'22''.2276$&	0.96 &	4.93 	&14.47 \\
		47&	\textbf{FraSCO-47}&$	05^{h}35^{m}26^{s}.3680$&	$-05 \tcdegree10'25''.6531$&	1.21 &	4.46 &	22.32 \\
		48&	\textbf{FraSCO-48}&$	05^{h}35^{m}26^{s}.1940$&	$-05 \tcdegree10'29''.7729$&	0.65 &	1.82 	&2.20 \\ 
		49&	FraSCO-49&$	05^{h}35^{m}24^{s}.7303$&	$-05 \tcdegree10'30''.1727$&	1.28 &	55.09& 	74.26\\
		50&	\textbf{FraSCO-50}&$	05^{h}35^{m}27^{s}.9954$&	$-05 \tcdegree10'28''.2880$&	2.27 &	1.83 	&26.09 \\
		51&	FraSCO-51&$	05^{h}35^{m}26^{s}.8767$&	$-05 \tcdegree11'07''.5346$&	0.44 &	1.89& 	1.08\\			
		\hline
	\end{tabular}
\end{table}

\begin{table}
	\centering
	\caption{Summary of multi-wavelength counterparts with respect to the identified 1.3\,mm continuum sources.}
	\label{cont-source2}
	\scriptsize
\begin{tabular}{l|ccccccccr} 
		\hline
		\hline
		Source name & 5\,cm$^{1}$ & 3.6\,cm$^{2}$ & 3\,cm$^{1}$ & 9\,mm$^{3}$ & 7\,mm$^{1}$ &  3\,mm$^{4}$  & 870\,$\mathrm{\mu m} ^{3}$ &  Infrared$^{5, 6, 7}$ & Class$^{5}$  \\
		\hline
		FraSCO-4	&&&& detected & & & detected &	HOPS-66B &flat\\
		FraSCO-6	&&&& detected && P10	& detected & HOPS-66A/MIR20 & flat\\
		FraSCO-9	&&& & && D4 & &Spitzer disk source&II \\
		FraSCO-16	& VLA 11 & VLA  11 &VLA11&detected	 & VLA11 & P1	& detected &HOPS-370 / MIR21 & I\\
		FraSCO-23	&MIPS2297 &&MIPS2297&detected & MIPS2297 &D23		 & detected & MIR23  & II\\
		FraSCO-28	& && & &  &U7	\\
		FraSCO-32	&HOPS-64&& HOPS-64&detected &HOPS-64  & P12 & detected & HOPS-64/MIR24  & I\\
		FraSCO-35	&&& &	&&  U17	\\
		FraSCO-37	&HOPS-108 & VLA 12&HOPS-108  &detected &HOPS-108 & P15 & detected & HOPS-108 & 0\\
		FraSCO-39	& VLA 16 && VLA 16& detected & VLA 16  &	U8 & detected  \\ 
		FraSCO-40	& VLA 15 & &VLA 15 & detected & VLA 15  & U2	& detected  \\
		FraSCO-45	&&&& detected &   &	P22 & detected & HOPS-369/MIR27 & flat\\
		FraSCO-46	&&&& &  	& U20 \\
		FraSCO-49	&&VLA 13&& detected &  &  P9 &  detected &	HOPS-368/MIR28 & I\\
		FraSCO-51	&&&& & & D30  & &Spitzer disk source&II\\
		\hline 
	\end{tabular}
	\tablenotetext{1}{VLA 5\,cm, 3\,cm, 7\,mm \citep{osorio2017}}
	\tablenotetext{2}{VLA 3.6\,cm \citep{reipurth1999}}
	\tablenotetext{3}{VLA 9\,mm and ALMA 870$\,\mathrm{\mu m}$ \citep{tobin2019} }
	\tablenotetext{4}{ALMA 3\,mm sources as disk source (D), as protostar (P), and as no IR detected (U)  \citep{vanterwisga2019}}
	\tablenotetext{5}{HOPS sources \citep{furlan2016}}
	\tablenotetext{6}{mid-infrared sources \citep{nielbock2003}}
	\tablenotetext{7}{Spitzer disk sources \citep{megeath2012}}
\end{table}

\subsubsection{Dust mass}
\label{result-cont-mass}
Assuming that the 1.3\,mm continuum emission comes from optically thin dust emission and the temperature distribution of the continuum source is uniform, we can estimate the lower limit of the dust mass as
\begin{equation}
    M_{\mathrm{dust}} = \frac{F_{\mathrm{\lambda}}d^2}{\kappa_{\mathrm{\lambda}}B_{\mathrm{\lambda}}(T_{\mathrm{dust}})},
\end{equation}
where $\kappa_{\mathrm{\lambda}}$ is the mass-absorption coefficient for dust grains,
$B_{\mathrm{\lambda}} $ is the Planck function for the dust temperature $T_{\mathrm{dust}}$, $F_{\mathrm{\lambda}}$
is the total flux density for the continuum source emission, and $d$ is the distance to the source.
The dust masses are listed in Table\,\ref{cont-source3}, adopting a dust opacity of $\kappa_{\mathrm{\lambda}} = 0.899\,\mathrm{cm^2 g^{-1}}$  \citep{ossenkopf1994}, dust temperatures of $T_{\mathrm{dust}} = 15\,\mathrm{K}$ \citep{tatematsu2016, zhang2020, Li2013, Kirk2017, sadavoy2016, Mason2020}, and d=400\,pc \citep{groschedl2018, tobin2020b}.
Assuming that the gas to dust mass ratio is 100 to 1 \citep{hayashi1981} and that the geometrical structure of the identified source is a sphere, the mean molecular hydrogen number densities for the sources are $n_{\mathrm{H_2}} = M_{\mathrm{H_2}}/((4/3)\pi r^3\mu m_{\mathrm{H}})$.
Here, $M_{\mathrm{H_2}}$, $r$, $\mu$, and $m_{\mathrm{H}}$ are the gas mass of the source, the continuum source radius, the mean molecular weight ($\mu = \,$2.33), and the hydrogen mass, respectively. 
Table\,\ref{cont-source3} also describes the mean molecular hydrogen number densities for the sources.

The 1.3\,mm continuum emission might be contaminated by the free-free emission associated with the YSO ionized jet \citep{osorio2017}.
To estimate the contribution from the free-free emission at the 1.3\,mm wavelength, we used the spectral index, $\alpha$ ($F_{\mathrm{\nu}} \propto \nu^{\alpha}$), obtained between 1.3\,cm and 5\,cm \citep{osorio2017}. 
\cite{osorio2017} detected seven centimeter sources associated with protostars or YSOs and three centimeter sources considered to be jet knots. 
With the ALMA 12-m array, we detected six 1.3\,mm continuum sources associated with the protostars or YSOs detected by \cite{osorio2017}. 
We estimated $\alpha$ for the centimeter sources to be between  0.28 and 1.35 with a mean value of 0.7 using the number reported in \cite{osorio2017},
consistent with the mean $\alpha$ of 0.6 estimated for the ionized jet by \cite{anglada1998}. 
Using $\alpha$, we concluded that up to $7.0\,\%$ of the 1.3\,mm emission can be attributed to the free-free emission, but the majority of the emission originates from the thermal dust. 

\begin{table}
	\centering
	\caption{Dust mass of the identified sources for $T_{\mathrm{dust}} =$ 15 K ($M_{\mathrm{dust}}$) and the mean molecular hydrogen number density ($n_{\mathrm{H_2}}$, dust to gas mass ratio of 1 : 100). The radius [au] here was calculated from the radius [arcsec] in Table\,\ref{cont-source1} with $d = 400$ pc.}
	\label{cont-source3}
	\footnotesize
	\begin{tabular}{lrcc} 
	\hline
		source name &radius [au] & $M_{\mathrm{dust}}$ [\msun] (15K)& $n_{\mathrm{H_2}} \mathrm{[cm^{-3}]}$ (15K) \\
		\hline
		\hline
		FraSCO-1&	143 &	3.8E-05&	4.7E+07\\  
		FraSCO-2&	319 &	2.4E-04&	2.6E+07\\
		FraSCO-3&	228 &	8.5E-05&	2.6E+07\\
		FraSCO-4&	281 &	6.3E-04&	1.0E+08\\
		FraSCO-5&	390 &	3.0E-04&	1.9E+07\\
		FraSCO-6&	398 &	1.6E-03&	9.0E+07\\
		FraSCO-7&	355 &	9.8E-04&	7.9E+07\\
		FraSCO-8&	218 &	3.6E-04&	1.2E+08\\
		FraSCO-9&	447 &	1.3E-03&	5.1E+07\\
		FraSCO-10&	240 &	3.2E-04&	8.3E+07\\
		FraSCO-11&	146 &	4.1E-05&	4.8E+07\\
		FraSCO-12&	237 &	3.0E-04&	8.2E+07\\
		FraSCO-13&	153 &	5.7E-05&	5.7E+07\\
		FraSCO-14&	194 &	2.6E-04&	1.3E+08\\
		FraSCO-15&	187 &	3.0E-04&	1.7E+08\\
		FraSCO-16&	499 &	1.1E-02&	3.3E+08\\
		FraSCO-17&	150 &	3.9E-05&	4.2E+07\\
		FraSCO-18&	309 &	1.6E-04&	1.9E+07\\
		FraSCO-19&	158 &	9.6E-05&	8.8E+07\\
		FraSCO-20&	239 &	1.3E-04&	3.4E+07\\
		FraSCO-21&	249 &	2.0E-04&	4.6E+07\\
		FraSCO-22&	257 &	1.7E-04&	3.6E+07\\
		FraSCO-23&	195 &	9.3E-05&	4.5E+07\\
		FraSCO-24&	215 &	1.3E-04&	4.8E+07\\
		FraSCO-25&	434 &	4.2E-04&	1.9E+07\\
		FraSCO-26&	488 &	7.2E-04&	2.2E+07\\
		FraSCO-27&	208 &	1.9E-04&	7.8E+07\\
		FraSCO-28&	801 &	1.9E-03&	1.3E+07\\
		FraSCO-29&	252 &	1.3E-04&	2.9E+07\\
		FraSCO-30&	341 &	5.5E-04&	5.0E+07\\
		FraSCO-31&	244 &	1.0E-04&	2.6E+07\\
		FraSCO-32&	280 &	8.0E-04&	1.3E+08\\
		FraSCO-33&	308 &	6.5E-04&	8.0E+07\\
		FraSCO-34&	447 &	7.9E-04&	3.2E+07\\
		FraSCO-35&	982 &	2.6E-03&	1.0E+07\\
		FraSCO-36&	169 &	1.3E-04&	9.7E+07\\
		FraSCO-37&	313 &	1.0E-03&	1.2E+08\\
		FraSCO-38&	571 &	2.3E-03&	4.5E+07\\
		FraSCO-39&	279 &	5.6E-04&	9.3E+07\\
		FraSCO-40&	514 &	2.8E-03&	7.6E+07\\
		FraSCO-41&	358 &	5.7E-04&	4.5E+07\\
		FraSCO-42&	170 &	5.0E-05&	3.6E+07\\
		FraSCO-43&	251 &	3.2E-04&	7.2E+07\\
		FraSCO-44&	338 &	5.7E-04&	5.3E+07\\
		FraSCO-45&	264 &	3.0E-04	&5.9E+07\\
		FraSCO-46&	384 &	7.4E-04	&4.7E+07\\
		FraSCO-47&	483 &	1.1E-03	&3.6E+07\\
		FraSCO-48&	259 &	1.1E-04	&2.3E+07 \\
		FraSCO-49&	512 &	3.8E-03	& 1.0E+08 \\
		FraSCO-50&	909 &	1.3E-03	& 6.4E+06 \\
		FraSCO-51&	176 &	5.5E-05	& 3.6E+07 \\
\hline 
	\end{tabular}
\end{table}

\subsubsection{Jeans Analysis}
\label{result-cont-jeans}
As described in Section \ref{result-cont-mass}, the detected 1.3\,mm sources have dust masses and gas number densities in the range $3.8 \times 10^{-5}$--$ 1.1 \times 10^{-2}$\,\msun\,and $6.4 \times 10^{6}$--$3.3 \times 10^{8}\,\mathrm{cm^{-3}}$, respectively.
To investigate the gravitational stability, we performed a Jeans analysis.
Assuming an infinite and homogeneous medium, the Jeans length is described as follows \citep{jeans1902}:
\begin{equation}
    \lambda_{\mathrm{frag}}= \sqrt{\frac{\pi c_{\mathrm{s}}^2}{G \rho_0}},
    \label{j-1}
\end{equation}
where $G$ is the gravitational constant, $\rho_0$ is the mean density, and $c_{\mathrm{s}}$ is the sound speed.
The relation between the sound speed and the gas temperature is described as $c_{\mathrm{s}}=\sqrt{k T_{\mathrm{gas}}/\mu m_{\mathrm{H}}}$, where $T_{\mathrm{gas}}$ is the gas temperature and $k$ is the Boltzmann constant.
From equation\,(\ref{j-1}), the critical number density is derived as follows:
\begin{equation}
    n_{\mathrm{H_2}}(\lambda_{\mathrm{frag}}, T_{\mathrm{gas}}) = \frac{\pi k}{G(\mu m_{\mathrm{H}})^2}\frac{T_{\mathrm{gas}}}{\lambda_{\mathrm{frag}}^2}.
    \label{j-2}
\end{equation}

Figure 6 shows the relation between the $\mathrm{H_2}$ gas number density and radius for the continuum sources. 
Using a solid curve, we also plotted the Jeans critical number density as a function of radius, with $r = \lambda_{\mathrm{frag}}/2$ and a gas temperature of 15\,K (e.g., \citealt{tatematsu2016}).
Here, we assumed that the gas temperature is equal to the dust temperature (see Section \ref{result-cont-mass}).
The critical number density determines whether the identified sources are gravitationally bound,
with the sources being bound when the gas number density is above the critical number density. 

It should be noted that we assumed that the turbulent / non-thermal motions are acting isotropically and therefore can be treated as thermal-like support. Recent interferometric observations toward the OMC-2/3 region show that the non-thermal velocity dispersion of the spatially resolved cores is 0.12 $\pm$ 0.05 \kms, whereas the mean thermal velocity dispersion for $\mathrm{H_2}$ gas is estimated to be 0.23 \kms\, at the gas temperature of 15\,K \citep{zhang2020}. This results indicate that the detected cores are thermal dominant. 
Moreover, \cite{takahashi2013} and \cite{teixeira2016} presented that the fragmentation length within the OMC filaments is rather consistent with the Jeans length, implying that the region is thermal dominant. These facts also strengthen the scenario that, in the OMC-2/3 region, turbulence is dissipated in the size scale of less than a few 1000\,au, where individual star formation is taken place.
We should also consider another effect; the magnetic field. Star-formation activities are actually confirmed in the FIR\,3, 4, and 5 regions (e.g., \citealt{furlan2016}), which indicates that the magnetic field may not be enough strong to suppress the star formation.
Note that we do not have direct measurements of the magnetic field in the observed size scale here, hence it is difficult to discuss the effects of the magnetic field in detail.
Another factor in a full virial analysis that is not included here is the external pressure from the material surrounding the FraSCO sources which can help to bind them, including turbulent pressure (e.g., \citealt{Pattle2015}) and cloud weight pressure (e.g., \citealt{Lada2008, Kirk2017}). Estimates of these pressures are not available for FraSCO sources, however we expect the latter pressure to be the more significant factor, since turbulence appears to be largely dissipated around FraSCO sources as mentioned above.

\begin{figure}
  \centering
  \includegraphics[width=15cm]{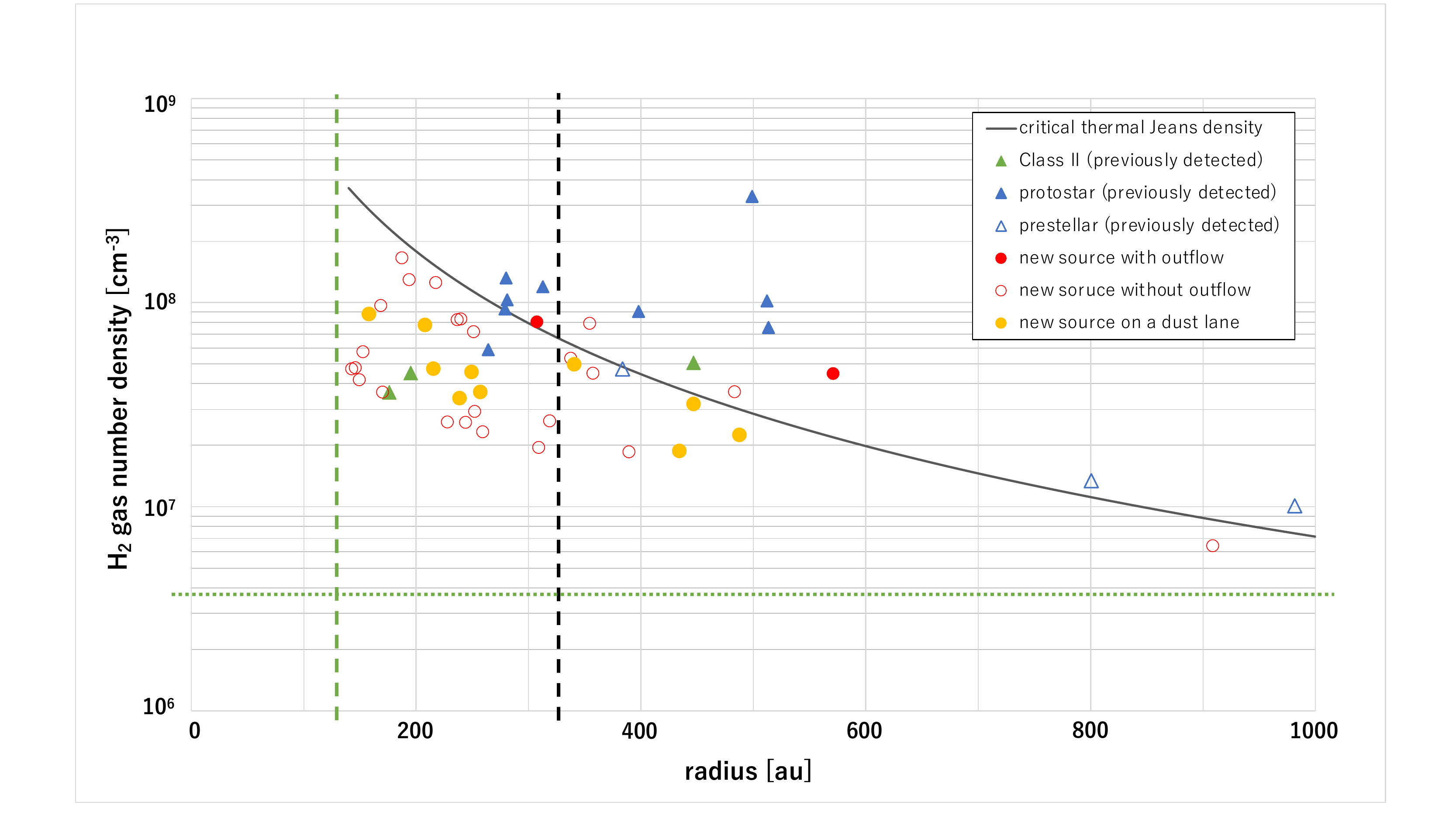}
   \caption{$\mathrm{H_2}$ gas number density for $T_{\mathrm{dust}} = 15$ K vs. source radius. Both parameters are estimated from our ALMA 12-m array 1.3\,mm continuum emission following the method described in Section \ref{result-cont-mass} and Section \ref{result-cont-jeans}. The different data point symbols indicate the category of the source. Triangles represent 1.3\,mm continuum sources associated with previously identified sources. Protostars denoted by filled blue triangles are 1.3\,mm continuum sources associated with previously identified Class 0, Class I, and Class flat sources, while pre-main-sequence sources denoted by filled green triangles are 1.3\,mm continuum sources associated with previously identified Class \two\,sources \citep{davis2009, megeath2012, furlan2016}. Prestellar sources denoted by open blue triangles are 1.3\,mm continuum sources associated with 3\,mm sources not identified at infrared wavelengths \citep{kainulainen2017, vanterwisga2019}. Circles represent 1.3\,mm continuum sources identified for the first time in this study. Filled red circles are 1.3\,mm continuum sources considered to be outflow driving sources, while open red circles indicate sources not considered to drive outflows. Filled yellow circles are 1.3\,mm continuum sources detected on the dust lane discussed in Section \ref{dis-out}. The black solid curve represents critical thermal Jeans density for 15 K. The radius for the Jeans density is defined as $\lambda_{\mathrm{frag}}/2$. The dashed green and black lines represent half of the minor axes of the synthesized beam sizes achieved from our 1.3\,mm continuum ($\theta \sim 0''.33 \sim$ 130\,au) and 3\,mm continuum ($\theta \sim 0''.8 \sim$ 320\,au; \citealt{vanterwisga2019}). The lower threshold of the $\mathrm{H_2}$ gas number density calculated from the 3$\sigma$ mass limit is $3.7\times10^6\,\mathrm{cm^{-3}}$ for our study (the green dotted line), while the threshold for \cite{vanterwisga2019} is $1.9 \times 10^{8}\,\mathrm{cm^{-3}}$. }
 \label{jeans-1}
 \end{figure}

Figure\,\ref{jeans-1} shows that the majority ($\sim 80\,\%$) of the previously identified sources, denoted by triangles, have $\mathrm{H_2}$ gas number densities above the Jeans critical number density regardless of whether they are pre- or proto-stellar sources, while three sources (FraSCO-23, FraSCO-45, and FraSCO-51, also known as MIR 23, HOPS-369/ MIR 27, and a Spitzer disk source) have $\mathrm{H_2}$ gas number densities below the threshold.
Among these three sources, FraSCO-23 and FraSCO-51 are associated with Class \two \,sources, while FraSCO-45 is known to be a flat spectrum source, HOPS-369 \citep{furlan2016}, which is considered to be between the Class I and Class II evolutionary stages. Therefore, these three sources are considered to be relatively evolved and the dust emission originating from the core and  envelope are mostly dissipated.

In addition, Figure\,\ref{jeans-1} shows that a low percentage of newly identified sources ($\sim 14\,\%$) have $\mathrm{H_2}$ gas number densities above the Jeans critical number density (i.e., the sources are considered to be gravitationally bound).
Among them, two continuum sources are associated with two outflows newly detected in this study, flow-5 and flow-7 (see Section \ref{result-line-outflow}).
The rest of the newly identified sources, about 86$\,\%$, have $\mathrm{H_2}$ gas number densities below the critical density and are considered to be gravitationally unbound.
A majority of the sources newly detected in this study show sizes comparable or smaller than the best angular resolution previously achieved with the ALMA mosaic mapping studies in this region (i.e., $\theta \sim 0''.8$ by \citealt{vanterwisga2019}). 
Note that ten of the newly identified sources with number densities below the Jeans critical number density may be on a dust lane heated by outflow driven by HOPS-370, flow-3, filled yellow circles in Figure\,\ref{jeans-1} (further explanation in Section \ref{dis-out}).
This means that the origin of dust concentrations may not be dominated by thermal fragmentation, but rather is affected by outflow shocks. 
Excluding these ten sources, the percentage of newly detected sources that are gravitationally unbound is reduced to 80$\,\%$. Finally, we found that gravitationally unbound sources are located more or less uniformly across the region. 
In the Jeans analysis, no clear trend in terms of the core nature and evolutionary stage were found between the FIR\,3, 4, and 5 regions.  

\subsection{CO\,($J$ = 2--1) and SiO\,($J$ = 5--4) Line Emissions}
\label{result-line}

Outflows are a significant part of the star formation process.
Outflows are important for efficiently transporting angular momentum from pre- and proto-stellar cores as part of the process of cores evolving to stars \citep{arce2007, machida2007, machida2008}.
Outflow gas can be traced by CO line emissions on the protostellar scale of $\sim$ 0.01--0.1 pc \citep{snell1980, bontemps1996, bachiller1996}.
Interactions between outflows and surrounding gas are often observed with shocked gas tracers such as SiO line emissions.
SiO in the gas phase is believed to be formed through sputtering dust grains \citep{ziurys1989, caselli1997} 
and SiO emission is considered to be a tracer of collimated high velocity outflows \citep{zhang2002, zapata2006, hirano2010, matsushita2019, liu2021}.
SiO emission is also considered to be an extended bow shock from protostellar outflow \citep{gueth1998, shimajiri2008}.

\begin{figure}
\centering
\includegraphics[width=17cm]{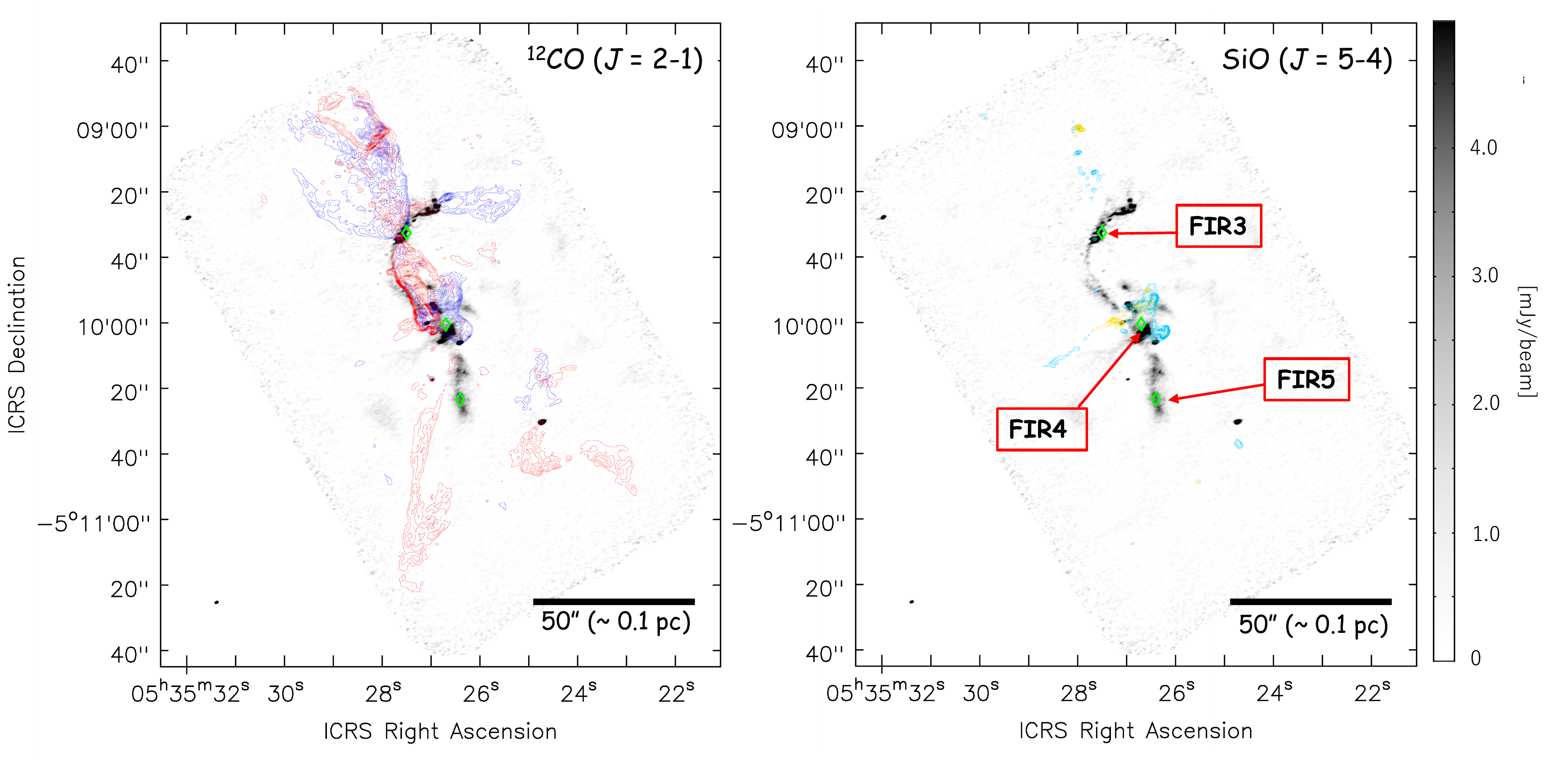}
\caption{\textbf{Left}: CO blue- and red-shifted integrated intensity maps obtained from the ALMA 12-m array. The red contours represent the integrated intensity map using the red-shifted CO components in the velocity range \vlsr = 15--95\,\kms . The red contour levels are [10, 15, 20, 25, 30, 35, 40, 45, 50, 55, 60, 65] $\times 1\sigma$ ($1\sigma = 0.17$\,\jykms ). The blue contours represent the integrated intensity map using the blue-shifted CO components in the velocity range \vlsr = $-$100--10\,\kms . The blue contour levels are [10, 15, 20, 25, 30, 35, 40, 45, 50, 55, 60, 65, 70, 75, 80] $\times 1\sigma$ ($1\sigma = 0.23$\,\jykms ). \textbf{Right}: SiO blue- and red-shifted integrated intensity map obtained from the ALMA 12-m array. The yellow contours represent the integrated intensity map using the red-shifted SiO components in the velocity range \vlsr = 12--29\,\kms .  The yellow contour levels are  [10, 15, 20, 25, 30, 35, 40, 45, 50, 55] $\times 1\sigma$ ($1\sigma = 0.015$\,\jykms ). The cyan contours represent the integrated intensity map using the blue-shifted SiO components in the velocity range \vlsr = $-$30--11\,\kms . The cyan contour levels are [10, 15, 20, 25, 30, 35, 40, 45, 50, 55, 60, 70, 80, 90, 100, 120, 140, 160, 180, 200] $\times 1\sigma$ ($1\sigma = 0.02$\,\jykms ). In both panels, the gray scale represents our 1.3\,mm continuum image obtained from the ALMA 12-m array. The green diamonds denote the locations of FIR\,3, 4, and 5 identified by single-dish 1.3\,mm continuum observations \citep{chini1997}. The synthesized beam sizes of the CO, SiO, and 1.3\,mm continuum emission are $1.''19 \times 0.''74$, $1.''26 \times 0.''79$, and $1.''13 \times 0.''65$, respectively. The black ellipse at the bottom-left corner represents the largest synthesized beam sizes among them, i.e., SiO.}
\label{LINE-1}
\end{figure}

Figure\,\ref{LINE-1} shows integrated intensity maps obtained from CO\,($J$ = 2--1) (left panel) and SiO\,($J$ = 5--4) (right panel) datasets. 
The images indicate that the gas distributions traced by CO and SiO are very different. The CO emission traces extended structures, while the SiO emission traces compact structures. The CO\,($J$ = 2--1) emission traces the molecular outflows well, and their spatial and velocity distributions are consistent with previous CO studies \citep{takahashi2008, tobin2019, tanabe2019, feddersen2020}.
 Our CO observations achieved eight times higher angular resolution compared to recent interferometric CO observations (e.g., \citealt{feddersen2020}).
 We successfully resolved internal structures toward previously reported molecular outflows. 
 The SiO\,($J$ = 5--4) emission was intensely detected toward the FIR\,4 region. 
 While the strongest SiO component detected in the FIR\,4 region was previously reported in the lower transition of the SiO\,($J$ = 2--1) observations \citep{shimajiri2008}, our observations detected some additional SiO components. 
 Extended and relatively strong components are located around the FIR\,4 region, and several compact SiO components likely originate from the local shocked regions. Individual structures traced by CO\,($J$ = 2--1) and SiO\,($J$ = 5--4) emissions are described in the following subsections. In Section \ref{result-line-outflow}, we focus on identifying molecular outflows using the CO\,($J$ = 2--1) image. In Section \ref{result-line-shock}, we identify shocked regions (i.e., shocked gas not directly produced from outflow or jet components). The identification of the shocked regions was based on the SiO\,($J$ = 5--4) image. 
 Note that we adopted a systemic velocity (\vsys) of 11\,$\mathrm{km}\,\mathrm{s^{-1}}$ in our observational mapping regions based on multiple line survey observations toward the FIR\,4 region \citep{lopez2013, tobin2019}    .

\begin{figure}
    \centering
    \includegraphics[width=17cm]{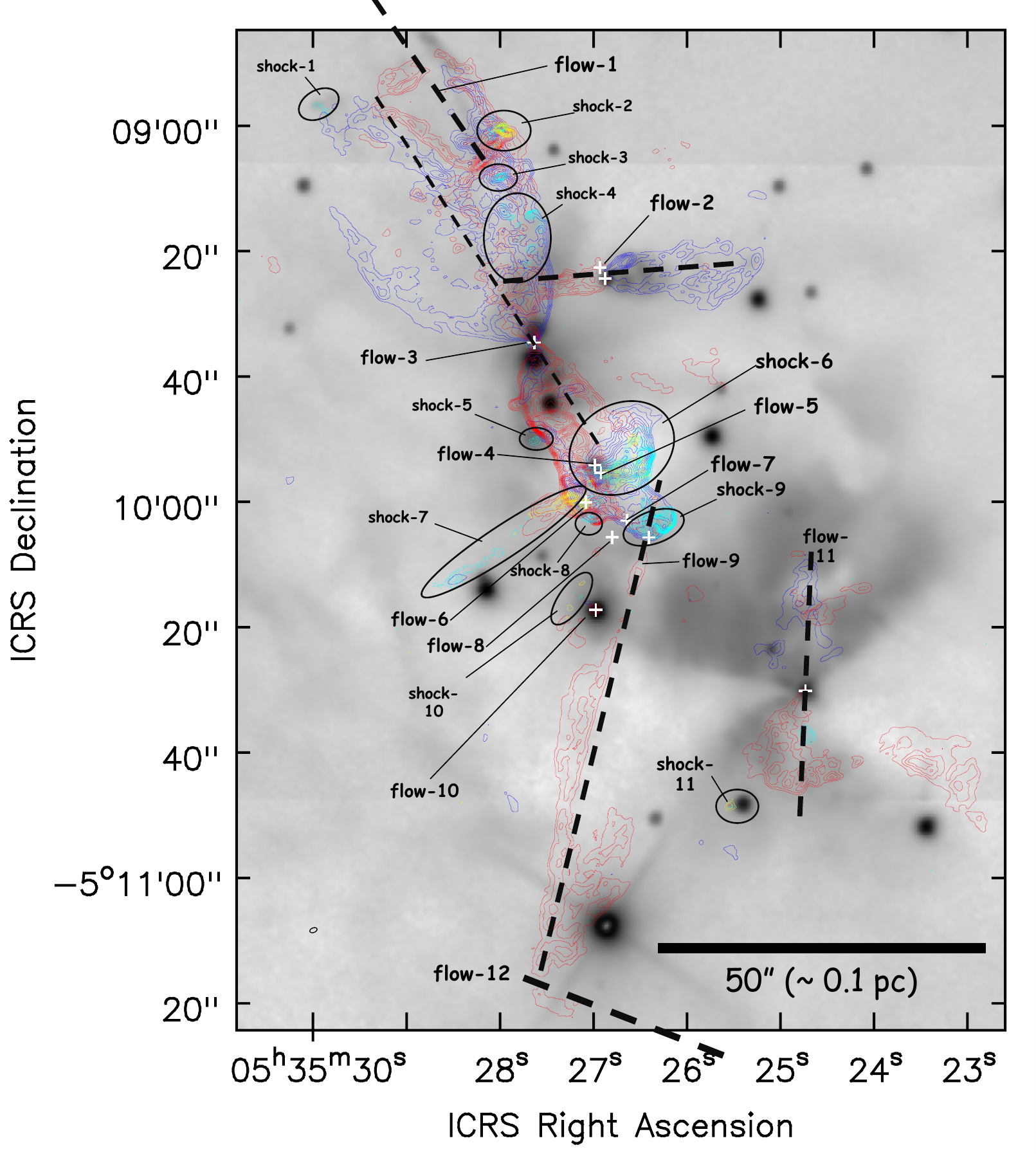}
    \caption{Identification of CO outflows and SiO shocked structures, summarized in Table\,\ref{out-1} and \ref{interact-2}, respectively. The white crosses correspond to the locations of outflow driving 1.3\,mm continuum sources identified in Section \ref{result-cont-id} using dendrograms (see Table\ref{cont-source1}), although the driving sources of flow-1 and flow-12 are outside our observation mapping image. The red, blue, yellow, and cyan contours are the same as those in Figure\,\ref{LINE-1}. The gray background images show the 2.2\,\micron\,$Ks$ band image obtained from SIRIUS/IRSF  \citep{takahashi2008}. Some expected outflow axes are represented by the black dashed lines. Note that flow-12 is not visible here because the integrated intensity map is not optimized for the velocity range of this outflow. The image of flow-12 with an optimized velocity range can be found in Figure\,\ref{flow-12}. The black open ellipse at the bottom-left corner is the synthesized beam size of SiO (same as that in Figure\,\ref{LINE-1}).}
      \label{line-2}
\end{figure}

\subsubsection{Outflow Identifications}
\label{result-line-outflow}
To identify the molecular outflow, we set the following five criteria: 
\textbf{(1)} CO emission is detected at greater than 10$\sigma$ in the integrated intensity map (where the velocity range of the integrated intensity map is optimized for individual regions)\footnote{We decided to use $10\sigma$ threshold to identify outflows instead of $5\sigma$. This threshold was decided after carefully checking the channel maps (i) not to miss any faint and high-speed outflow-related emission and at the same time (ii) not to pick up faint and extended emission originating from the molecular cloud around the systemic velocity. We confirmed that outflow identification with the threshold as $5\sigma$ and $10\sigma$ does not change the number of detected outflows.
}, \textbf{(2)} the extension of the CO emission is larger than the beam size, \textbf{(3)} the CO emission has a collimated structure, \textbf{(4)} the gas velocity of the CO emission is greater than 5\,$\mathrm{km}\,\mathrm{s^{-1}}$ with respect to the systemic velocity, and \textbf{(5)} the driving source candidate is identified from the 1.3\,mm sources listed in our FraSCO catalogue or sources previously catalogued in other studies. 

We then identified molecular outflow for which the CO emission satisfies the five criteria. We also categorized the identified outflows into three groups, ``Clear'', ``Probable'', and ``Marginal'' (C, P, and M in Table\,\ref{out-1}), to indicate the confidence level of the outflow detection. 
The categorization criteria are as follows. \textbf{Clear}: Emissions have both localized blue- and red-shifted lobes with clear bipolarity in the CO\,($J$ = 2--1) integrated intensity map.
\textbf{Probable}: Emissions show a monopolar structure of either a localized blue- or red-shifted lobe in the CO\,($J$ = 2--1) integrated intensity maps.
\textbf{Marginal}: Emissions have either blue- or red-shifted components associated with a candidate driving source.  However, it is hard to completely identify the molecular outflow separate from the ambient gas, due to contamination from the extended gas around the systemic velocity. 

\begin{figure}
    \centering
    \includegraphics[width=17cm]{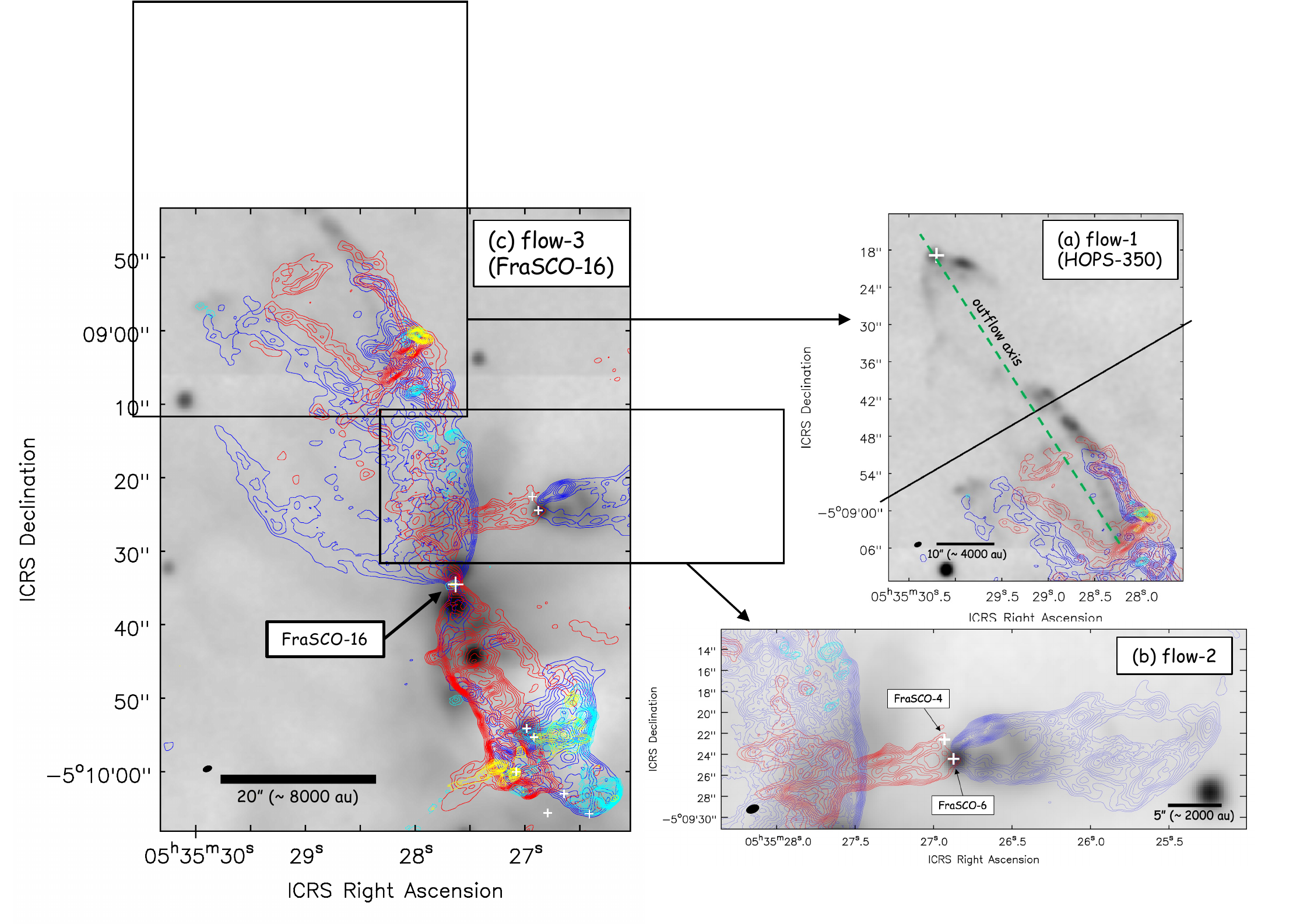}
    \caption{flow1--flow3. The red, blue, yellow, and cyan contours represent the integrated intensities obtained from the ALMA 12-m array of red-shifted CO, blue-shifted CO, red-shifted SiO, and blue-shifted SiO line emissions, respectively. The gray background images show the 2.2\,\micron\,$Ks$ band image obtained from SIRIUS/IRSF  \citep{takahashi2008}. The white crosses are the positions of the outflow driving sources in all the figures. The black ellipse at the bottom-left corner is the same as that in Figure\,\ref{LINE-1}.
    \textbf{(a)}: flow-1 (Probable). The red contour levels are [20, 30, 40, 50, 60, 70, 80, 90, 100, 110, 120, 130, 140] $\times 1\sigma$ ($1\sigma = 0.08$\,\jykms, \vlsr = 12--30\,\kms). The blue contour levels are [20, 30, 40, 50, 60, 70, 80, 90, 100] $\times 1\sigma$ ($1\sigma= 0.08$\,\jykms, \vlsr = 0--11\,\kms). The yellow contour levels are [10, 15, 20, 25, 30, 35, 40, 45, 50, 55] $\times 1\sigma$ ($1\sigma = 0.015$\,\jykms, \vlsr = 12--30\,\kms). The cyan contour levels are  [5, 10, 15, 20] $\times 1\sigma$ ($1\sigma = 0.03$\,\jykms, \vlsr = 0--11\,\kms). The black solid line is the boundary of our observation mapping. The green dashed line represents the outflow axis. 
    \textbf{(b)}: flow-2 (Clear). This outflow is considered to be driven by two sources (FraSCO-4 and FraSCO-6). The red contour levels are [8, 10, 12, 14, 16, 18, 20, 22, 24, 26, 28, 30, 32, 34, 36, 38, 40, 42, 44, 46, 48, 50, 52, 54, 56, 58, 60, 62, 64]$\times 1\sigma$ ($1\sigma = 0.2$\,\jykms, \vlsr = 15--35\,\kms). The blue contour levels are [8, 11, 14, 17, 20, 23, 26, 29, 32, 35, 38, 41, 44, 47, 50]$\times 1\sigma$ ($1\sigma= 0.2$\,\jykms, \vlsr = $-$70--10\,\kms). The cyan contour levels are [4, 5, 6, 8, 10, 12, 16, 20] $\times 1\sigma$ ($1\sigma = 0.07$\,\jykms, \vlsr = $-$30--11\,\kms).
    \textbf{(c)}: flow-3 (Clear). The driven source candidates is considered to be FraSCO-16. The red contour levels are [10, 15, 20, 25, 30, 35, 40, 45, 50, 55, 60, 65] $\times 1\sigma$ ($1\sigma = 0.17$\,\jykms, \vlsr = 12--85\,\kms). The blue contour levels are [10, 15, 20, 25, 30, 35, 40, 45, 50, 55, 60, 65, 70, 75, 80] $\times 1\sigma$ ($1\sigma = 0.23$\,\jykms, \vlsr = $-$40--11\,\kms). The yellow contour levels are [10, 15, 20, 25, 30, 35, 40, 45, 50, 55, 60, 70, 80, 90, 100, 120, 140, 160, 180, 200] $\times 1\sigma$ ($1\sigma = 0.015$\,\jykms, \vlsr = $-$30--11\,\kms). The cyan contour levels are [10, 15, 20, 25, 30, 35, 40, 45, 50, 55, 60, 70, 80, 90, 100, 120, 140, 160, 180, 200] $\times 1\sigma$ ($1\sigma = 0.02$\,\jykms, \vlsr = 12--30\,\kms).
    }
    \label{flow1-3}
\end{figure}

\begin{figure}
    \centering
    \includegraphics[width=18cm]{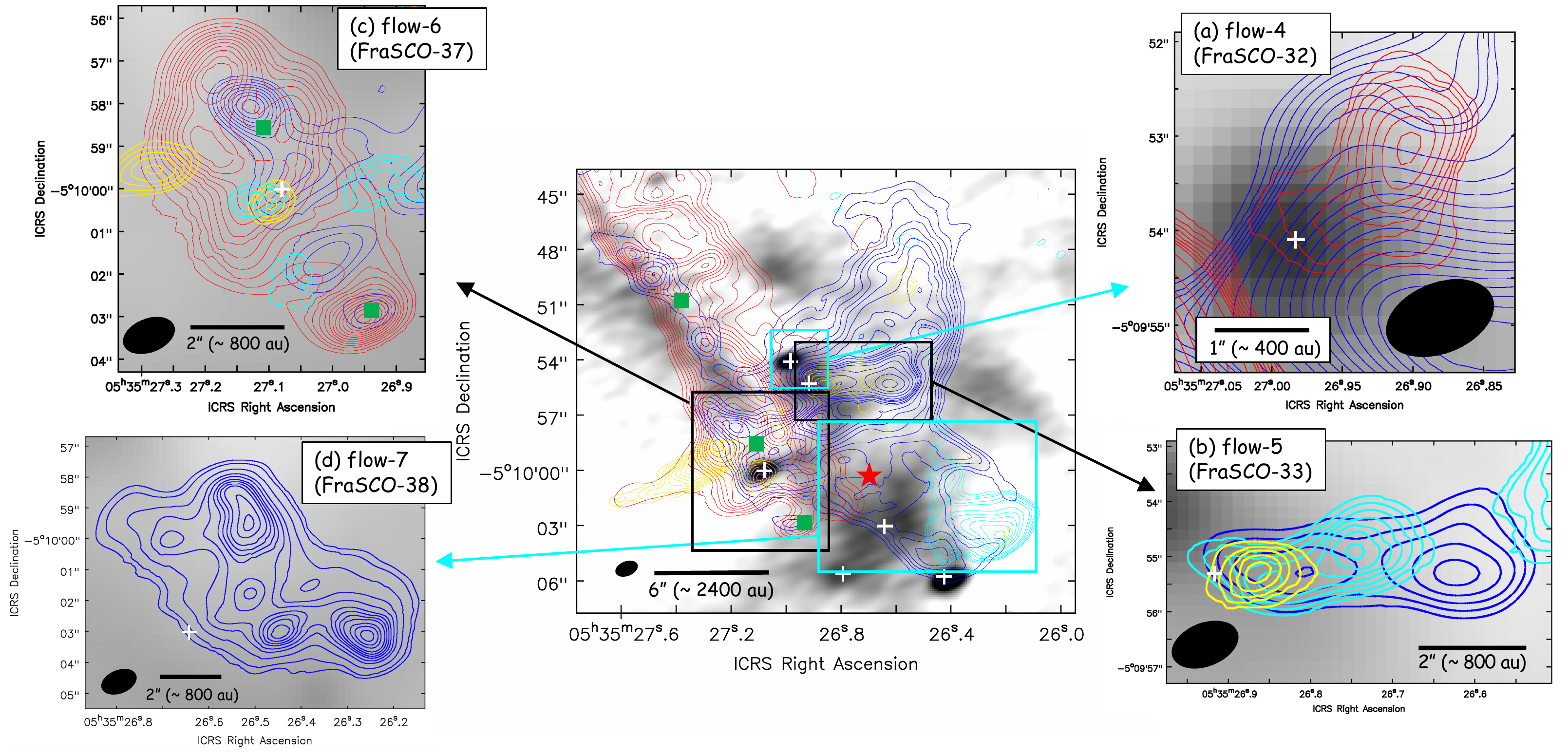}
    \caption{flow-4--flow-7. The red, blue, yellow, and cyan contours represent the integrated intensities obtained from the ALMA 12-m array of red-shifted CO, blue-shifted CO, red-shifted SiO, and blue-shifted SiO, respectively. The white crosses are the positions of the outflow driving sources in all the figures. The green squares are the positions of non-thermal radio knots detected by 5 cm continuum emission \citep{osorio2017}.
    \textbf{Central panel}: zoomed-out image around the position of FIR\,4 (red star, \citealt{chini1997}). The gray background image represents our 1.3\,mm continuum image. The red contour levels are [10, 20, 30, 40, 50, 60, 70, 80, 90, 100, 110]$\times 1\sigma$ ($1\sigma = 0.1$\,\jykms , \vlsr = 15--22\,\kms). The blue contour levels are [10, 15, 20, 25, 30, 35, 40, 45, 50, 52, 55, 57, 60, 65, 70, 75, 80]$\times 1\sigma$ ($1\sigma = 0.2$\,\jykms , \vlsr = 2--9\,\kms). The yellow contour levels are [3, 4, 5, 6, 7, 8, 9, 10, 11, 12, 13, 14, 15] $\times 1\sigma$ ($1\sigma = 0.04$\,\jykms , \vlsr = 15--20\,\kms). The cyan contour levels are [3, 4, 5, 7, 9, 15, 20, 25, 30, 35, 40] $\times 1\sigma$ ($1\sigma = 0.04$\,\jykms , \vlsr = 0--5\,\kms). Figure\,\ref{flow4-7}(a)--(d) are zoomed-in images overlaid on the 2.2\,\micron\,$Ks$ band image obtained from SIRIUS/IRSF  (gray background image; \citealt{takahashi2008}). 
    \textbf{(a)}: flow-4 (Probable). The red contour levels are [8, 10, 12, 14, 16, 18, 20, 22, 30, 50, 70, 90, 110, 130, 150, 170, 190, 210, 230, 250] $\times 1\sigma$ ($1\sigma = 0.04$\,\jykms, \vlsr = 1--23\,\kms). The blue contour levels are [8, 10, 12, 14, 16, 18, 20, 30, 40, 50, 60, 70, 80, 90, 100, 110, 120, 130, 140, 150, 160, 170, 180, 190, 200, 210, 220, 230, 235] $\times 1\sigma$ ($1\sigma = 0.04$\,\jykms, \vlsr = $-$1--5\,\kms). 
    \textbf{(b)}: flow-5 (Probable). The blue contour levels are [60, 65, 70, 75, 80]  $\times 1\sigma$ ($1\sigma = 0.2$\,\jykms, \vlsr = 2--9\,\kms). The yellow contour levels are [4, 6, 8, 10, 12, 14] $\times 1\sigma$ ($1\sigma = 0.015$\,\jykms , \vlsr = 14--16\,\kms). The cyan contour levels are [10, 12, 14, 16, 18, 20, 22] $\times 1\sigma$ ($1\sigma = 0.027$\,\jykms , \vlsr = 2--5\,\kms).
    \textbf{(c)}: flow-6 (Probable). The red contour levels are [4, 6, 8, 10, 12, 14, 16, 18, 20, 22, 24, 26, 28] $\times 1\sigma$ ($1\sigma = 0.2$\,\jykms, \vlsr =  25--50\,\kms). The blue contour levels are [8, 10, 12, 14, 16, 18, 20, 22, 24]$\times 1\sigma$ ($1\sigma= 0.2$\,\jykms, \vlsr = $-$7--5\,\kms). The yellow contour levels are [6, 7, 8, 9, 10, 11, 12, 13, 14, 15] $\times 1\sigma$ ($1\sigma = 0.04$\,\jykms , \vlsr = 12--18\,\kms). The cyan contour levels are [6, 8, 10, 12] $\times 1\sigma$ ($1\sigma = 0.01$\,\jykms , \vlsr = 1--3\,\kms). 
    \textbf{(d)}: flow-7 (Marginal). The blue contour levels are [3, 5, 10, 15, 18, 21, 23, 25, 27, 30, 32, 34] $\times 1\sigma$ ($1\sigma = 0.1$ \jykms, \vlsr = $-$5--5\,\kms ). In all the figures, the black ellipses at the bottom corners are the same as that in Figure\,\ref{LINE-1}.}
   \label{flow4-7}
\end{figure}

\begin{figure}
    \centering
    \includegraphics[width=10cm]{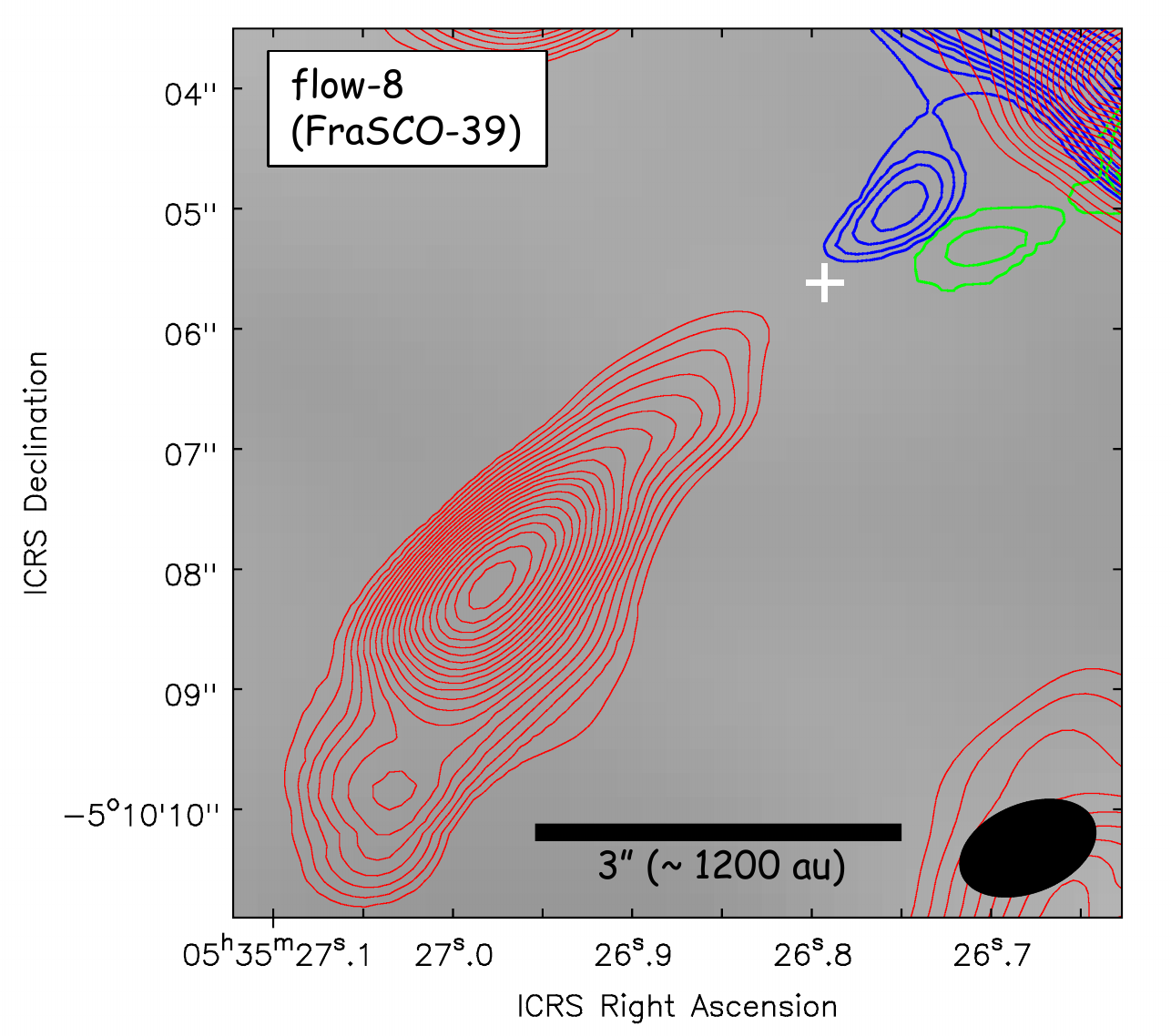}
    \caption{flow-8 (Clear). The gray background shows the 2.2\,\micron\,$Ks$ band image obtained from SIRIUS/IRSF  \citep{takahashi2008}. The white cross is the position of the outflow driving source (FraSCO-39). The red contours represent the integrated intensity of red-shifted CO components obtained from the ALMA 12-m array (\vlsr = 13--18\,\kms). The red contour levels are [10, 14, 18, 22, 26, 30, 34, 38, 42, 46, 50, 54, 58, 62, 66, 70, 74, 78, 82, 86]$\times 1\sigma$ ($1\sigma = 0.02$\,\jykms).  The blue contours represent the integrated intensity of blue-shifted CO components obtained from the ALMA 12-m array (\vlsr = 7--9\,\kms). The blue contour levels are [10, 12, 14, 16, 18, 20, 24, 28, 32, 36, 40, 44, 48, 52, 56, 60, 64, 68, 72, 76, 80, 84]$\times 1\sigma$ ($1\sigma = 0.03$\,\jykms). The green contours represent the integrated intensity of the SiO component obtained from the ALMA 12-m array (\vlsr = 10--12\,\kms). The green contour levels are [3, 4]$\times 1\sigma$ ($1\sigma = 0.03$\,\jykms). The black ellipse at the bottom-right corner is the same as that in Figure\,\ref{LINE-1}.}
    \label{flow-8}
\end{figure}

\begin{figure}
    \centering
    \includegraphics[width=8cm]{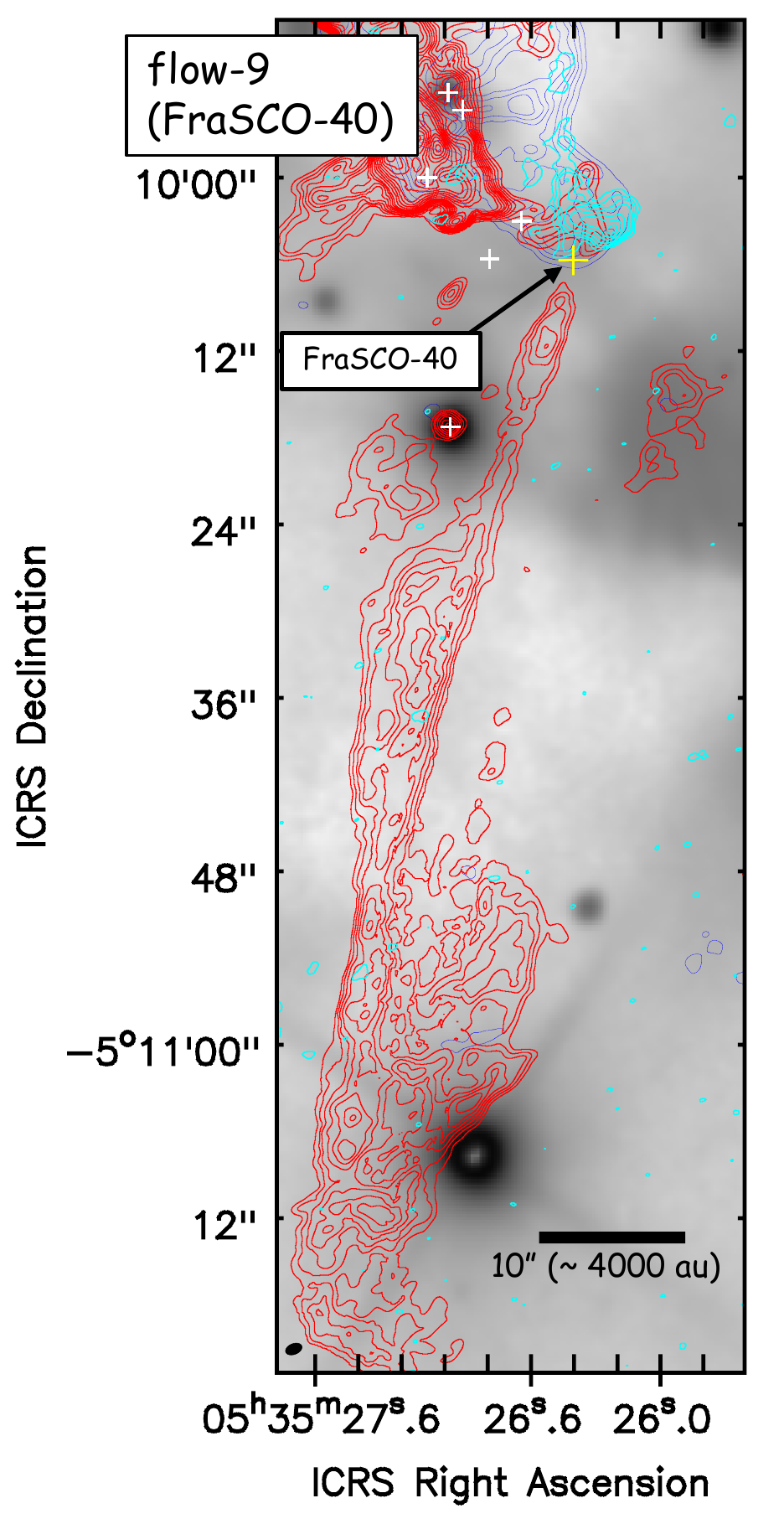}
    \caption{flow-9 (Probable). The gray background shows the 2.2\,\micron\,$Ks$ band image obtained from SIRIUS/IRSF  \citep{takahashi2008}. The white crosses and yellow cross represent the positions of outflow driving sources. The yellow cross is the position of the outflow driving source of flow-9 (FraSCO-40). The red contours represent the integrated intensity of red-shifted CO components obtained from the ALMA 12-m array (\vlsr = 13--60\,\kms). The red contour levels are [10, 15, 20, 25, 30, 35, 40, 45, 50, 60, 70, 80, 90, 100, 110, 120, 130, 140, 150, 160] $\times 1\sigma$ ($1\sigma = 0.1$\,\jykms).  The blue contours represent the integrated intensity of blue-shifted CO components obtained from the ALMA 12-m array (\vlsr = $-$5--10\,\kms). The blue contour levels are [10, 20, 30, 40, 50, 60, 70, 80] $\times 1\sigma$ ($1\sigma= 0.2$\,\jykms). The cyan contours represent the integrated intensity of the blue-shifted SiO component obtained from the ALMA 12-m array (\vlsr = $-$5--5\,\kms). The cyan contour levels are [10, 20, 30, 50, 90, 130, 170] $\times 1\sigma$ ($1\sigma = 0.005$\,\jykms). The black ellipse at the bottom-left corner is the same as that in Figure\,\ref{LINE-1}. }
    \label{flow-9}
\end{figure}

\begin{figure}
    \centering
    \includegraphics[width=10cm]{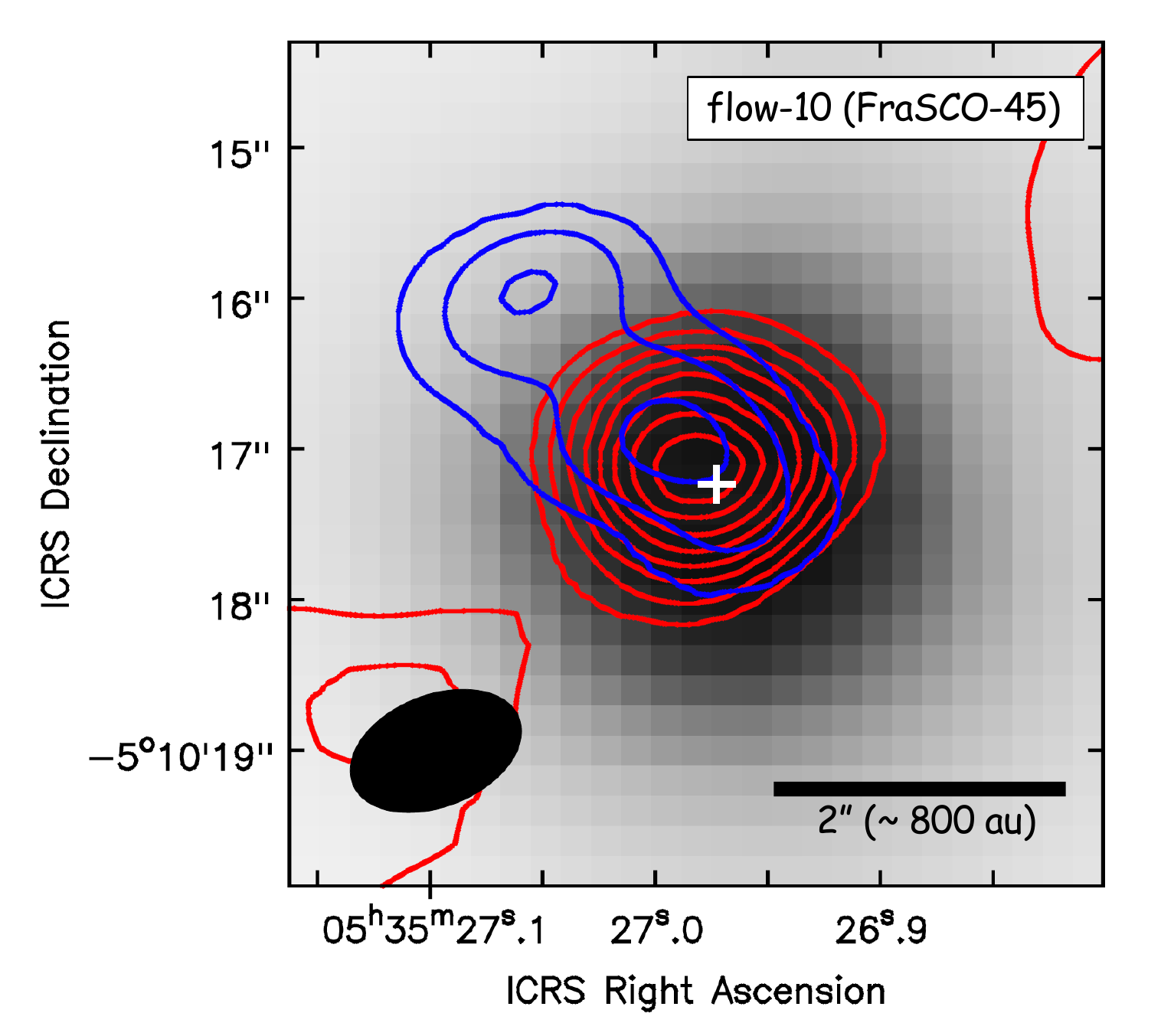}
    \caption{flow-10 (Clear). The gray background shows the 2.2\,\micron\,$Ks$ band image obtained from SIRIUS/IRSF  \citep{takahashi2008}. The white cross is the position of the outflow driving source (FraSCO-45). The red contours represent the integrated intensity of red-shifted CO components obtained from the ALMA 12-m array (\vlsr = 13--21\,\kms). The red contour levels are [10, 15, 20, 25, 30, 35, 40, 45] $\times 1\sigma$ ($1\sigma = 0.08$\,\jykms).  The blue contours represent the integrated intensity of blue-shifted CO components obtained from the ALMA 12-m array (\vlsr = 3--7\,\kms). The blue contour levels are [10, 15, 20] $\times 1\sigma$ ($1\sigma = 0.08$\,\jykms). The black ellipse at the bottom-left corner is the same as that in Figure\,\ref{LINE-1}.}
    \label{flow-10}
\end{figure}

\begin{figure}
    \centering
    \includegraphics[width=10cm]{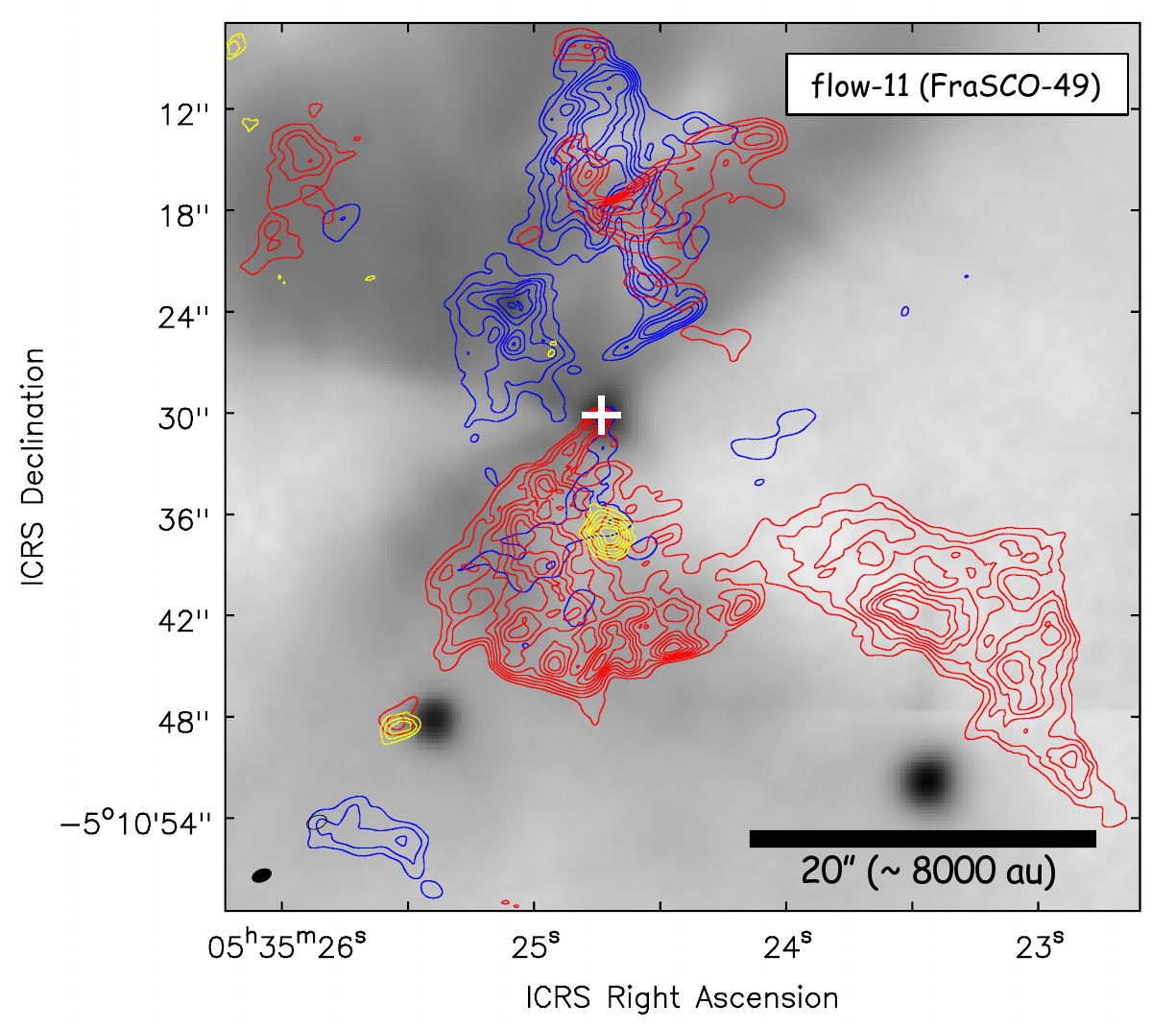}
    \caption{flow-11 (Clear). The gray background shows the 2.2\,\micron\,$Ks$ band image obtained from SIRIUS/IRSF  \citep{takahashi2008}. The white cross is the position of the outflow driving source (FraSCO-49). The red contours represent the integrated intensity of red-shifted CO components obtained from the ALMA 12-m array (\vlsr = 15--35\,\kms). The red contour levels are [10, 15, 20, 25, 30, 35, 40, 45]$\times 1\sigma$ ($1\sigma = 0.1$ \,\jykms).  The blue contours represent the integrated intensity of blue-shifted CO components obtained from the ALMA 12-m array (\vlsr = 0--5\,\kms). The blue contour levels are [10, 15, 20, 25, 30, 35]$\times 1\sigma$ ($1\sigma = 0.1$\,\jykms). The yellow contours represent the integrated intensity of the blue-shifted SiO component obtained from the ALMA 12-m array (\vlsr = 15--35\,\kms). The cyan contour levels are [4, 6, 8, 10, 12, 14, 16]$\times 1\sigma$ ($1\sigma = 0.05$\,\jykms). The black ellipse at the bottom-left corner is the same as that in Figure\,\ref{LINE-1}.}
    \label{flow-11}
\end{figure}

\begin{figure}
    \centering
    \includegraphics[width=12cm]{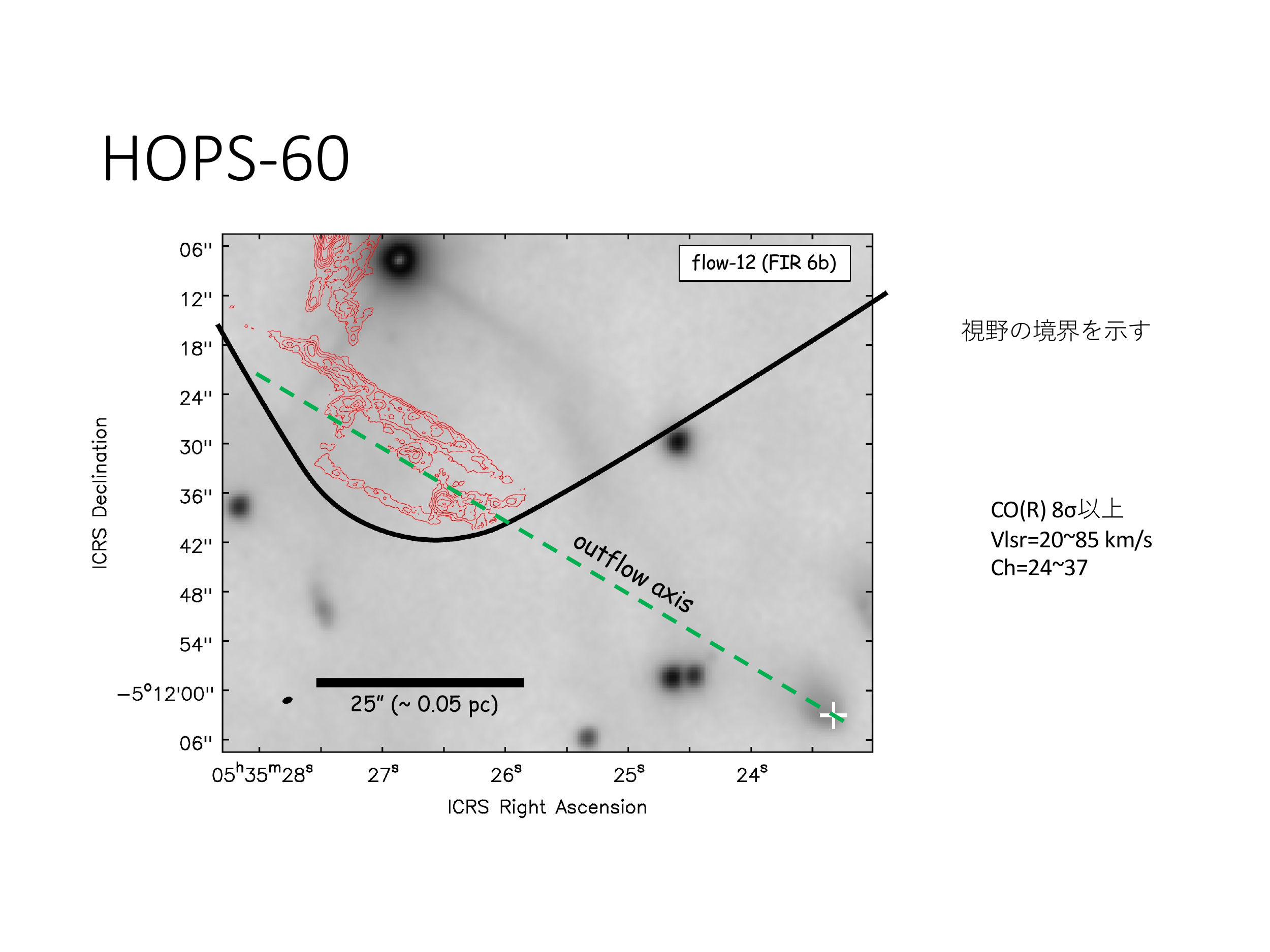}
    \caption{flow-12 (Clear). The gray background shows the 2.2\,\micron\,$Ks$ band image obtained from SIRIUS/IRSF  \citep{takahashi2008}. The white cross is the position of the outflow driving source (FIR\,6b, \citealt{Matsushita2021}). The red contours represent the integrated intensity of red-shifted CO components obtained from the ALMA 12-m array (\vlsr = 20--85\,\kms). The red contour levels are [8, 12, 16, 20, 24, 28, 32, 36, 40, 44]$\times 1\sigma$ ($1\sigma = 0.05$\,\jykms). The black solid line is the boundary of our observation mapping. The green dashed line represents the outflow axis. The black ellipse at the bottom-left corner is the same as that in Figure\,\ref{LINE-1}.}
    \label{flow-12}
\end{figure}

Based on the criteria, we have identified 12 CO molecular outflows as summarized in Figure\,\ref{line-2}. Zoomed-in images are presented in Figures\,\ref{flow1-3}--\ref{flow-12}. 
Of the 12 CO molecular outflows, we have newly identified six in the CO emission: flow-4, -5, -7, -8, -9, and -10. 
Table\,\ref{out-1} summarizes the outflow identifications and includes the names of the identified outflows, candidate driving sources, and comparisons with previous multi-wavelength studies, specifically the CO, SiO, 2.12\,\micron, $v$ = 1-0 $s$(1) $\mathrm{H_2}$ line emission, and centimeter continuum emission \citep{yu1997, reipurth1999, stanke2002, takahashi2008, shimajiri2008, osorio2017, tanabe2019, tobin2019, feddersen2020, kang2021, habel2021, Matsushita2021}. 
The outflow detection rate with respect to the FraSCO sources is estimated to be 24$\,\%$.
Outflow properties such as the outflow position angle and size, the maximum gas velocity, and the dynamical timescale for the outflow are summarized in Table\,\ref{out-2}. 
The position angle of the outflow (P.A.) was measured from the north 
and is positive when the rotation is counterclockwise ($-$180--180\,deg.). The projected outflow length, $L_{\mathrm{proj}}$, was measured using the integrated intensity maps (Figures\,\ref{flow1-3}--\ref{flow-12}). Along the position angle, $L_{\mathrm{proj}}$ was measured up to the lowest contour level presented  in each figure. 
The projected maximum gas velocity from the outflow, $v_{\mathrm{max,proj}}$, is taken from the difference between the absolute value of the maximum LSR velocity and \vsys. The maximum LSR velocity was determined based on the CO channel maps (see Appendix\,\ref{app-ch}) where the outflow emission is greater than 4$\sigma$. 
The projection uncorrected outflow dynamical timescale, $t_{\mathrm{dyn,proj}}$, is estimated from $L_{\mathrm{proj}}$/$v_{\mathrm{max,proj}}$. 
The projection corrected outflow dynamical timescale, $t_{\mathrm{dyn,corr}}$, is estimated from ($L_{\mathrm{proj}}$/cos$\theta$)/($v_{\mathrm{max,proj}}$/sin$\theta$). 
Here, $\theta$ = 30 deg. was adopted as a representative inclination angle of the outflow. This inclination angle was taken from a recent study by \cite{tobin2020b} measuring the disk inclination angle peaking at $\sim$ 60\,deg. (30\,deg.  in terms of the outflow inclination defined here). Note that there is no clear correlation between the elongation of the dust filamentary structures and the position angle of the detected outflows. 
Below, we present detailed results of the identified individual outflows.

\textbf{flow-1} (Probable; Figure\,\ref{flow1-3}a): A blue- and red-shifted cavity-like structure colliding with flow-2 blue-shifted emission, observed around R.A. = $05^{h}35^{m}28^{s}.535$, Dec. = $-05 \tcdegree 08'57''.687$. 
The emission was detected in the LSR velocity range 0--30\,$\mathrm{km}\,\mathrm{s^{-1}}$. Due to the limited observed imaging area, our image appears to only cover half of the lobe extending in the south-west direction with respect to the candidate driving source. 
After checking previous wide-field observations \citep{stanke2002}, the outflow is most likely driven by CRW FIR\,3 (also known as HOPS-350). 
Further evidence of mass ejection phenomena was reported from previous $\mathrm{H_2}$ line (2.12\,\micron, v=1-0 s(1)) observations by \cite{stanke2002}. Their $\mathrm{H_2}$ line image shows an ``S''-shaped structure, which is also seen in the $Ks$ image presented in Figure\,\ref{flow1-3}. The detected CO cavity-like structure spatially correlates well with a part of the S-shaped structure. Assuming that the CO outflow is driven by HOPS-350, the projected outflow length of the south-west lobe is $\sim 58''$ ($\sim$23200\,au).

\textbf{flow-2} (Clear; Figure\,\ref{flow1-3}b): flow-2 is driven by a binary system consisting of FraSCO-4 and FraSCO-6 (also known as HOPS-66B and HOPS-66A). The identified outflow elongates along the east--west direction. 
The CO blue-shifted emission extends to $\sim 26''$ ($\sim$10400 au)  and the red-shifted emission extends to $\sim 19''$ ($\sim$7600\,au). The CO emission is detected in the LSR velocity with a range of $-$70--35\,$\mathrm{km}\,\mathrm{s^{-1}}$.
The red-shifted component elongated in the east direction clearly shows the emission driven from both FraSCO sources (channel maps at \vlsr = 20, 25, and 30\,km\,s$^{-1}$ in the online journal). 
This component collides and penetrates through a blue-shifted outflow originating from another outflow, flow-3, located $\sim 10''$ east of the binary system. The blue-shifted emission elongated to the west seems to be mainly driven by FraSCO-6 (i.e., the southern component of the binary system). The blue-shifted component shows complicated internal structures such as blobs and wiggly structures within the lobe. 
Episodic mass ejection is suggested in the western lobe where the blue-shifted CO emission is located \citep{habel2021}.

\textbf{flow-3} (Clear; Figure\,\ref{flow1-3}c): flow-3 is driven by a bright 1.3\,mm source, FrasCO-16 (also known as HOPS-370). The CO outflow is elongated to the north-east and south-west direction. The CO emission is detected in the LSR velocity range $-$40--85\,\kms. 
The lobes are bright in both the red- and blue-shifted emissions, and hence the outflow is considered to be aligned closer to the plane of the sky. 
Adopting an inclination angle of 10 deg. with respect to the plane of the sky, the corrected dynamical timescale is three times shorter than the corrected dynamical timescale for an inclination of 30 deg listed in Table\,\ref{out-2}. 
 The north-east lobe extends to $\sim 48''$ ($\sim$19200\,au) and shows a U-shaped outflow lobe, particularly clear in the blue-shifted emission, while the red-shifted emission is mainly distributed in the right side of the north-east lobe. Note that the cavity-like structure from another outflow, flow-1, overlaps with this red-shifted component. 
 The lobe located in the south-west part of FraSCO-16 extends to $\sim 37''$ ($\sim$14800\,au) and is slightly compact compared to the other side of the lobe. 
 In addition, the south-western lobe appears to be more collimated than the north-east lobe. 
 The shape of the eastern edge of the south-west lobe overlaps identified 1.3\,mm continuum sources aligned within the dust filamentary structure (see Section \ref{dis-veri} for a possible interpretation).
 The left side of the south-west lobe is bright in both the red- and  blue-shifted emissions, while the right side is bright mainly in the red-shifted emission from the gas in the root of the outflow, and bright in the blue-shifted emission at the tip of the outflow lobe. 
 A compact SiO emission was detected at FraSCO-16. It extends to $\sim 3''$ ($\sim$ 1200\,au). 
 The emission was detected in an LSR velocity range of $-$1--16\,\kms. 
 The SiO emission is elongated perpendicular to the CO outflow axis and shows a velocity gradient across the major axis of the SiO component.
 This SiO emission might be not related to the CO outflow, but a rotational envelope. 
 
 \textbf{flow-4} (Probable; Figure\,\ref{flow4-7}a): This compact and elongated red-shifted CO emission is likely associated with FraSCO-32 and hence we identify this as a probable candidate. The emission is distributed to the north-west direction with respect to FraSCO-32 (also known as HOPS-64). It extends to $\sim 2''.7$ ($\sim$ 1080\,au) and shows a velocity range of \vlsr = 18--30\,$\mathrm{km}\,\mathrm{s^{-1}}$. There is an extended blue-shifted emission in this region; however, no clear collimated lobe-like structure associated with the driving source candidate was detected.  

\textbf{flow-5} (Probable; Figure\,\ref{flow4-7}b): A collimated monopolar outflow associated with the compact 1.3\,mm source FraSCO-33, which was detected as a millimeter source for the first time by our study. The CO emission is distributed in the west direction with respect to FraSCO-33. 
The outflow extends to  $\sim 6''$ ($\sim$2400\,au) and the blue-shifted emission velocity reaches the LSR velocity of 2\,\kms. 
Collimated SiO emission is detected in both the blue- and red-shifted emissions. 
The blue- and red-shifted SiO emission is elongated in the westward direction by $\sim 3''.8$ ($\sim 1520$\,au) and $\sim 2''$ ($\sim 800$\,au), respectively with respect to FraSCO-33. 
The blue- and red-shifted velocities reach \vlsr = 2\,$\mathrm{km}\,\mathrm{s^{-1}}$and \vlsr = 16\,\kms.
This region is strongly affected by the shocked gas (shock-6, as discussed in Section\,\ref{result-line-shock}), and hence it is difficult to disentangle the emission from the outflow and shocked components. 
However, the detected CO and SiO emissions show compact localized emission peaks associated with FraSCO-33.
The first peaks associated with FraSCO-33 (Figure\,\ref{flow4-7}b) most probably originate from flow-5, while some of more extended emissions are possibly associated with shock-originated extended gas (i.e., shock-6 as explained in Section\,\ref{result-line}).  
Therefore, we identified the emission as an outflow.

\textbf{flow-6 } (Probable; Figure\,\ref{flow4-7}c): The driving source of flow-6 is considered to be FraSCO-37 (also known as HOPS-108). This region shows localized CO blue- and red-shifted emissions. Although the emission peaks are not directly associated with FraSCO-37, the emission shows elongated structures in the north-east and south-west directions with a projected length of $\sim 7''$ ($\sim$2800\,au) centered at FraSCO-37. The red- and blue-shifted emissions reach \vlsr = 50\,$\mathrm{km}\,\mathrm{s^{-1}}$ and \vlsr = $-$7\,\kms, respectively. A compact SiO emission, extending to $\sim 1''.2$ ($\sim$ 480\,au), associated with FraSCO-37 is also detected in the LSR velocity ranges 1--3 and 12--18\,\kms. The spatial distribution is not consistent with the CO emission, but is perpendicular to the elongation of the CO emission. 
\cite{osorio2017} detected centimeter continuum emission from a non-thermal origin. They reported two emission peaks, denoted by green filled squares in Figure\,\ref{flow4-7}c, which spatially coincide with locations of one of the blue- (north-east) and red-shifted (south-west) CO lobes identified as flow-6.

\textbf{flow-7} (Marginal; Figure\,\ref{flow4-7}d): This blue-shifted emission shows a butterfly wing-like structure in the integrated intensity  image. The blue-shifted emission reaches \vlsr = $-$14\,\kms. 
The possible driving source is FraSCO-38, which was detected as a millimeter source for the first time by our study. The compact blue-shifted emission is located in the most complex area. It is difficult to distinguish whether the blue-shifted component is actually associated with FraSCO-38 or originates from the surrounding environment. No compact localized red-shifted emission is detected around FraSCO-38.

\textbf{flow-8} (Clear; Figure\,\ref{flow-8}):
flow-8 is a collimated bipolar CO outflow associated with FraSCO-39 (known as VLA 16). 
The blue- and red-shifted emissions are located in the north-west and the south-east directions with respect to FraSCO-39, extending with a projected length of $1''.25$ ($\sim$ 500\,au) and $6''.5$  ($\sim$ 2600\,au), respectively.
The blue- and red-shifted emissions are detected in the LSR velocity ranges of 8--9\,$\mathrm{km}\,\mathrm{s^{-1}}$ and 13--17\,\kms, respectively.
A compact SiO emission was also detected just next to the CO blue-shifted emission (LSR velocity range of 10--12\,\kms). However, the emission is associated with neither the candidate driving source nor the CO blue-shifted emission, and hence the SiO emission is not likely associated with the outflow.

\textbf{flow-9} (Probable; Figure\,\ref{flow-9}):
 An elongated structure with a length of $\sim 80''$ ($\sim$ 32000\,au) was detected in the red-shifted emission with an LSR velocity range of 13--60\,\kms.
 The structure is prominent in the observed region, and the corresponding features were also detected in the 2.12\,\micron\,$\mathrm{H_2}$ line image \citep{stanke2002}. 
Assuming that the emission originates from a strongly collimated red-shifted CO outflow, FraSCO-40 (known as VLA 15), located at the northern end of the collimated emission, is the only candidate driving source. Checking the other side, we also found a blue-shifted emission although the emission is difficult to disentangle from the extended emission originating from the cloud velocity. 
Interestingly, we also detected a very collimated knot-like SiO blue-shifted emission with an LSR velocity range of $-$5--5\,$\mathrm{km}\,\mathrm{s^{-1}}$ (see channel maps in the online journal). This emission extends to the north with $L_{\mathrm{proj}} \sim  14''$ ($\sim$ 5600\,au) and is associated with bright components (more than 0.2\,\jy) of the blue-shifted CO emission with an LSR velocity in the range $-$1--3\,\kms. No counterpart (i.e., red-shifted SiO emission extending to the south) was detected from our observations. 

\textbf{flow-10} (Clear; Figure\,\ref{flow-10}): 
A compact bipolar outflow driven by FraSCO-45 (also known as HOPS-369). 
The blue-shifted emission has two peaks and is elongated in the north-east direction with an LSR velocity range of 3--7\,\kms, extending with a projected length of $2''$ ($\sim$ 800\,au). 
The red-shifted emission peaks at the location of FraSCO-45. It has an LSR velocity range of 13--21\,\kms and extends $L_{\mathrm{proj}} \sim 2''$ ($\sim$ 800\,au). This is the most compact outflow identified in our study. 

\textbf{flow-11} (Clear; Figure\,\ref{flow-11}):
 A clear bipolar CO outflow driven by FraSCO-49 (also known as HOPS-368 and VLA 13). 
 The blue-shifted emission with $L_{\mathrm{proj}} \sim 20''$ ($\sim$ 8000\,au) and the red-shifted emission with $L_{\mathrm{proj}} \sim 15''$ ($\sim$ 6000\,au) extend to the north and south with respect to FraSCO-49, respectively. The CO emission traces a cone shaped outflow cavity structure, which was previously detected in the $Ks$ image (e.g., \citealt{takahashi2008}). The blue- and red-shifted emissions have LSR velocities up to 0\,$\mathrm{km}\,\mathrm{s^{-1}}$and 30\,\kms, respectively. A compact SiO red-shifted emission, extending to $L_{\mathrm{proj}} \sim 3'' (\sim$ 1200\,au) with \vlsr = 15--35\,\kms, was detected $\sim 10''$ south of the driving source. This emission is considered to trace shocked gas within the outflow. 
 Note that another compact SiO red-shifted emission was detected $\sim 20''$ ($\sim 8000$\,au) south-east of the driving source. This emission is not associated with CO emissions originating from flow-11, thus we identified this emission as shocked gas not originating outflow, shock-11 (see Section\,\ref{result-line-shock}).

\textbf{flow-12} (Clear; Figure\,\ref{flow-12}):
This is a known CO outflow driven by FIR\,6b also known as HOPS-60 (e.g., \citealt{takahashi2008, feddersen2020, kang2021, Matsushita2021}). Due to the limited imaging, we have imaged only the northern half of the red-shifted CO lobe. The red-shifted emission extends to $\sim 90''$ ($\sim$ 36000\,au) in the north-east direction with respect to the driving source. A cavity-like structure appears with an LSR velocity range of 25--35\,\kms\,and a collimated structure appears within the cavity-like structure with an LSR velocity range of 60--85\,\kms.
Our results are consistent with those in a recent study \citep{matsushita2019}.

\begin{table}
	\centering
	\caption{Outflow properties. In the identification columns, C, P, and M indicate structures identified as clear, probable, and marginal, respectively. A check mark indicates that the corresponding line/continuum emission has been detected  in previous works.} 
	\label{out-1}
	\footnotesize
	\begin{tabular}{l|cccc|cccc} 
	& \multicolumn{2}{c}{Identification} & \multicolumn{2}{c}{Driving source candidate} &  \multicolumn{4}{c}{Counterpart}\\
	name &  CO  & SiO & HOPS* & FraSCO  & CO & SiO & $\mathrm{H_2} $ & 3\,cm, 5\,cm  \\ 
		\hline
		 flow-1 & P &M & HOPS-350 & - &$\surd$ & & $\surd$ \\
		 flow-2 &C &-  &HOPS-66A/B & FraSCO-4/6 & $\surd$ & & $\surd$ \\
		 flow-3 &C &C** &HOPS-370 & FraSCO-16  & $\surd$ & & $\surd$ &   $\surd$ \\
		 flow-4 &P &-  & HOPS-64& FraSCO-32 & &  &$\surd$ &$\surd$\\
		 flow-5 &P & P & - & FraSCO-33  &\\
		 flow-6 &P & P & HOPS-108 & FraSCO-37&    $\surd$ & &  & $\surd$ \\
		 flow-7 &M & - & - & FraSCO-38 &    \\
		 flow-8&C & M & - &FraSCO-39 & && $\surd$\\
		 flow-9 & P & P  &- & FraSCO-40  &    &$\surd$ & & $\surd$  \\
		 flow-10 &C & -  &HOPS-369& FraSCO-45 &  & & $\surd$  & \\
		 flow-11 &C & M  &HOPS-368& FraSCO-49  & $\surd$ & & $\surd$ & $\surd$  \\
		 flow-12 &C &-  &HOPS-60& - &$\surd$ & & $\surd$  \\
	\end{tabular}
	\tablenotetext{*}{Infrared source \citep{megeath2012, furlan2016}}
	\tablenotetext{**}{The compact SiO emission is elongated perpendicular to the CO outflow axis.}
\end{table}

\begin{table}
	\centering
	\caption{Physical properties of the identified CO outflows. R and B correspond to the measured value in the red-shifted and the blue-shifted components, respectively. }
	\label{out-2}
	\footnotesize
	\begin{tabular}{lc|ccccc} 
	name & R or B & P.A. [deg]&$L_{\mathrm{proj}}$  [au] & $v_{\mathrm{max,proj}}$ [\kms] & $t_{\mathrm{dyn,proj}}$ [yr]  & $t_{\mathrm{dyn,corr}}$ [yr]\\
	\hline
	\hline
	flow-1 &R & $-148$& 22400 & 19 & 5.6E+04	& 3.2E+04\\
	flow-1 &B &  $-150$& 23200 & 11 &1.0E+05	&5.8E+04\\
	\hline
	flow-2 &R & 104 & 7600 & 24 & 1.5E+04&	8.7E+03\\
	flow-2 &B & $-76$& 10400& 81 & 6.1E+03	& 3.5E+03\\
	\hline
	flow-3 (north-east) & R & 27 & 19200 & 34 &2.7E+04&	1.6E+04\\
	flow-3 (north-east) & B &37 & 19200&  51& 1.8E+04	&1.0E+04 \\
	flow-3 (south-west) & R & $-154$&  12000 & 74 & 7.7E+03	 & 4.5E+03\\
	flow-3 (south-west) & B & $-144$  & 14800 & 16 & 4.4E+04&	2.5E+04\\
	\hline
	flow-4 &R & $-48$& 1080& 19 & 2.7E+03&	1.6E+03\\
	flow-4 &B & - & - & -& -& -\\
	\hline
	flow-5 &R & - & - & -& -& -\\
	flow-5 &B & $-90$&2400 & 9 & 1.3E+04	&7.3E+03\\
	\hline
	flow-6 &R & $-150$& 2800& 39 &3.4E+03	&2.0E+03\\
	flow-6 &B & $-150$  & 2600 & 18 &6.9E+03	&4.0E+03\\
	\hline
	flow-7 &R & - & - & -& -& -\\
	flow-7 &B &$-5$5&2000& 25 & 3.8E+03	&2.2E+03\\
	\hline
	flow-8 &R &135&2600& 6 & 2.1E+04	 & 1.2E+04\\
	flow-8 &B &  135 & 500 & 5 &4.8E+03&	2.7E+03\\
	\hline
	flow-9 &R & 155& 32000 &  49 & 3.1E+04	&1.8E+04 \\
	flow-9 &B  &-& - & - & - & -\\
	\hline
	flow-10 &R &44&800& 8 & 4.8E+03	&2.7E+03\\
	flow-10 &B &44 & 800 & 10 & 3.8E+03&	2.2E+03\\
	\hline
	flow-11 &R & 139&6000 &  19 &1.5E+04	& 8.7E+03 \\
	flow-11&B & $-5$ & 8000 & 11 & 3.5E+04	&2.0E+04\\ 
	\hline
	flow-12 &R & 60& 36000& 74 & 2.3E+04&	1.3E+04\\
	flow-12 &B& - & - & -& -& -\\
	\end{tabular}
\end{table}

\subsubsection{Non Outflow Emission Originating from Shocked Gas}
\label{result-line-shock}
In addition to identifying outflows, we identified SiO shocked gas structures that do not originate from outflows.
To identify shocked gas structures, we set the following three criteria: \textbf{(1)} SiO emission detected at greater than $10\sigma$ in the integrated intensity map (Figure\,\ref{line-2}), \textbf{(2)} SiO emission shows a gas velocity greater than 2\,\kms\,with respect to the systemic velocity, and \textbf{(3)} SiO gas structure is not associated with the identified outflows in Section \ref{result-line-outflow}.

\begin{figure}
    \centering
    \includegraphics[width=18cm]{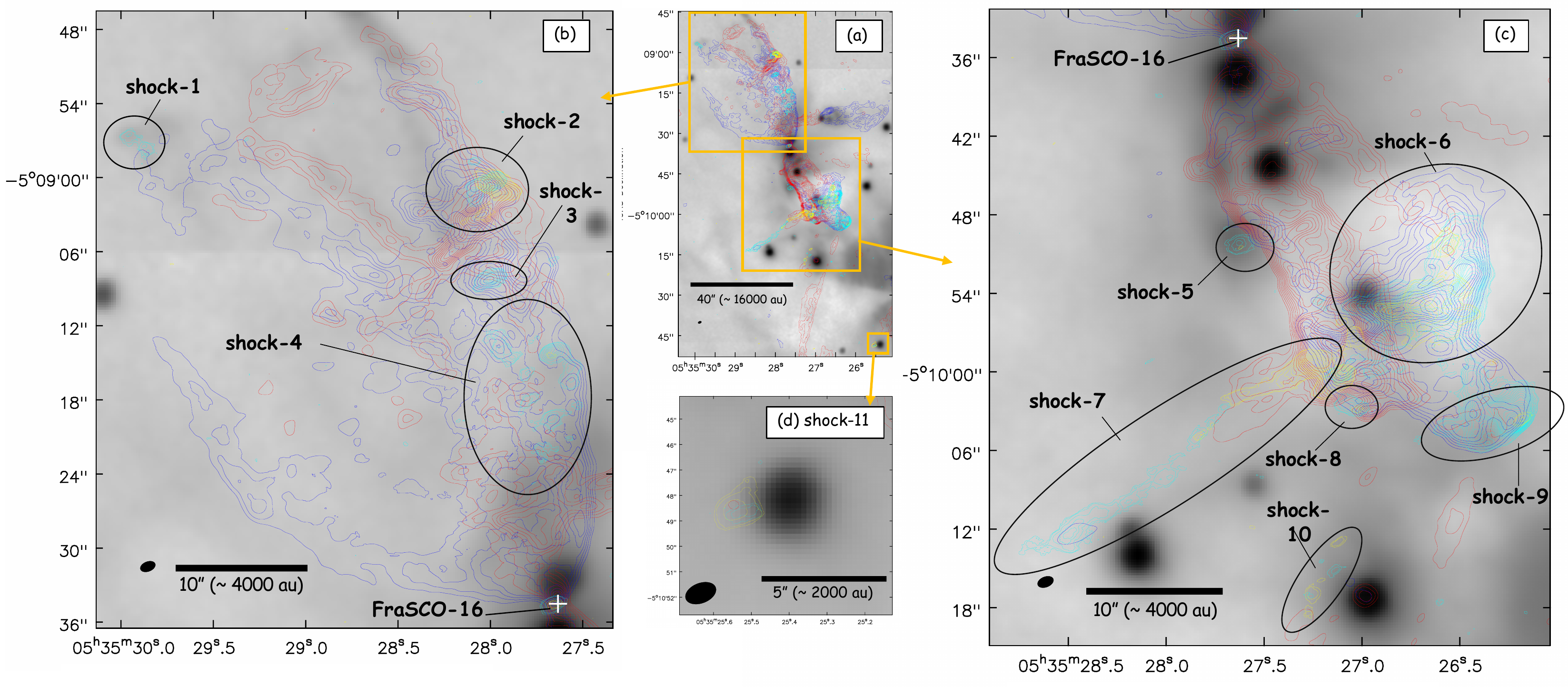}
    \caption{Interacting structure identification. The data and contour levels are the same in each panel. The white cross is the position of the flow-3 driving source (FraSCO-16). The gray background images show the 2.2\,\micron\,$Ks$ band image obtained from SIRIUS/IRSF  \citep{takahashi2008}. The red and blue contours are the same as those in Figure\,\ref{LINE-1}. The yellow contours represent the integrated intensity map using the red-shifted SiO components with a velocity range of \vlsr = 12--29\,\kms.  The yellow contour levels are  [8, 10, 15, 20, 25, 30, 35, 40, 45, 50, 55] $\times 1\sigma$ ($1\sigma = 0.015$\,\jykms ). The cyan contours represent the integrated intensity map using the blue-shifted SiO components with a velocity range of \vlsr = $-$30--11\,\kms. The cyan contour levels are [8, 10, 15, 20, 25, 30, 35, 40, 45, 50, 55, 60, 70, 80, 90, 100, 120, 140, 160, 180, 200] $\times 1\sigma$ ($1\sigma = 0.02$\,\jykms ). In all the figures, the black ellipses at the bottom-left corner are the same as that in Figure\,\ref{LINE-1}.}
    \label{interact-1}
\end{figure}

Based on these criteria, we identified 11 shocked gas structures. 
Overview and zoomed-in images are presented in Figure\,\ref{interact-1}. 
Table\,\ref{interact-2} lists the sizes of the shocked regions measured using the integrated intensity map (Figure\,\ref{interact-1}) showing SiO emission detection greater than the 8$\sigma$ level. 
The maximum velocity, $v_{\mathrm{max}}$, is defined as the difference between the maximum LSR velocity and the systemic velocity, for SiO emission greater than 4$\sigma$ in the SiO channel maps.
Below, we provide detailed results for the individual identified SiO shocked gas structures.

\textbf{shock-1} (Figure\,\ref{interact-1}b): 
shock-1 was detected at the tip of the blue-shifted outflow, flow-3, driven by FraSCO-16. The SiO emission extends to $\sim$ 800\,au, showing a blue-shifted component with a velocity range of $-$4--10\,\kms. Since the emission is detected at the edge of the outflow, the emission might originate from the interacting region between the outflow and surrounding ambient gas.
Faint emission associated with SiO emission was detected in the $Ks$ band, including a 2.12\,\micron $\,\mathrm{H_2}$ shock originated emission.

\textbf{shock-2} (Figure\,\ref{interact-1}b): 
shock-2 was detected $\sim 34''$ north of FraSCO-16.  The SiO emission shows both blue- and red-shifted emissions with an LSR velocity range of $-$5--23\,\kms\,. The spatial distribution of the SiO emission seems to spatially correlate with the tip of a cavity like structure, possibly originating from HOPS-350 (flow-1), which seems to interact with a blue-shifted outflow emission originating from FraSCO-16, flow-3.
No obvious $Ks$ band emission associated with the SiO emission was detected in shock-2. 

\textbf{shock-3}  (Figure\,\ref{interact-1}b): 
 shock-3 was detected $\sim 27''$ north of FraSCO-16. The SiO emission extends to $\sim 2''.5$ and was only detected in the blue-shifted component with a velocity range of $-$1--6\,\kms. The origin of the emission is uncertain, but it may trace the post shock region produced by the outflow interaction between flow-1 and flow-3, although the spatial correlation between the outflow cavity and the detected emission is not as clear as shock-1.
No obvious $Ks$ band emission associated with the SiO emission was detected in shock-3.

\textbf{shock-4}  (Figure\,\ref{interact-1}b): 
shock-4 was detected $\sim 15''$ north of FraSCO-16.
It consists of several compact components with a typical size of $\sim 2''$ ($\sim$ 800\,au) and distributed over $\sim 9''$ ($\sim$ 3600\,au).
The SiO emission was only detected in the blue-shifted component with a velocity range of 0--13\,\kms. 
The detected SiO emission is located in the northern area of the region where flow-2 and flow-3 collide, and hence the origin of the shocked SiO gas might be related to the interaction between those two outflows.
No obvious $Ks$ band emission associated with the SiO emission was detected in shock-4.

\textbf{shock-5} (Figure\,\ref{interact-1}c): 
 shock-5 was detected $\sim 16''$ south of FrasCO-16. The SiO emission was mainly detected as a blue component with a velocity range of 9--12\,\kms, while a red-shifted component was detected marginally with a 3$\sigma$ emission level in the integrated intensity map. 
 The structure is as compact as $\sim 2''.5$ ($\sim$ 1000\,au).  A faint emission associated with the SiO emission was detected in the $Ks$ band. This compact SiO emission is located just outside the outflow lobes of flow-3 detected with both red- and blue-shifted CO components.
This SiO emission seems to show a local shock created by the outflow lobe from flow-3 interacting with ambient dense gas.
Indeed, the CO lobe from flow-3 detected for both blue- and red-shifted gas shows a clear anti-correlation with the dust lane traced by the 1.3\,mm continuum emission (further discussion in Section\,\ref{dis-out}). 
The SiO emission is located between the CO outflow and dust lane. This also supports  the suggestion that the local shock is produced by the interaction between the outflow and the dust lane.

\textbf{shock-6} (Figure\,\ref{interact-1}c): 
shock 6 was detected $\sim 23''$ south-west of FraSCO-16. The detected SiO emission is the most extended and brightest component in the observed area. The emission extends to $\sim 10''$ ($\sim$ 4000\,au). The SiO emission was previously detected in SiO\,($J$ = 2--1) by \cite{shimajiri2008} with a velocity range of 4--13\,\kms. 
Our observations also confirmed this detection with a higher transition of SiO\,($J$ = 5--4). The emission was detected with both blue- and red-shifted velocity and the LSR velocity range is 0--13 \kms. 
The spatial distribution of the SiO blue-shifted emission is well correlated with that of the CO blue-shifted emission.
Their velocity coverage is the same. 
In contrast, there is no clear spatial correlation between the SiO blue-shifted emission and the CO red-shifted emission, but rather an anti-correlated distribution is suggested between CO (mainly tracing the flow-3 lobe) and SiO (shock originated gas distributed at the edge of outflow lobe).
The interaction between flow-3 and dense gas in the FIR\,4 region was suggested by \cite{shimajiri2008} based on previous SiO observations and also \cite{nakamura2019} using multi molecular line observations. 
Our SiO result, showing an arc-like structure (particularly recognized around the systemic velocity), geometrically suggests an interaction between flow-3 and dense gas from the FIR\,4 region. 
Our result does not conflict with the proposed scenario by \cite{shimajiri2008}.
It is important to note that two outflows (flow-4 and flow-5; see Figure\,\ref{flow4-7}) are identified within the FIR\,4 region, where strong CO and SiO emissions are located at shock-6.
It is difficult to disentangle the CO and SiO blue-shifted emissions associated with shock-6 and flow-5, because they overlap spatially and also in the velocity domain.
Moreover, another outflow, flow-9, shows a chain-like collimated outflow (see Figure\,\ref{flow-9}). 
The distribution of the blue-shifted collimated SiO emission reaches shock-6.
Our results show very complicated SiO and CO emission distributions within a protocluster star forming environment. Further discussion and interpretation will be presented in Section \ref{dis-veri} and \ref{dis-frag}.

\textbf{shock-7} (Figure\,\ref{interact-1}c):
The SiO emission for shock-7 shows an elongated structure with an overall length of $\sim 28''$ ($\sim$11200\,au). This structure is located south-east of shock-6. The SiO emission shows an LSR velocity range of 5--21\,\kms\, with a smooth velocity gradient. The red-shifted velocity component seems to be connected to the velocity component observed in shock-6, and hence both emissions may be related. The elongated structure showing both blue and red-shifted components seems to be explained by the molecular outflow; however no driving source candidate was found in the region searching with multi-wavelength images.

\textbf{shock-8} (Figure\,\ref{interact-1}c):
shock-8 was detected $\sim 30''$ south of FrasCO-16. The SiO emission was mainly detected in the blue-shifted component with a velocity range of 3--12\,\kms, while a red-shifted component was detected marginally with a 3$\sigma$ emission level in the integrated intensity map.
The structure is as compact as $\sim 2''$ ($\sim$ 800\,au). 
This compact SiO emission seems to be located at the tip of flow-3
and seems to show a local shock created by the outflow lobe from flow-3 interacting with ambient dense gas.

\textbf{shock-9} (Figure\,\ref{interact-1}c): 
shock-9 was detected $\sim 35''$ south-west of FraSCO-16. 
The blue-shifted component of the SiO emission extends $\sim 8''$ ($\sim$ 3200\,au) in the east--west direction with an  LSR velocity range of  $-$14--16\,\kms. 
The spatial distribution of the blue-shifted component is well correlated with that of the CO blue-shifted emission, as shown in Figure\,\ref{interact-1}c.
A compact red-shifted component of SiO emission is associated with  the blue-shifted component at R.A. = $05^{h}35^{m}26^{s}.179$, Dec. = $-05 \tcdegree 10'03''.638$.
Note that the SiO emission originating from flow-9 overlaps  with the blue-shifted component of shock-9 in the south--north direction (Figure\,\ref{flow-9}).
shock-9 was newly resolved as an isolated feature in an SiO emission.
shock-9 seems to show a local shock created by the outflow lobe from flow-3 interacting with ambient dense gas.

\textbf{shock-10} (Figure\,\ref{interact-1}c): 
shock-10 was detected at R.A. = $05^{h}35^{m}27^{s}.193$, Dec. = $-05 \tcdegree 10'15''.172$ and consists of a few blobs with a typical size of $\sim 1''.5$ ($\sim$ 600\,au).
The LSR velocity of the detected blobs has a range of  10--14\,\kms, which is close to the systemic velocity.
The blobs are distributed along the north-west to south-east direction.
The origin of the SiO emission potentially tracing shocked gas is not clear, because there is no direct evidence of gas interaction caused by an outflow.

\textbf{shock-11} (Figure\,\ref{interact-1}d):
shock-11 was detected at R.A. = $05^{h}35^{m}25^{s}.527$, Dec. = $-05 \tcdegree 10'48''.637$.
It extends $\sim 2''.1$ ($\sim$ 840\,au) with an LSR velocity range of 10--15\,\kms.
This compact emission is also detected in the CO emission with the same velocity range.
Around the detected SiO emission, there is a 2MASS source located $\sim 2''$ west of shock-11.
This SiO emission may be explained by an outflow from this source.
However, the 2MASS source is detected in the $J$, $H$, and $Ks$ bands \citep{nielbock2003},
suggesting the source is evolved and not likely associated with the dense envelope.
In addition, the SiO emission is only detected with the relative velocity of $\le$ 4\,\kms.
Therefore, the emission likely does not originate from a molecular outflow driven by this 2MASS source.
The origin of this emission is uncertain.

\begin{table}
	\centering
	\caption{Physical properties of the identified SiO shocked gas. We did not assume inclination angles to determine the sizes and velocities. Only shock-6 has been detected in previous SiO($J$=2--1) line observations.}
	\label{interact-2}
	\footnotesize
	\begin{tabular}{l|rrcc} 
	name &  projected size [au]  & \vlsr\,range  [\kms]   & comment  \\ 
		\hline
		shock-1 &  800&  -4 -- 10    &- \\
		shock-2  & 1400& -5 - 23   & -\\
		shock-3  & 1200&-1 -- 6   & -\\
		shock-4 &  800& 0 -- 13  & size is typical one\\
		shock-5 &  1000& 9 -- 12  &-  \\
		shock-6 & 4000  & 0 -- 13   & previously detected  \\
		shock-7 &  11200&  5 -- 21  & -\\
		shock-8 & 800 & 3 -- 12   &  -\\
		shock-9& 3200&  -14 -- 16  &-  \\
		shock-10 & 600& 10 -- 14 & size is typical one \\
		shock-11 & 840 & 10 -- 15  & - \\
	\end{tabular}
\end{table}

\section{Discussion}
\label{dis}
In this section, we discuss the star forming environment in the FIR\,4 region, which is presented in Figure\,\ref{cont-2}b.
The FIR\,4 region ($L = 1000$\,\lsun; \citealt{crimier2009}) is known to be the most bright and centrally concentrated dust condensation within the embedded protocluster \citep{chini1997, lis1998, johnstone1999}. 
Previous millimeter and sub-millimeter interferometric observations have suggested possible star formation activity within the FIR\,4 region \citep{takahashi2008, shimajiri2008, lopez2013, osorio2017, fontani2017, favre2018, tobin2019, evans2021, nakamura2019}. 
Multi-wavelength infrared and centimeter observations have also revealed that the region contains several sources, indicating the presence of protostellar candidates \citep{megeath2012, furlan2016, reipurth1999, osorio2017}. 
Our ALMA 12-m array observations, with an angular resolution of $\sim 1''$, spatially resolved substructures within the FIR\,4 region in the 1.3\,mm continuum emission.
In addition, the spatial distributions of outflow and shock-originated gas were traced by CO and SiO emissions.
In Section \ref{dis-veri} we use this dataset to discuss three previously proposed scenarios to explain the star formation environment in the FIR\,4 region.
In Section \ref{dis-frag}, we compare timescales to assess the previously proposed triggered star formation scenario within the FIR\,4 region. 
Finally, in Section \ref{dis-hub}, we discuss a hub-filament system in the FIR\,4 region.

\subsection{Star Formation Environment in the FIR\,4 Region}
\label{dis-veri}
Three possible scenarios have been proposed in previous studies to explain the origin of the protocluster in the FIR\,4 region.
The first scenario is the collision of an energetic outflow, flow-3, driven from FraSCO-16 (known as HOPS-370) in the FIR\,3 region, with the dust condensation, FIR\,4 \citep{shimajiri2008, tobin2019, nakamura2019}. 
\cite{shimajiri2008} found 11 dust condensations embedded within the FIR\,4 region. 
Since these condensations are located around the south-west tip of flow-3, it was proposed that the interaction triggered a fragmentation process within the protocluster and formed next generation protostars in the FIR\,4 region. 
The second scenario is that the large $L_{\mathrm{bol}}$ indicated in the FIR\,4 region ($L = 1000$\,\lsun) originates from an internal source \citep{lopez2013}. 
\cite{lopez2013} proposed that the FIR\,4 region could be an HII region powered by a B3--B4 type young star.
The third scenario is that the FIR\,4 region is irradiated by an external source but not by flow-3 \citep{fontani2017, favre2018, evans2021}. 
These three scenarios have been mainly examined with regard to (sub)millimeter interferometric observations with a single pointing.
However, some of the observations have an insufficient angular resolution of $1''.5$--$6''$ to spatially resolve substructures and shocked gas within FIR\,4 region \citep{shimajiri2008, lopez2013, favre2018, evans2021}.
ALMA 0.87 mm observations by \cite{tobin2019} and ALMA 1.3\,mm observations by \cite{tobin2020b} have a high angular resolution of $\sim 0''.25$ 
and can image dust emission and several molecular lines from sources embedded in the FIR\,4 region,
though their fields of view do not cover the whole FIR\,4 region, and therefore cannot be used to investigate each proposed scenario.
Our ALMA 12-m array observations covered a large spatial area to map the entire FIR\,3 and FIR\,4 regions with a sufficiently high angular resolution of $\sim 1''$ to spatially resolve substructures within each region.

In summary, we have spatially resolved the 1.3\,mm continuum, CO\,($J$ = 2--1), and SiO\,($J$ = 5--4) emissions across the FIR\,3 and FIR\,4 regions for the first time. 
Our dataset enables us to discuss the three previously proposed scenarios to explain the protocluster environment in the FIR\,4 region.
We compared the spatial locations between the shocked gas traced by SiO\,($J$ = 5--4), the outflow originating emission traced by CO\,($J$ = 2--1), and the dust filament traced in the 1.3\,mm continuum emission. 
Our results are most consistent with an interaction model between the energetic flow-3 and the dense condensations within the FIR\,4 region that is described in Section \ref{dis-out}. 
The schematic picture of the model is summarized in Figure\,\ref{dis-3}. 
In the following, we discuss the three proposed scenarios.

\begin{figure}
\begin{minipage}[b]{0.3\linewidth}
    \centering
    \includegraphics[width=16cm]{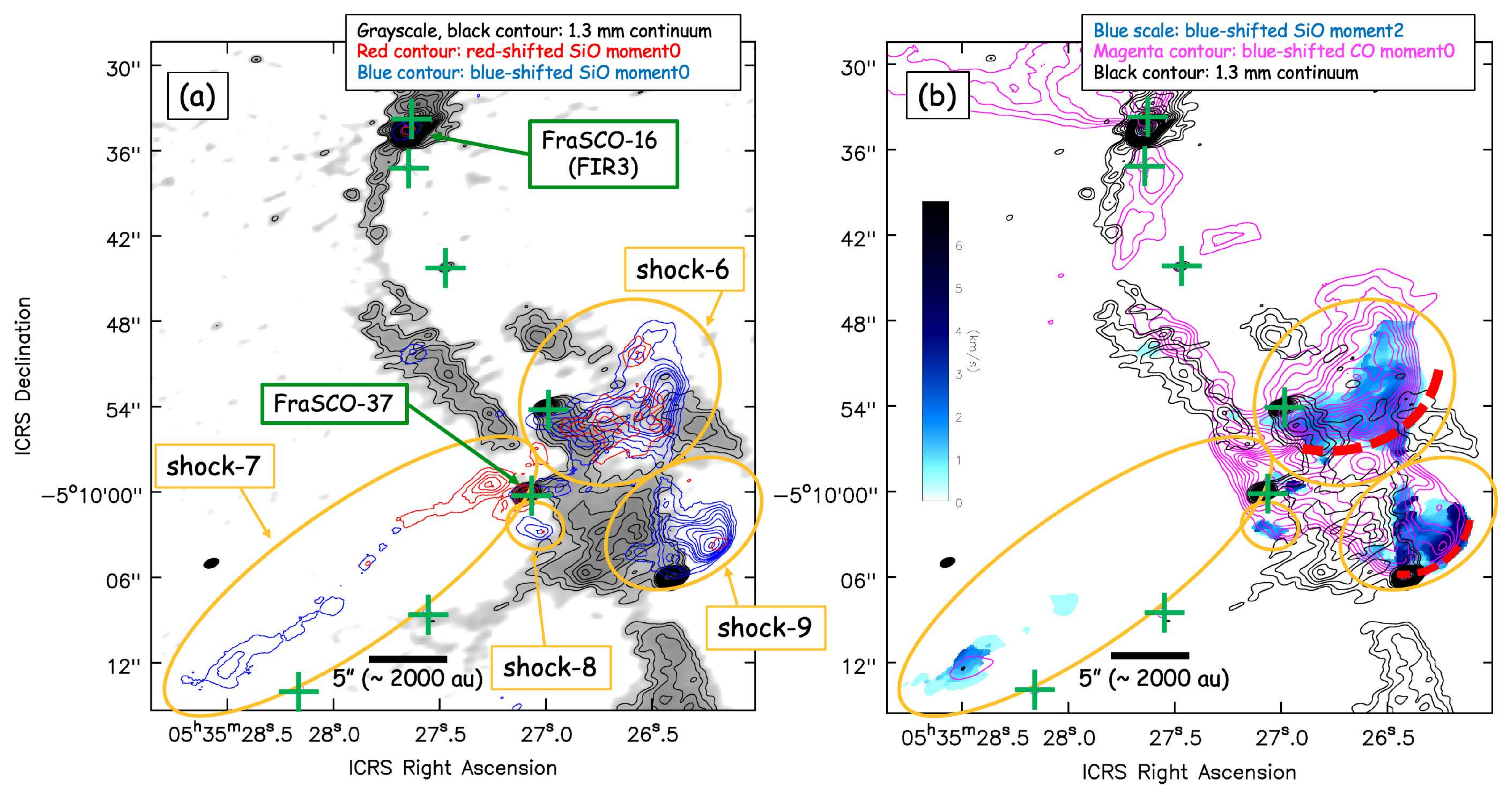}
\end{minipage}\\
\begin{minipage}[b]{0.3\linewidth}
    \centering
    \includegraphics[width=16cm]{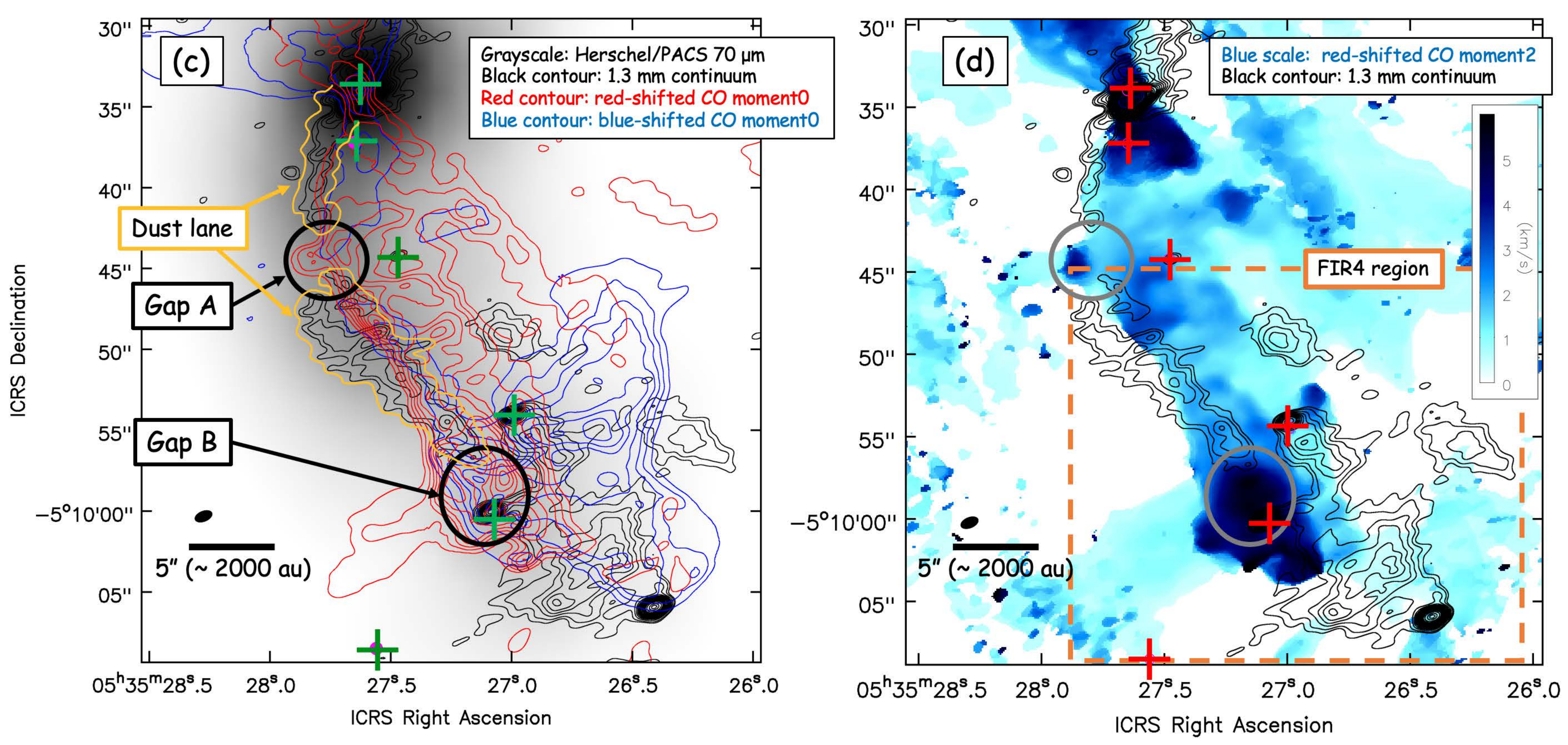}
\end{minipage}\\
    \caption{Spatial distribution comparisons between the 1.3\,mm continuum, CO\,($J$ = 2--1), SiO\,($J$ = 5--4), and Herschel/PACS 70\,\micron\, emissions. In all figures, the black contours represent the 1.3\,mm continuum emission and the contour levels are [7, 10, 15, 20, 25, 30, 35, 40, 45, 50, 60, 70, 80, 90, 100, 120, 150, 170, 200, 300, 400, 500, 600, 700]$\times 1\sigma$ (1$\sigma$ = 0.22\,\mjy). The black filled ellipses in the bottom-left corner show the synthesized beam size of the continuum image. The green and red crosses show the positions of previously identified infrared sources \citep{nielbock2003, furlan2016}. 
    \textbf{(a)}: The 1.3\,mm continuum image overlaid with the SiO\,($J$ = 5--4) moment 0 images separated with red- (\vlsr\,= 12--30\,\kms) and blue-shifted (\vlsr\,= $-$30--10\,\kms) components. The blue contour levels are [10, 20, 30, 40, 50, 60, 80, 100, 120, 160, 200]$\times 1\sigma$ (1$\sigma$ = 0.02\,\jy). The red contour levels are [10, 20, 30, 40, 50]$\times 1\sigma$ (1$\sigma$ = 0.015\,\jy). Locations of the identified shocked regions (shock-6, shock-7, shock-8, and shock-9) are denoted by yellow ellipses. 
    \textbf{(b)}: The 1.3\,mm continuum image overlaid with a moment 2 image obtained from the SiO\,($J$ = 5--4) blue-shifted component (\vlsr\,= $-$30--10\,\kms) and a moment 0 image obtained from the CO\,($J$ = 2--1) blue-shifted component (\vlsr\,= $-$10--10\,\kms). The magenta contour levels are [10, 15, 20, 25, 30, 35, 40, 45, 50, 55, 60, 65, 70, 75, 80]$\times 1\sigma$ (1$\sigma$ = 0.23\,\jy). The two red dashed lines represent U-shaped structures are mentioned in the main text (indicating interacted regions between outflows and surrounding dense materials). Locations of the identified shocked regions (shock-6, shock-7, shock-8, and shock-9) are denoted by yellow ellipses. 
    \textbf{(c)}: Spatial distribution comparisons between the Herschel/PACS 70\,\micron\,(grayscale), 1.3\,mm continuum (black contours), and CO\,($J$ = 2--1) moment 0 images (red and blue contours). The CO\,($J$ = 2--1) moment 0 images were produced with the red- (\vlsr\,= 15--85\,\kms) and blue-shifted (\vlsr\,= $-$10--10\,\kms) components separately. The red contour levels are [10, 20, 30, 40, 50, 60, 70, 80, 90]$\times 1\sigma$ (1$\sigma$ = 0.17\,\jy). The blue contour levels are [10, 20, 30, 40, 50, 60, 70]$\times 1\sigma$ (1$\sigma$ = 0.23\,\jy). The dust lane is marked with a orange solid line. The two black circles represent the locations of Gap A and Gap B.
    \textbf{(d)}: Moment 2 image obtained from the CO\,($J$ = 2--1) red-shifted component (\vlsr = 15--85\,\kms). The two gray circles represent the locations of Gap A and Gap B. The orange dashed square corresponds to the FIR\,4 region.}
    \label{dis-1}
\end{figure}

\subsubsection{Outflow Interaction Scenario}
\label{dis-out}
Shock originating SiO ($J$ = 2--1) emission was previously detected in the FIR\,4 region and interpreted as an interaction between the molecular outflow (flow-3) and a dense clump \citep{shimajiri2008}. 
Our ALMA 12-m array observations, with a higher transition of SiO\,($J$ = 5--4), spatially resolved further detailed structures to strengthen the outflow--dense clump interaction scenario. 
The SiO\,($J$ = 5--4) emission was detected in four regions within the FIR\,4 region (shock-6, shock-7, shock-8, and shock-9 in Figure\,\ref{dis-1}a). 
Among the four components, shock-6 and shock-9 show very clear U-shaped structures, denoted by red dashed lines in Figure\,\ref{dis-1}(b).
These two shocks also seem to be located downstream of the south-west lobe of flow-3.
shock-6 was previously detected in SiO ($J$ = 2--1) observations by \cite{shimajiri2008} with a similar velocity range, whereas shock-9 is newly detected in this study.

As presented in Figure\,\ref{dis-1}(b), both shock-6 and shock-9 show large velocity dispersion in the range $\sim$ 2--5\,\kms\,in the SiO\,($J$ = 5--4) emission. 
For shock-6, the SiO gas distribution partially overlaps the blue-shifted CO emission.
The 1.3\,mm continuum emission is distributed downstream of shock-6. 
The positional relationship between the 1.3\,mm dust, CO, and SiO emissions can be explained by an interaction between flow-3 and dust condensations located in the FIR\,4 region. 
The SiO emission from shock-9 also seems to spatially correlate with the south-west tip of the outflow lobe traced by the CO blue-shifted emission, as described in Figure\,\ref{dis-1}(b). 
Unlike the case of shock-6, we did not detect dust condensations downstream of shock-9. 
A possible explanation for this is that the interacting condensations are not dense enough to be traced by our ALMA 12-m array continuum observations ($n_{\mathrm{H_2}} \lesssim 3.9 \times 10^6\,\mathrm{cm^{-3}}$).
Alternatively, their structure is rather extended and the emission is not detected with our 12-m array observations ($\gtrsim$ 5200\,au). 
Indeed, an extended continuum emission was detected in that region both in the 1.3\,mm continuum emission with our ACA 7-m array and by previous single-dish (sub)millimeter continuum observations \citep{chini1997, lis1998, johnstone1999}.

We also found a south-west outflow lobe of flow-3, clearly shaped by the surrounding dense material. 
The eastern edge of the CO outflow lobe, which is detected in both blue- and red-shifted CO emissions, shows a tight spatial correlation with the 1.3\,mm continuum emission (denoted as ``dust lane'' in Figure\,\ref{dis-1}c, which is bright all along the edge). 
The CO contours are very steep at that edge of the outflow, indicating that the outflow is clearly compressed along the entire eastern edge.
The CO second moment map presented in Figure\,\ref{dis-1}(d) also shows that a large CO velocity dispersion (up to 10\,\kms) associated with the eastern edge of the CO outflow lobe has a tight correlation with the 1.3\,mm continuum emission. 
These tight correlations support a scenario in which the outflow lobe interacts with filamentary dense structures traced by the 1.3\,mm continuum emission, distributed just next to the outflow lobe. 
Note that the south-west lobe of flow-3 has a smaller opening angle compared with the north-east lobe of flow-3, as described in Section \ref{result-line-outflow} (Figure\,\ref{flow1-3}c). 
This implies that the surrounding material traced by the continuum emission prevents gas at the eastern edge of the south-west outflow lobe from expanding freely. 
\cite{tobin2019} pointed out that the mid-infrared emission (24\,\micron\,and 70\,\micron\,band images), indicating warm dust, is bright not only at the location of FraSCO-16 (HOPS-370) where a protostar is located, but also along the dust filament extending to the south ($\sim 30''$). 
We found that the elongated direction of the warm dust corresponds to the region where a strong interaction occurs between the outflow lobe and dust lane, as presented in Figure\,\ref{dis-1}(c). 
Spatial correlations between the outflow lobe, dust lanes, and the large velocity dispersion of the CO gas clearly support a scenario where the dust emission from the dust lane arises from shock-originating warm dust produced by the interaction between flow-3 and the dust lanes.

It is interesting to note that there is spatial anti-correlation between the 1.3\,mm continuum emission and the red-shifted CO emission tracing the east edge of the south-west outflow lobe, as presented in Figure\,\ref{dis-1}(c). 
Red-shifted CO gas appears to be leaking from gaps in the dust lanes traced by the 1.3\,mm continuum emission, denoted by Gap\,A and Gap\,B in Figure\,\ref{dis-1}(c). 
The red-shifted CO components of flow-3 are flowing with P.A. of 210\,deg. At the positions of both Gap\,A and Gap\,B, a part of the red-shifted CO emission suddenly changes P.A. to $\sim$ 120\,deg., i.e., P.A. of the flow changes 90\,deg. clockwise.  
Furthermore, the CO second moment map shows local spots at Gap\,A and Gap\,B, which show an increased velocity dispersion up to 5.0\,\kms\,and 9.7\,\kms, respectively (Figure\,\ref{dis-1}d). 
This value is 1.5--2 times larger than that of the typical velocity dispersion measured in other parts of the outflow lobe. 
This can be interpreted as showing that the CO outflow collides with surrounding dense materials (i.e., observed as the dust condensation) and is compressed at the locations of Gap\,A and Gap\,B, resulting in a part of the outflow being changed by 90\,deg. in P.A., and the red-shifted gas is a hint of leaking through a low-density region of the dust lane. 

Furthermore, two SiO components identified as shock-8 and shock-7 are detected at the eastern edge of the south-west lobe of flow-3 (see Figure\,\ref{dis-1}a). 
In particular, shock-8 and the red-shifted component of shock-7 with a large velocity dispersion of $\sim$5\,\kms\, are detected at the region where the eastern edge of the red-shifted CO outflow lobe changes the position angle with a local increment of the velocity dispersion up to 11\,\kms. 
This is considered to be another local point having a strong interaction between the outflow and dense material traced by the 1.3\,mm continuum emission.

In summary, we revealed detailed spatial distributions of shock originating gas traced by higher transition SiO\,($J$ = 5--4) emissions. 
We confirmed that the shocked region associated with previously detected shocked gas has a U-shaped structure at the colliding surface between flow-3 and the condensations within the FIR\,4 region.
In addition, another interaction region was newly detected further downstream of flow-3.
Furthermore, our observations revealed that the eastern side of the flow-3 south-west lobe interacts with surrounding dense dust material. 
The interacting material shows a narrow dust lane and the velocity dispersion of flow-3 increases around the dust lane.
These results are clear evidence that a prominent outflow driven by FraSCO-16, flow-3, significantly interacts with surrounding material and possibly affects the star formation environment in the FIR\,4 region. 
Thus, our results strongly support the outflow interaction scenario.

\subsubsection{Internal Heating Source Scenario}
\label{dis-internal}
\cite{lopez2013} obtained $\mathrm{CH_3OH}$, $\mathrm{DCO^+}$, $\mathrm{C^{34}S}$, DCN, and $\mathrm{NH_3}$ line images with an angular resolution of $\sim 2''.7$ centered at FIR\,4 ($\sim35''$ FoV).  
Based on the observations, they proposed an internal heating scenario from an embedded B star. 
Their conclusion was drawn from (i) the large bolometric luminosity
estimated in the FIR\,4 region (1000\,\lsun;\citealt{crimier2009}) and (ii) their interferometric observations, showing a bright peak in the 2\,mm continuum emission and $\mathrm{CH_3OH}$ at the center of the FIR\,4 region.
The source position coincides with a previously detected 3.6\,cm source, VLA 12, identified by \cite{reipurth1999}, which is considered to be a free-free jet.
A more recent centimeter wavelength study by \cite{osorio2017} spatially resolved VLA 12 into three components, VLA 12C, HOPS-108 (associated with FraSCO-37), and VLA 12S using multiwave bands between 5\,cm and 0.7\,cm. 
Their proper motion study concluded that only HOPS-108 originates from the heating source embedded within the FIR\,4 region, while VLA 12C and VLA 12S are likely associated with radio jets driven by HOPS-370 located within the FIR\,3 region.
In addition, an infrared wavelength study with higher angular resolution resolving individual sources within the FIR\,4 region by \cite{furlan2016} re-estimated the bolometric luminosity of HOPS-108 as 38.3\,\lsun. 
Assuming that the bolometric luminosity of 38.3\,\lsun\,originates from the stellar luminosity at a stellar age of $\sim 10^5$\,yr, the spectral type of the star is expected to be G0--G5 based on the relation between the stellar luminosity and effective temperature \citep{palla1993}.
In fact, HOPS-108 is classified as a Class 0 source from its spectral energy distribution \citep{furlan2016}.
Therefore, most of the bolometric luminosity likely originates from accretion luminosity (not stellar-internal luminosity).
This indicates that the stellar luminosity should be much less than 38.3\,\lsun, i.e., the embedded source is less massive than G0--G5 stars. 
Finally, the FWHM for FraSCO-37 associated with HOPS-108 is measured to be $\sim0''.8$ from a 2D Gaussian fitting to our 1.3\,mm continuum image.
The gas mass of FraSCO-37 is also estimated to be 0.18\,\msun\,from the 1.3\,mm continuum observations, which is an order of magnitude less than typical B-type stellar masses (2.68--17.7\,\msun; \citealt{pecaut2013}).
Updated observational results since \cite{lopez2013} provide further support that HOPS-108 (considered to be a main source in the FIR\,4 region) is likely to be an ordinary Class 0 source, but not an embedded B-type source.
Hence our study does not strongly support this internal heating scenario and indicates that the large bolometric luminosity estimated in the FIR\,4 region is mainly caused by the interaction between the outflow and the dust condensation of FIR\,4.

\subsubsection{Irradiation by an FUV Field Scenario}
\label{dis-irrad}
\cite{fontani2017} made $\mathrm{HC_3N}$ and $\mathrm{HC_5N}$ observations centered at FIR\,4 (FoV $\sim 60''$ and $\theta \sim 6''$), in a framework of the IRAM/NOEMA Large Program ``SOLIS''.
They found that the $\mathrm{HC_3N/HC_5N}$ abundance ratio is smaller ($\le 10$) in the eastern region of the FIR\,4 region than in the western region. 
According to their chemical models, this small ratio can be reproduced only when the cosmic-ray ionization rate is as large as $\sim 4 \times 10^{-14} \mathrm{s^{-1}}$ in the eastern region.
This value is $\sim$1000 times higher than that of the interstellar medium, implying that the FIR\,4 region is strongly irradiated.
Hence, they proposed that the FIR\,4 region is irradiated by energetic cosmic-ray particles from an interior embedded source, following the internal heating source scenario proposed by \cite{lopez2013}. 
As discussed in Section \ref{dis-internal}, our observations do not support the scenario by \cite{lopez2013}.
A follow-up study by \cite{favre2018} performed c-$\mathrm{C_3H_2}$ observations centered at FIR\,4 (FoV $\sim 60''$ and $\theta \sim 6''$) in the same framework as \cite{fontani2017}. 
They derived the excitation temperature distribution within the FIR\,4 region using chemical models. 
They noted that if the outflow interaction scenario proposed by \cite{shimajiri2008} is appropriate, there should be some physically induced effects such as temperature gradients along the outflow axis of flow-3 as evidence of the interaction between flow-3 and the FIR\,4 region.   
However, their c-$\mathrm{C_3H_2}$ observations does not show a temperature gradient along the outflow axis, suggesting that there is no evidence of direct physical interaction between flow-3 and objects within the FIR\,4 region. 
Their c-$\mathrm{C_3H_2}$ observations instead show that the cosmic-ray ionization rate in the eastern side of the FIR\,4 region is higher than in the western side and the value of $\sim 4 \times 10^{-14} \mathrm{s^{-1}}$ is high enough to show that the eastern region is irradiated, which is consistent with the case of $\mathrm{HC_5N}$ observations by \cite{fontani2017}.
Furthermore, they noted that previous Herschel observations show that there is a tenuous cloud between OMC-2 illuminated by an FUV field.
They concluded that the FIR\,4 region is bathed in an FUV field and is irradiated by energetic particles.

These previous studies have investigated the chemical properties in the FIR\,4 region, but did not directly investigate the dynamical properties of flow-3 itself, which would be relevant to the discussion of the interaction between flow-3 and objects within the FIR\,4 region.
Comparing a map of the $\mathrm{HC_3N/HC_5N}$ abundance ratio derived by \cite{fontani2017} with our CO image, the area where the abundance ratio is small is spatially in agreement with the tip of the red-shifted CO outflow lobe. 
From another comparison of the temperature distribution derived by \cite{favre2018} and our CO second moment map, we found that the location of the c-$\mathrm{C_3H_2}$ temperature peak is located downstream of Gap\,B. 
Furthermore, the position angle of the red-shifted CO outflow after colliding with objects within the FIR\,4 region ($\sim$ 120\,deg.) is well aligned with the temperature gradient perpendicular to the outflow axis obtained from the c-$\mathrm{C_3H_2}$ presented in Figure\,4 of \cite{favre2018}. 
These spatial correlations imply that the interaction between flow-3 and objects within the FIR\,4 region creates a shock, heating the surrounding materials and decreasing the abundance ratio. 
Although our dataset does not contradict the scenario where the FIR\,4 region is likely bathed in an FUV field, proposed by \cite{fontani2017} and \cite{favre2018}, our study indicates that the heating source causing the high cosmic-ray rate and low abundance rate of the $\mathrm{HC_3N/HC_5N}$ may originate from the interaction between the outflow and the dust condensation of FIR\,4.

\begin{figure}
    \centering
    \includegraphics[width=15cm]{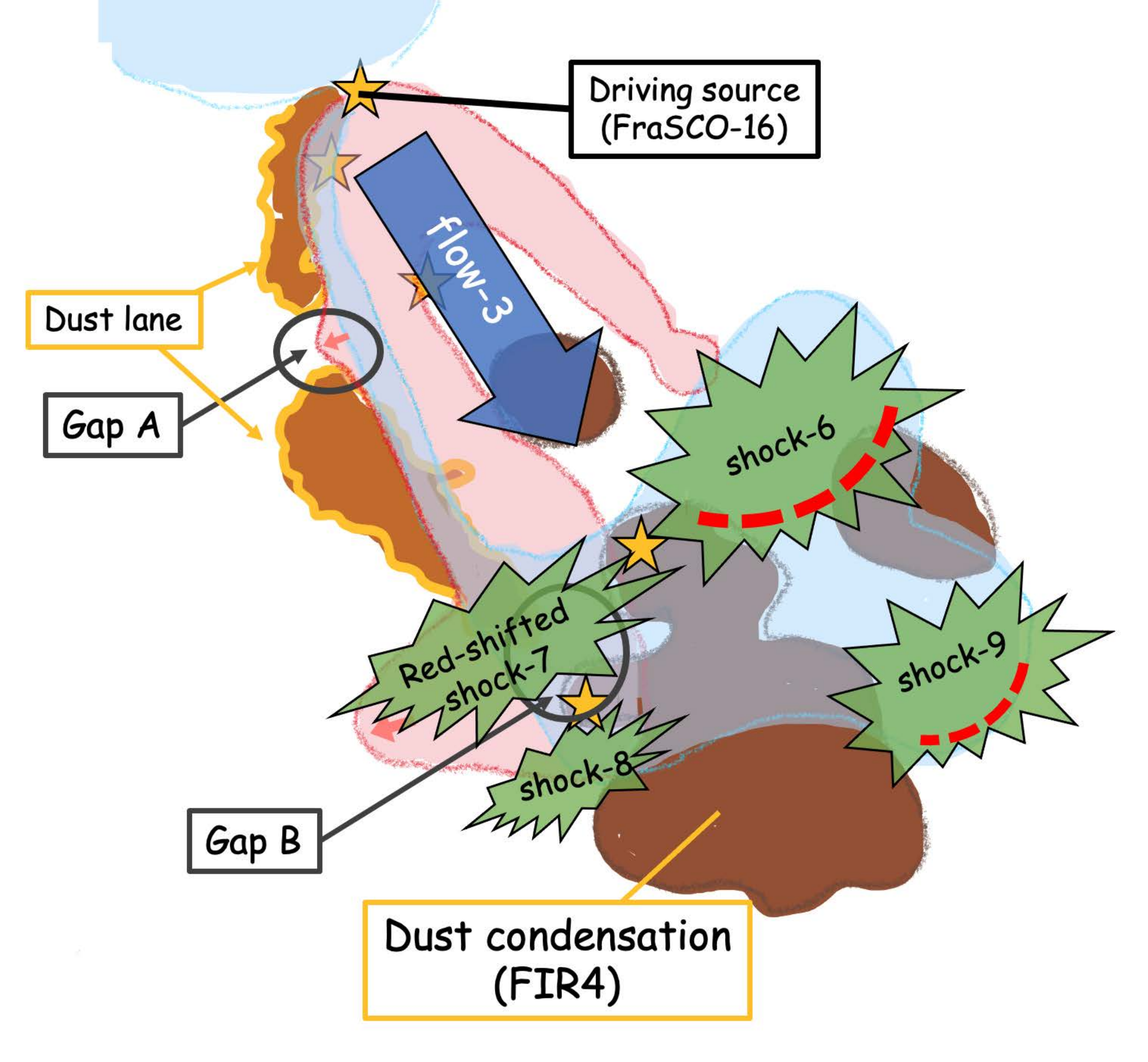}
    \caption{Schematic image of the interaction between flow-3 and condensations within the FIR\,4 region. The yellow star symbols show the locations of FraSCO sources associated with infrared sources \citep{nielbock2003, furlan2016}. The red and blue colored sketches represent the red- and blue-shifted CO gas of flow-3 from FraSCO-16, respectively. The brown colored sketches represent the condensations within the FIR\,4 region. The blue arrow presents the proceeding directions of flow-3. The four green shock-shaped sketches denote shock-6, shock-7, shock-8, and shock-9, indicating interactions of flow-3 with condensations within the FIR\,4 region (brown area). The two red dashed lines represent U-shaped structures observed in the SiO\,($J$ = 5--4) emission at the locations of shock-6 and shock-9 presented in Figure\,\ref{dis-1}b.}
    \label{dis-3}
\end{figure}

\subsection{Star Formation Triggered by Molecular Outflow}
\label{dis-frag}
 In the previous section, we presented evidence of the interaction between flow-3 and condensations within the FIR\,4 region. 
 \cite{shimajiri2008} analyzed fragmentation within the FIR\,4 region, considered to occur due to an interaction between flow-3 and a clump within the FIR\,4 region. 
 They identified 11 cores within the FIR\,4 region and estimated a fragmentation timescale ($\tau_{\mathrm{frag}}$) of $3.8 \times 10^4$\,yr by measuring the separations between the identified cores.
 We compare the peak positions of the 11 cores with those of our identified FraSCO sources in Appendix\,\ref{app-comp}. 
\cite{shimajiri2008} also estimated the dynamical timescale ($\tau_{\mathrm{dyn}}$) of the outflows associated with flow-3 as $1.4 \times 10^4$\,yr with the CO\,($J$ = 3--2) line emission. 
 Assuming that $\tau_{\mathrm{dyn}}$ is similar to an interaction timescale ($\tau_{\mathrm{interact}}$), they interpreted $\tau_{\mathrm{dyn}} \sim \tau_{\mathrm{interact}} \sim \tau_{\mathrm{frag}}$. 
 In addition, they found that mid-infrared sources detected by \cite{nielbock2003} were located at the root of the outflow and along the lobe. 
 From these estimates and previous IR observations, \cite{shimajiri2008} proposed a triggered star formation scenario within the FIR\,4 region by the following four steps.
 
 (i) An embedded star associated with a mid-infrared source was born within the FIR\,3 region and drove the outflow (i.e., HOPS-370 drove flow-3).

(ii) The outflow driven by the embedded star started interacting with the FIR\,4 region.

(iii) The interaction caused fragmentation of the FIR\,4 region into 11 cores.

(iv) These cores will form stars.

\cite{megeath2012} and \cite{furlan2016} developed a model of the spectral energy distribution fitting for protostars in this region.
They classified two sources within the FIR\,3 and FIR\,4 regions as Class I sources. 
This indicates that star formation within both these regions started at almost the same time, which is inconsistent with the triggered star formation scenario proposed by \cite{shimajiri2008}. 
Therefore, the scenario requires further consideration.	

 We estimated the fragmentation timescale ($\tau_{\mathrm{frag}}$) of FraSCO sources within the FIR\,4 region assuming that $\tau_{\mathrm{frag}}$ is a sound crossing timescale, following Equations (\ref{frag-1}) to (\ref{frag-3}) described in \cite{shimajiri2008}. 
 First, we estimate the average three dimensional separation among FraSCO sources located within the FIR\,4 region by
 
 \begin{equation}
     \Delta l = 2(\frac{1}{\gamma^{1/3}} - 1)r, 
     \label{frag-1}
 \end{equation}
 \begin{equation}
     \gamma = \frac{n_{\mathrm{leaf}} V_{\mathrm{leaf}}}{V_{\mathrm{clump}}},
     \label{frag-2}
 \end{equation}
 where $\gamma$, $r$, $n_{\mathrm{leaf}}$, $V_{\mathrm{clump}}$, and $V_{\mathrm{leaf}}$ are the volume filling factor, average radius of the FraSCO sources, number of sources, and average volume of the FIR\,4 region and a single source, respectively.
 Note that only $V_{\mathrm{clump}}$ was estimated from our ACA 7-m array data and the remaining values were estimated from our ALMA 12-m array data.
 Then, we can estimate $\tau_{\mathrm{frag}}$ by the following formula
 
\begin{equation}
    \tau_{\mathrm{frag}} = \frac{ \Delta l}{C_{\mathrm{eff}}},
    \label{frag-3}
\end{equation}
where $C_{\mathrm{eff}}$ is the effective sound speed estimated from the line. 
Here, we employed the same value of $C_{\mathrm{eff}}$ used by \cite{shimajiri2008}, because our observations do not include any line data tracing dense material detected with the 1.3\,mm continuum emission. 
We conservatively estimated $\tau_{\mathrm{frag}}$ $\sim$ (2.5--3.2) $\times 10^4$\,yr. 
Note that we estimated the minimum timescale of $2.5 \times 10^4$\,yr using six sources (FraSCO-28, -32, -35, -37, -39, and -40), which are all gravitationally bound.
A maximum timescale of $3.2 \times 10^4$\,yr was estimated using eighteen sources (FraSCO-23 to -40), which include both gravitationally bound and unbound sources.
Note the latter timescale is longer despite the additional leaves within the same clump volume, because the average leaf volume is much smaller (the mean radius of FraSCO sources is approximately 2.7 times smaller).
The dynamical timescale for flow-3 ($\tau_{\mathrm{dyn}}$) was also estimated to be in the range $\sim$ (0.46--2.54) $\times 10^4$\,yr from CO\,($J$ = 2--1) emissions in Section \ref{result-line-outflow}.
$\tau_{\mathrm{frag}}$ and $\tau_{\mathrm{dyn}}$ estimated from our observations are roughly consistent with the timescales previously estimated by \cite{shimajiri2008}, however the errors of our estimated timescales are so large that it is difficult to determine whether $\tau_{\mathrm{frag}}$ is similar to $\tau_{\mathrm{dyn}}$ with our data. 
Furthermore, the dynamical timescale for outflow driven within the FIR\,4 region was estimated to be in the range $\sim$ (0.2--1.7) $\times 10^4$\,yr, which is similar to that for outflow driven within the FIR\,3 region, $\sim$ (0.4--2.5) $\times 10^4$\,yr.

In summary, fragmentation within the FIR\,4 region is considered to have started when flow-3 started interacting with the FIR\,4 region or before this interaction. This conclusion does not strongly support the triggered star formation scenario proposed by \cite{shimajiri2008}.

Finally, we should note  that the time scale discussion described above using the outflow dynamical timescale may be limited by the current mass ejection event. Previous studies suggest a possibility of episodic mass ejection in the protostellar phase (e.g., \citealt{plunkett2015, zhang2019}). Indeed, HOPS-370 is identified as a Class I source, having the stellar age of $\sim 10^5$ yr. In comparison, the time scale for readsorption of evaporated SiO on dust grains is estimated as $\sim 8\times 10^3$\,yr in case of the density of $4\times 10^5$\,cm$^{-3}$ \citep{mikami1992}, implying previous mass ejection phenomena (if exist any) is not detectable with our SiO observations.

\subsection{Hub-filament System in the FIR\,4 Region}
\label{dis-hub}
In this section, we discuss the morphology of our 1.3\,mm continuum image in the FIR\,3, 4, and 5 regions, comparing its spatial distribution with the hub-fiber system previously identified in this region with $\mathrm{N_2H^+}$\,($J$ = 1--0) observations \citep{hacar2018, zhang2020}. Note that we focus on the spatial distribution of dense material rather than the dynamics of the gas, i.e., we focus on filaments detected in the 1.3\,mm continuum image. Investigating velocity structures within the hub-fiber system would be a future topic for high angular resolution molecular observations.  

Filamentary molecular clouds are considered to play a critical role in forming dense cores.  Recent studies have shown that dense cores tend to be located around a hub in hub-filament systems (e.g., \citealt{clarke2020}). It has also been reported that column densities and velocity gradients increase along the filaments toward a hub \citep{hill2011, tanaka2013, kirk2013, yuan2018, trevino2019, ren2021, cao2022}. This implies possible gas inflows toward the hub and enhanced star formation activity in the hub \citep{schneider2010}. 

\begin{figure}
    \centering
    \includegraphics[width=17cm]{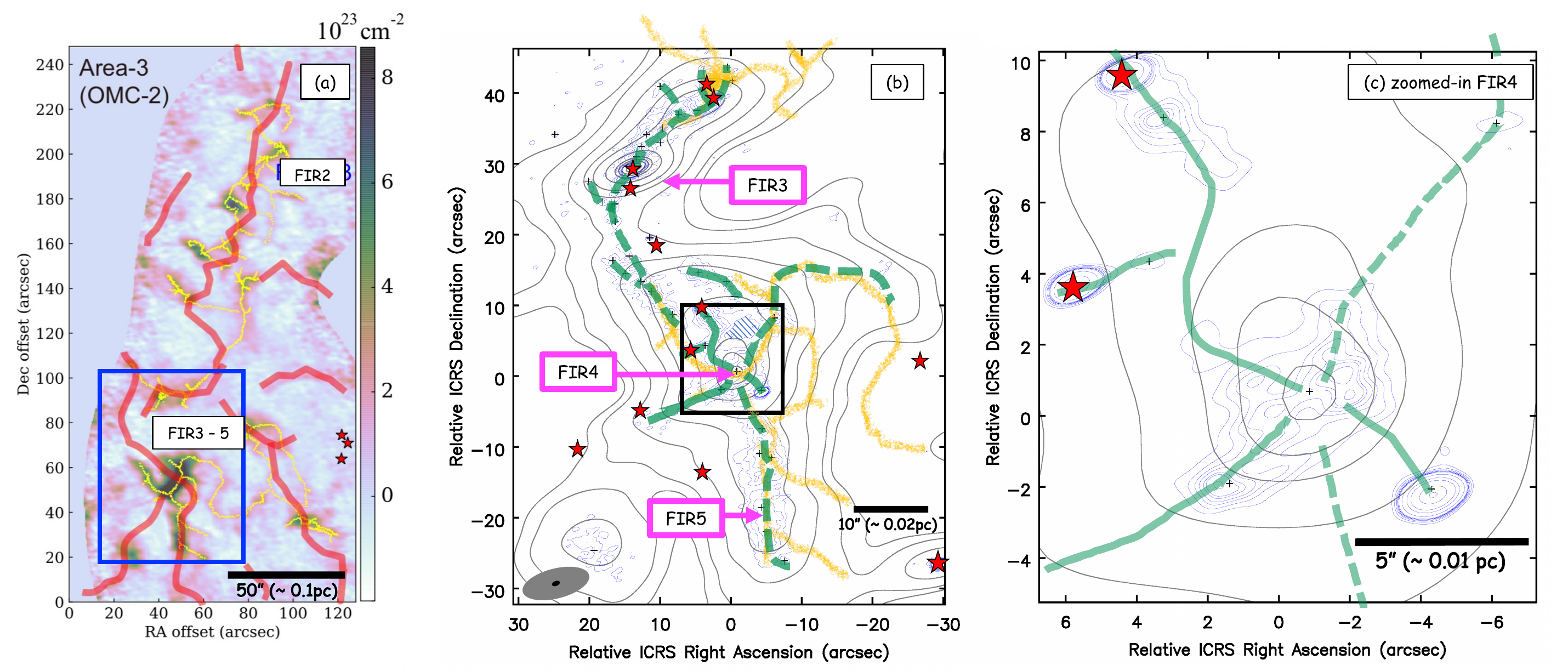}
    \caption{\textbf{(a)}: $\mathrm{N_2H^+}$\,($J$ = 1--0) column density map, as Figure\,3 of \cite{zhang2020}. The yellow and red lines represent $\mathrm{N_2H^+}$ fibers identified by \cite{zhang2020} and \cite{hacar2018}, respectively. The blue frame shows the area of Panel (b).
    \textbf{(b)}: Fiber-like structures identified with 1.3\,mm continuum image with the ALMA 12-m array (blue contours). Our identified three fiber-like structures (green dashed lines) are consistent with some of the $\mathrm{N_2H^+}$ fibers (yellow lines, as in Panel (a)), whereas the six fiber-like structures denoted by the green solid lines were newly identified. The red star symbols represent the locations of previous detected infrared sources \citep{nielbock2003, furlan2016}. The locations of FIR\,3, 4, and 5 \citep{chini1997} are indicated by pink arrows. The blue contour levels are [5, 10, 15, 20, 25, 30] $\times 1\sigma$ ($1\sigma = 0.22$\,\mjy). The gray contours represent the 1.3\,mm continuum emission with the ACA 7-m array. The gray contour levels are [10, 20, 50, 100, 150, 230, 290, 310, 325, 340] $\times 1\sigma$ ($1\sigma = 0.78$\,\mjy). The black frame shows the area of Panel(c).
    \textbf{(c)}: Zoomed-in image of the FIR\,4 region. The gray contour levels are [160, 240, 290, 310, 325] $\times 1\sigma$ ($1\sigma = 0.78$\,\mjy). The blue contour levels are [16, 20, 23, 25, 27, 28, 30, 40, 60, 80, 130, 170] $\times 1\sigma$ ($1\sigma = 0.22$\,\mjy). The red star, black crosses, and green lines are the same as those in Panel (b).}
    \label{dis-4}
\end{figure}

As presented in Figure\,\ref{dis-4}a, previous $\mathrm{N_2H^+}$\,($J$ = 1--0) observations have identified several velocity coherent structures, i.e., fibers, in FIR\,3, 4, and 5 in the OMC-2 regions \citep{hacar2018, zhang2020}. In particular, \cite{zhang2020} identified three fibers intersecting at FIR\,4, showing a hub-fiber system and reported that dense cores tend to be located around the hub.

Figure\,\ref{dis-4}b shows the 1.3\,mm  continuum image of the FIR\,3, 4, and 5 regions obtained from this study. Our ALMA 12-m array 1.3\,mm continuum observations showed a factor of $\sim$3 improvement in the spatial resolution compared with previous $\mathrm{N_2H^+}$\,($J$ = 1--0) observations by \cite{hacar2018} and \cite{zhang2020}. 
In order to identify filaments from our ALMA 12-m array 1.3\,mm continuum image, we assume that filaments are elongated structures within which FraSCO sources are continuously located, and then the filaments are connected toward the peak of the FIR\,4 imaged with the ACA 7\,m-array data.
With these two criteria, we identified eight filaments (green solid and dashed lines in Figure\,\ref{dis-4}b). 
Three of the filaments identified by our ALMA 12-m array 1.3\,mm observations (green dashed lines in Figure\,\ref{dis-4}b) are approximately consistent with previously identified fibers (yellow lines in the same panel) in the $\mathrm{N_2H^+}$\,($J$ = 1--0) emission reported by \cite{zhang2020}, whereas the other five filaments (green solid lines in the same panel) are identified for the first time in our high-angular resolution ALMA 12-m array continuum image. 
The newly identified filaments are located within FIR\,4, where a hub was identified in the previous study by \cite{zhang2020}. The hub was also determined to be a single peaked condensation in FIR\,4 with our ACA 7-m array image (gray contours in Figure\,\ref{dis-4}b). 

The newly identified filaments are estimated to have $\mathrm{H_2}$ gas column densities in the range 1--17 $\times 10^{23}\,\mathrm{cm^{-2}}$ \footnote{In order to estimate the average column density of each filament, we fitted the elongated filaments using the 2D Gausian.}.
These values are denser by one order of magnitude than those measured by \cite{hacar2018} and denser by a factor of two than those measured by \cite{zhang2020}. Hence, the filaments identified from our ALMA 12-m array 1.3\,mm continuum image are considered to trace inner dense parts compared with the previously identified $\mathrm{N_2H^+}$\,($J$ = 1--0) fibers. 
The locations of newly identified filaments with higher density indirectly support a scenario of gas inflow motion to accumulate material onto the hub, as previously reported for other protocluster regions (e.g., \citealt{hill2011, kirk2013, trevino2019}). 

More interestingly, the ALMA 12-m array continuum image denoted by blue contours in Figure\,\ref{dis-4}c does not exhibit a single-peaked structure within the hub, as the lower resolution ACA 7-m array continuum image shows (gray contours in Figure\,\ref{dis-4}c), but rather exhibits the highly complex substructures within. A Jeans analysis suggested that a central extended substructure within the hub was gravitationally bound, whereas no star formation activity such as outflows or disks was clearly detected within the central substructure. 
The results signify that the central substructure within the hub was possibly present in the prestellar phase and may be deemed as transient. 
We should note that the feedback of outflows driven within the FIR\,4 region could also affect the star formation within the FIR\,4 region. We detected six outflows driven by FraSCO sources located within FIR\,4 region: flow-4 – flow-9 identified in Section\,\ref{result-line-outflow}. These outflows could mix the surrounding material within the FIR\,4 region and affect the star formation there.

To test a scenario of massive core formation through the OMC-2 hub-filament system, the gas inflow motion along the filaments should be determined. This dynamical information will enable us to constrain how much material can be accumulated onto the hub by estimating the gas inflow rates, and will eventually determine the final mass of the forming stars. 

\section{Summary}
\label{summary}
To reveal the star forming environment in one of the nearest embedded protoclusters, FIR\,3, 4, and 5 in the OMC-2 region, we obtained observations of the 1.3\,mm continuum, CO\,($J$ = 2--1) line, and SiO\,($J$ = 5--4) line emissions with ALMA. The main results and conclusions are summarized as follows. 
\begin{enumerate}
   \item Using the 1.3\,mm continuum image obtained with the ALMA 12-m array, we identified 51 dense sources. Among them, 36 sources were newly identified in this study. The dust masses and projected sizes of all the 51 sources are in the ranges $3.8 \times 10^{-5}$--$ 1.1 \times 10^{-2}$\,\msun\,and 28--1964\,au, respectively. Their $\mathrm{H_2}$ gas number densities are estimated to be $6.4 \times 10^{6}$--$3.3 \times 10^{8}\,\mathrm{cm^{-3}}$. Most of the previously identified 15 sources ($\sim 80 \,\%$) have number densities above the critical Jeans number density regardless of whether they are pre- or proto-stellar sources, whereas the remaining sources, consisting of three protostellar sources, have number densities below the critical density. Among the newly identified sources, $\sim 14\,\%$ have a number density above the critical Jeans number density. 

    \item Using the CO\,($J$ = 2--1) line image with the ALMA 12-m array, we identified six clear, five probable, and one marginal outflow in total. In addition, seven of the identified outflows have associated SiO emission. This study newly identified six CO outflows consisting of two clear, three probable, and one marginal outflows. These are associated with newly identified 1.3\,mm continuum sources in the FIR\,4 region. In addition to the molecular outflow, we detected 11 extended and compact SiO emissions that do not originate from molecular outflows. The SiO emissions are mainly distributed in the FIR\,4 region and along lobes of an energetic outflow driven by HOPS-370.
    
    \item We discussed three previously proposed scenarios to explain the origin of the protocluster in the FIR\,4 region; (1) Outflow interaction, (2) internal heating source, and (3) irradiation by an FUV field. High angular resolution and high sensitivity observations of CO\,($J$ = 2--1) and SiO\,($J$ = 5--4) provided spatially resolved images showing direct evidence of an interaction between the dust condensation, FIR\,4, and an outflow (flow-3) driven from HOPS-370. Our observational results support the first scenario above, while no evidence was found to support the second and third scenarios.

    \item Based on the outflow interaction scenario described above, we discussed fragmentation of FraSCO sources within the FIR\,4 region by comparing the fragmentation timescale for the FraSCO sources with the dynamical timescale for flow-3. The fragmentation timescale was estimated to be $\sim$ (2.5--3.2) $\times 10^4$\,yr, which is similar to the dynamical timescale for flow-3, $\sim$ (0.46--2.54) $\times 10^4$ yr. Furthermore, the dynamical timescale for the outflow driven within the FIR\,4 region is estimated to be in the range $\sim$ (0.2--1.7) $\times 10^4$\,yr, which is similar to the dynamical timescale for the outflow driven within the FIR\,3 region, $\sim$ (0.4--2.5) $\times 10^4$\,yr. Therefore, fragmentation within the FIR\,4 region is considered to have started when flow-3 started interacting with the FIR\,4 region or before that interaction. This conclusion does not strongly support the triggered star formation scenario previously proposed by \cite{shimajiri2008}.
    
    \item Finally, using the 1.3\,mm continuum images with the ALMA 12-m array and the ACA 7-m array, we discussed the morphology of the hub-filament system located within the FIR\,3, 4, and 5 regions. We identified eight filaments intersecting at the central hub, i.e., the center of FIR\,4. Five of them were newly identified in this study and located within the FIR\,4 region. The $\mathrm{H_2}$ gas column densities of the filaments were estimated to be (1--17) $\times 10^{23}\,\mathrm{cm^{-2}}$, which is denser than those of the previously identified $\mathrm{N_2H^+}$ fibers. Interestingly, the ALMA 12-m array continuum image exhibits highly complex substructures within the hub. Based on the result that the central substructure within the hub is gravitationally bound with no star formation activity, the central substructure was possibly present in the prestellar phase and may be deemed as transient. To test a scenario of massive core formation through the OMC-2 hub-filament system, the gas inflow motion along the filaments should be determined. However, our observations have no velocity information about the detected hub-filament system, and hence investigation of the dynamical motion will be a topic for future study.
    
\end{enumerate}

\section{acknowledgments}
We deeply acknowledge the referee for the very careful reading and constructive comments that have helped to improve this manuscript.
We thank D. Johnstone for providing us the submillimeter continuum data taken by JCMT. 
We thank Y. Aso for helping us with creating the animations presented in this paper.
This paper makes use of the following ALMA data: ADS/JAO.ALMA\#2017.1.01353S. ALMA is a partnership of ESO (representing its member states), NSF (USA) and NINS (Japan), together with NRC (Canada), MOST and ASIAA (Taiwan), and KASI (Republic of Korea), in cooperation with the Republic of Chile. The Joint ALMA Observatory is operated by ESO, AUI/NRAO and NAOJ. 
This work was supported by NAOJ ALMA Scientific Research Grant Code 2022-22B.
The present study was supported by JSPS KAKENHI Grant (JP17H06360, JP17K05387, JP17KK0096, JP21H00046, JP21K03617: MNM).
L.A.Z. acknowledges financial support from CONACyT-280775 and UNAM-PAPIIT IN110618 grants, México.
This project has received funding from the European Research Council (ERC) under the European Union’s Horizon 2020 research and innovation programme (Grant agreement No. 851435).

\appendix
\section{Image comparisons with previous observations}
\label{app-comp}
\begin{figure}
    \centering
    \includegraphics[width=15cm]{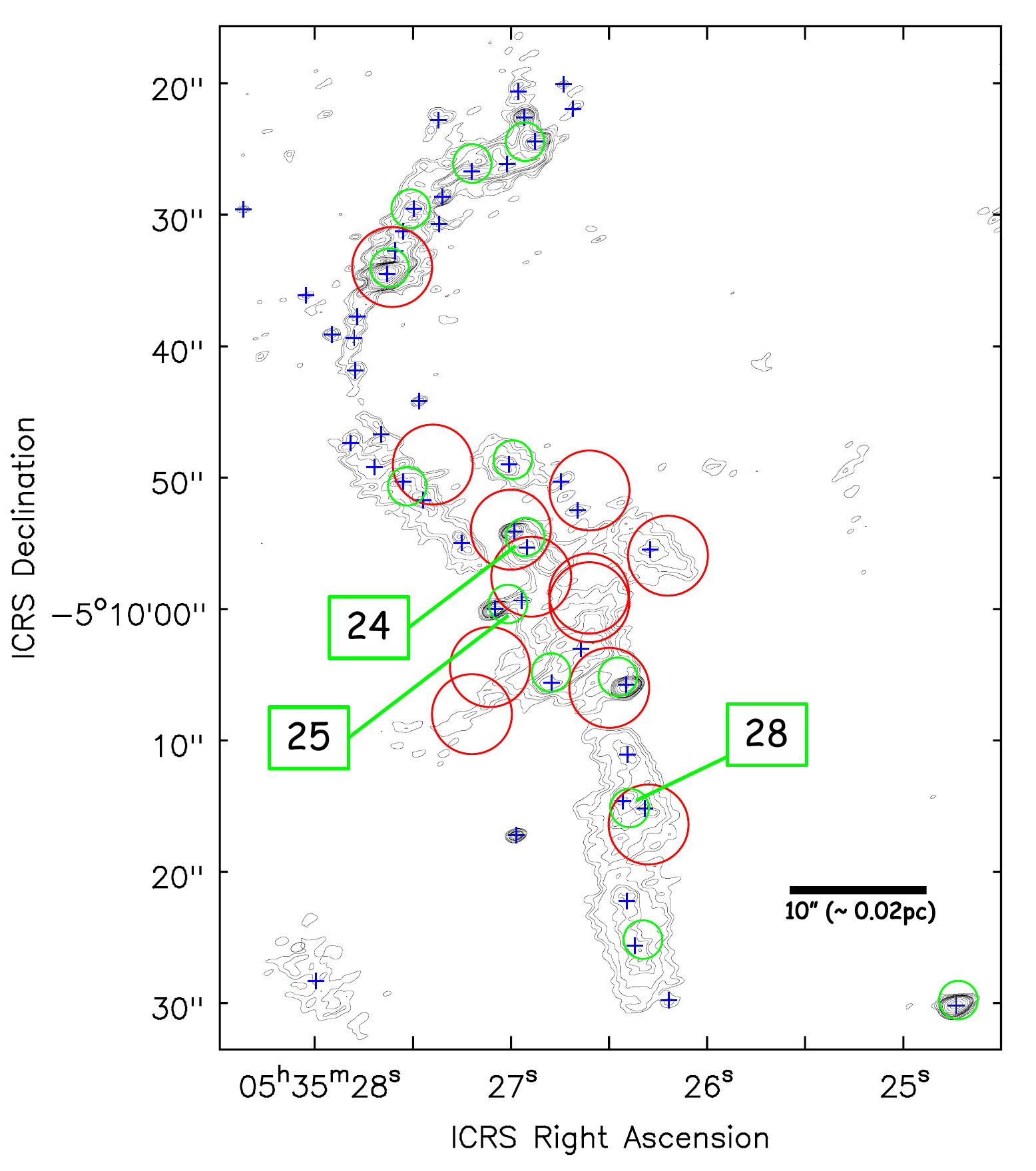}
    \caption{Comparison between 1.3\,mm FraSCO sources and previously identified millimeter dense cores. The black contours and blue crosses represent our 1.3\,mm continuum images obtained from the ALMA 12-m array and the positions of FraSCO sources, respectively. The red open circles and green open circles show the positions of 3.3\,mm dense cores identified by \cite{shimajiri2008} and 3\,mm dense cores identified by \cite{kainulainen2017}. The size of the red and green circles corresponds to the angular resolution of each observation, $6''$ and $3''$, respectively. The black contour levels are [5, 7, 10, 15, 17, 20, 25, 32, 40, 60, 100, 170, 300, 700] $\times 1\sigma$ ($1\sigma = 0.22$\,\mjy). ``FraSCO-16'' and ``FraSCO-32'' are name of FraSCO sources identified in this study. ``24'', ``25'', and ``28'' are name of the 3\,mm sources identified in by \cite{kainulainen2017}.}
    \label{appendix-comp}
\end{figure}

In Table\,\ref{cont-source2} of Section\,\ref{result-cont-id}, we compared the positions of FraSCO sources with those of previously identified sources located within the synthesized beam size for our 1.3\,mm continuum image obtained from the ALMA 12-m array ($\sim 1''$) from the positions of FraSCO sources. 
However, some of the millimeter sources identified by \cite{shimajiri2008} and \cite{kainulainen2017} are not associated with individual FraSCO sources due to differences in the angular resolutions and observational frequencies.
These spatial associations made it difficult to list their source positions in Table\,\ref{cont-source2}.
Therefore, in this section, we plot their positions over our 1.3\,mm continuum image obtained from the ALMA 12-m array in Figure\,\ref{appendix-comp} to spatially compare them with FraSCO sources.

\cite{shimajiri2008} carried out 3.3\,mm observations toward the FIR\,3 and FIR\,4 regions using the Nobeyama Millimeter Array (NMA) with an angular resolution of $\sim 6''$ and rms noise level of 1.4\,\mjy.
In Figure\,\ref{appendix-comp}, we compared positions of the sources identified in our 1.3mm ALMA observations and the 3.3\,mm NMA observations. Although beam sizes are very different between the two observations, we found that a few FraSCO sources are located within most of the 3.3\,mm continuum sources, denoted by the red open circles in Figure\,\ref{appendix-comp}. Positional inconsistency between FraSCO sources and the 3.3\,mm continuum sources could be explained that the sources identified with lower angular resolution break up into multiple components at higher resolution.

\cite{kainulainen2017} carried out 3\,mm continuum observations toward the OMC-2/3 region using the ALMA 12-m array and the ACA 7-m array data.
They combined these two data and obtained images with an angular resolution of $\sim 3''$ and rms noise level of 0.23\,\mjy.
In Figure\,\ref{appendix-comp}, the position of most 3\,mm continuum sources is located within $\sim 1''$ from the position of FraSCO sources, while three of them, named 24, 25, and 28 in \cite{kainulainen2017}, are located between positions of two FraSCO sources.
These positional inconsistency could be also explain that the three sources with lower angular resolution break up into two components at higher resolution, as mentioned above when we compared the results with \cite{shimajiri2008}.

\section{CO\,($J$ = 2--1) and SiO\,($J$ = 5--4) Channel maps}
\label{app-ch}
\begin{figure}
    \centering
    \includegraphics[width=15cm]{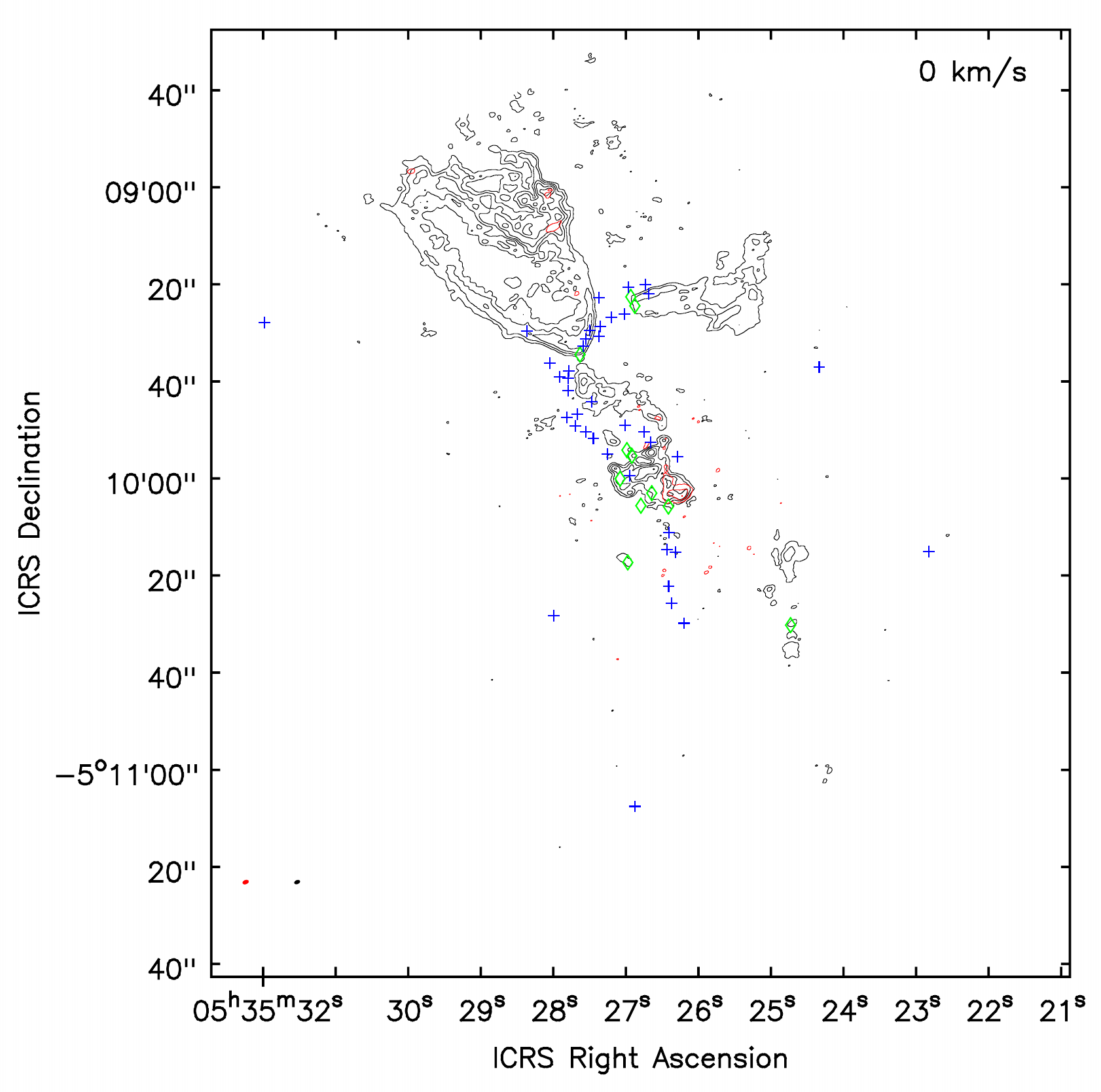}
    \caption{Channel map at \vlsr\,= 0$\,\mathrm{km}\,\mathrm{s^{-1}}$ for both CO\,($J$ = 2--1) and SiO\,($J$ = 5--4) line emissions obtained from the ALMA 12-m array with the velocity resolution of 5\,$\mathrm{km}\,\mathrm{s^{-1}}$ denoted by the black and red contours, respectively. The black contour levels are [4, 20, 50, 100, 180, 250] $\times 1\sigma$ ($1\sigma = 5.3$\,\mjy). The red contour levels are [5, 30, 70] $\times 1\sigma$ ($1\sigma = 4.8$\,\mjy). The symbols show the positions of the FraSCO sources: the green diamonds and blue crosses represent outflow driving sources and sources without outflow, respectively. The black and red ellipses at the bottom-left corner show the synthesized beam size of the CO and SiO images, respectively. The complete figure set including other channel maps (47 images) is available in the online journal.}
    \label{app-ch-5kms}
\end{figure}

\begin{figure}
    \centering
    \includegraphics[width=15cm]{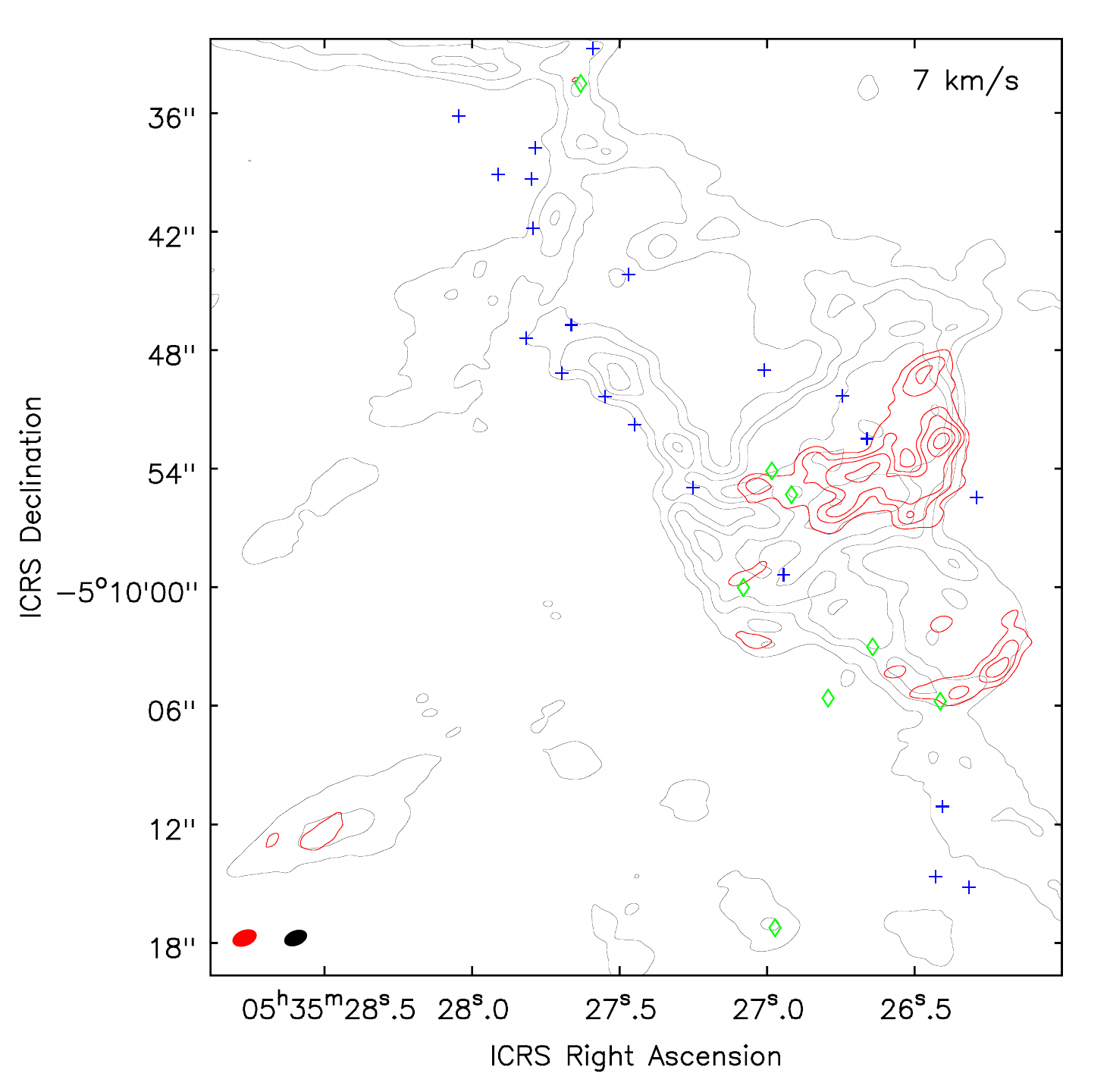}
    \caption{Channel map zoomed in the FIR\,4 region at \vlsr\,= 7\,\kms\,for the CO\,($J$ = 2--1) and SiO\,($J$ = 5--4) line emissions obtained from the ALMA 12-m array with the velocity resolution of 1\,$\mathrm{km}\,\mathrm{s^{-1}}$ denoted by the black and red contours, respectively. The black contour levels are [10, 50, 100, 150, 200, 250] $\times 1\sigma$ ($1\sigma = 10$\,\mjy). The red contour levels are [5, 10, 15, 25, 35, 45] $\times 1\sigma$ ($1\sigma = 9.0$\,\mjy). The symbols show the positions of FraSCO sources: the green diamonds and blue crosses represent outflow driving sources and sources without outflow, respectively. The black and red ellipses at the bottom-left corner show the synthesized beam size of the CO and SiO images, respectively. The complete figure set including other channel maps (24 images) is available in the online journal.}
    \label{app-ch-1kms}
\end{figure}

\begin{figure}
\begin{interactive}{animation}{221109_paper1_co_5kms_ch.mp4}
    \centering
    \includegraphics[width=15cm]{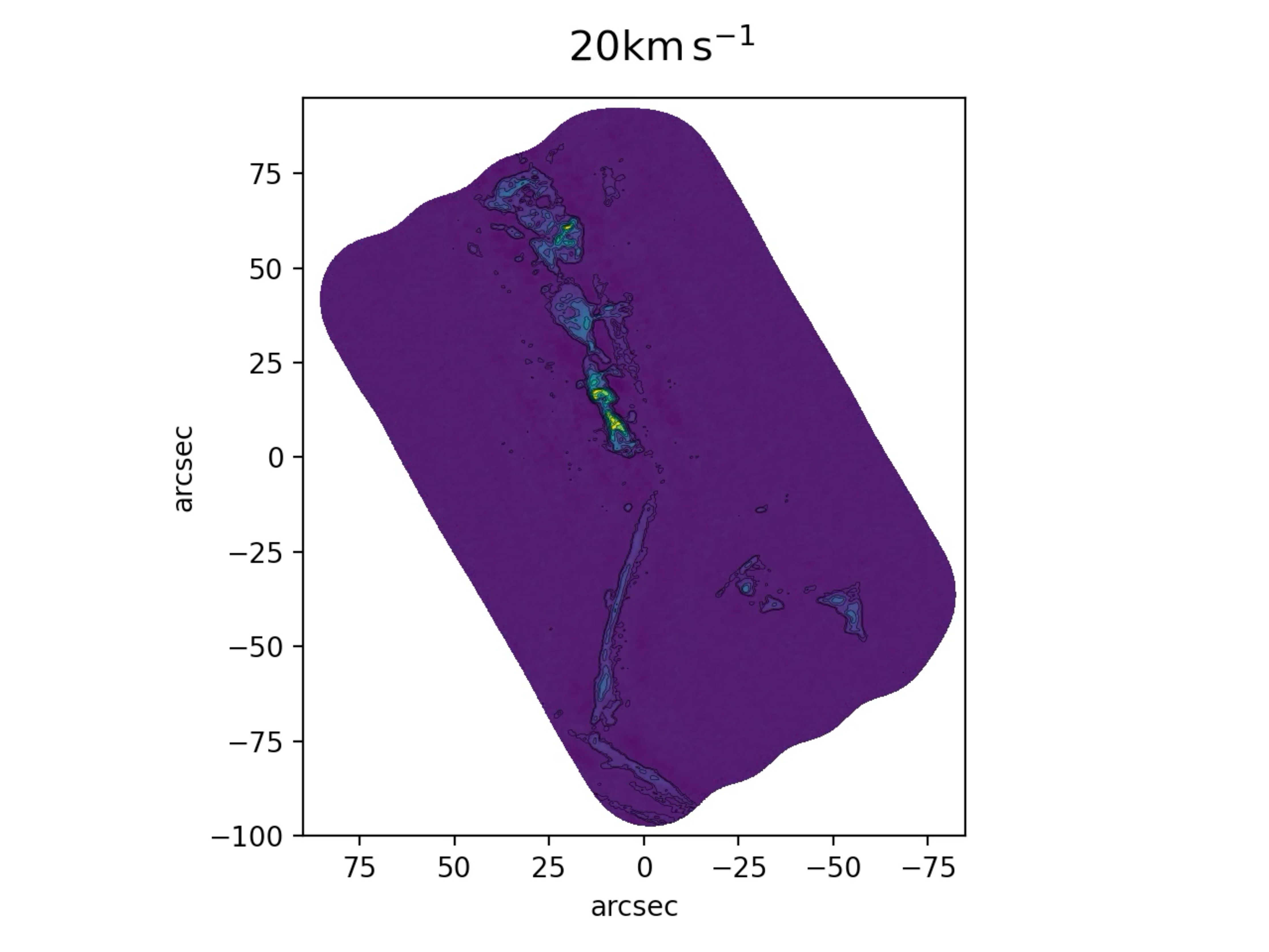}
\end{interactive}
    \caption{Channel map at \vlsr\,= 20$\,\mathrm{km}\,\mathrm{s^{-1}}$ for CO\,($J$ = 2--1) line emission obtained from the ALMA 12-m array with the velocity resolution of 5\,$\mathrm{km}\,\mathrm{s^{-1}}$ denoted by the color scale and black contours. The black contour levels are [5, 10, 30, 50, 100, 150, 200, 250] $\times 1\sigma$ ($1\sigma = 5.3$\,\mjy). This figure is available as an animation.}
    \label{app_co_ch_mp4}
\end{figure}

\begin{figure}
\begin{interactive}{animation}{221109_paper1_sio_1kms_ch.mp4}
    \centering
    \includegraphics[width=15cm]{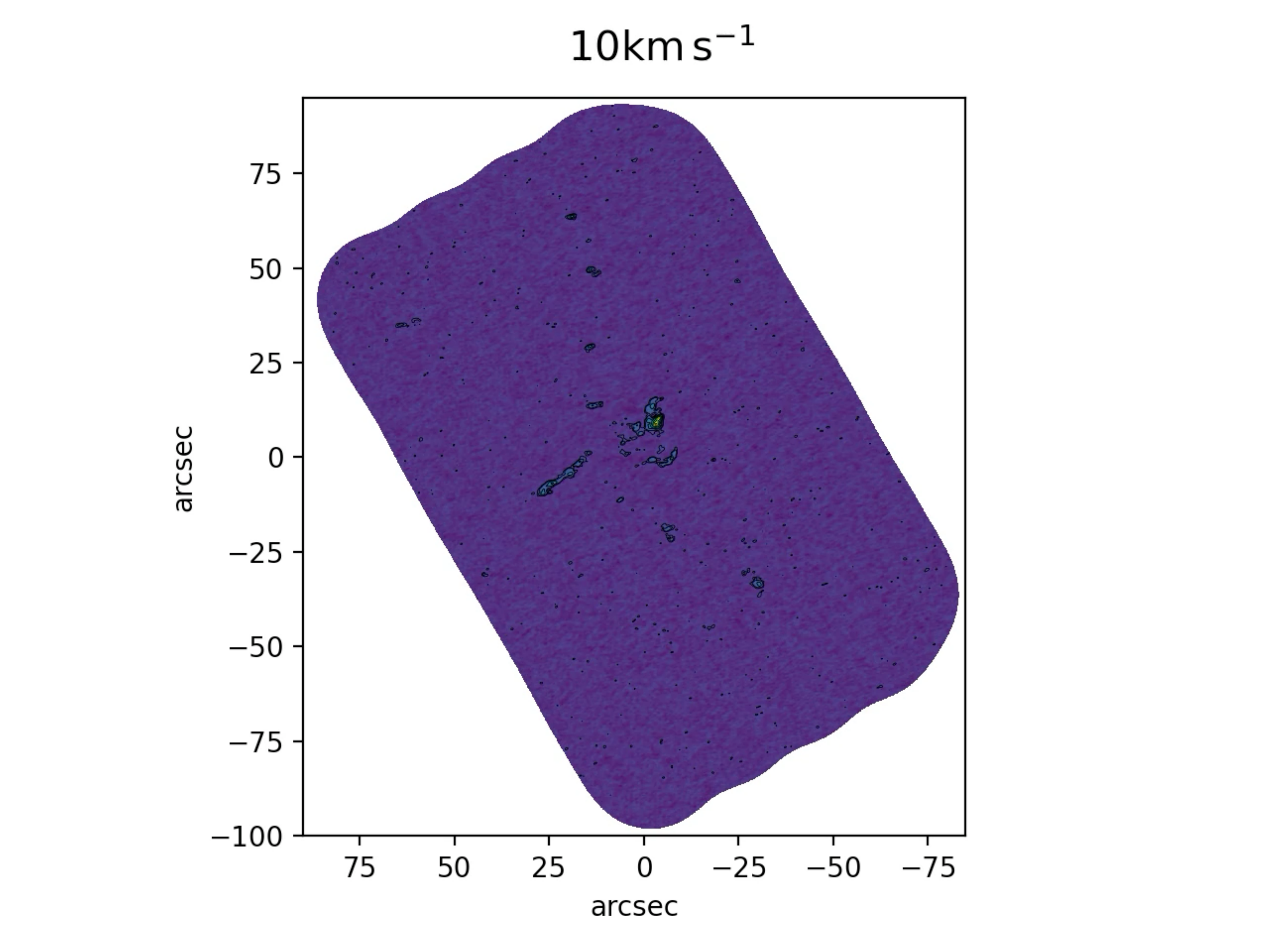}
\end{interactive}
    \caption{Channel map at \vlsr\,= 10$\,\mathrm{km}\,\mathrm{s^{-1}}$ for SiO\,($J$ = 5--4) line emission obtained from the ALMA 12-m array with the velocity resolution of 1\,$\mathrm{km}\,\mathrm{s^{-1}}$ denoted by the color scale and black contours. The black contour levels are [3, 5, 10, 15, 25, 35, 45] $\times 1\sigma$ ($1\sigma = 9.0$\,\mjy). This figure is available as an animation.}
    \label{app_sio_ch_mp4}
\end{figure}

In order to present detailed spatial distributions for CO\,($J$ = 2--1) and SiO\,($J$ = 5--4) emissions, we show channel maps for both line emissions in this section. 
Figures\,\ref{app-ch-5kms} shows an example channel map at \vlsr $\sim$ 0\,$\mathrm{km}\,\mathrm{s^{-1}}$ with the velocity resolution of 5\,$\mathrm{km}\,\mathrm{s^{-1}}$ for the CO and SiO line emissions.
The complete figure set including other channel maps (47 images) is available in the online material, named Fig. Set 1.
Figure\,\ref{app-ch-1kms} shows an example channel map at \vlsr= 7\,$\mathrm{km}\,\mathrm{s^{-1}}$ zooming in the FIR\,4 region with the velocity resolution of 1\,$\mathrm{km}\,\mathrm{s^{-1}}$ for the CO and SiO line emissions.
 The complete figure set including other channel maps (24 images) is available in the online material, named Fig. Set 2.
 Figure\,\ref{app_co_ch_mp4} shows a channel map with the velocity resolution of 5\,\kms\, for the CO line emission as a part of an animation.
 Figure\,\ref{app_sio_ch_mp4} also shows a channel map with the velocity resolution of 1\,\kms\, for the SiO line emission as a part of an animation.
 The animations of there figures are available in the online journal.

\figsetstart
\figsetnum{1}
\figsettitle{Channel maps at \vlsr\,= $-$75--$+$85\,\kms\,for both CO\,($J$ = 2--1) and SiO\,($J$ = 5--4) line emissions obtained from the ALMA 12-m array with the velocity resolution of 5\,$\mathrm{km}\,\mathrm{s^{-1}}$.}

\figsetgrpstart
\figsetgrpnum{figurenumber.16}
\figsetgrptitle{5kms-ch_co-sio_0kms}
\figsetplot{220417-5kms-co-sio-ch-20.eps}
\figsetgrpnote{Channel map at \vlsr\,= 0$\,\mathrm{km}\,\mathrm{s^{-1}}$ for both CO\,($J$ = 2--1) and SiO\,($J$ = 5--4) line emissions obtained from the ALMA 12-m array with the velocity resolution of 5\,$\mathrm{km}\,\mathrm{s^{-1}}$ denoted by the black and red contours, respectively. The black contour levels are [4, 20, 50, 100, 180, 250] $\times 1\sigma$ ($1\sigma = 5.3$\,\mjy). The red contour levels are [5, 30, 70] $\times 1\sigma$ ($1\sigma = 4.8$\,\mjy). The symbols show the positions of the FraSCO sources: the green diamonds and blue crosses represent outflow driving sources and sources without outflow, respectively. The black and red ellipses at the bottom-left corner show the synthesized beam size of the CO and SiO images, respectively.}
\figsetgrpend

\figsetend

\figsetstart
\figsetnum{2}
\figsettitle{Channel maps at \vlsr\,= $-$19--$+$27\,\kms\, for both CO\,($J$ = 2--1) and SiO\,($J$ = 5--4) line emissions obtained from the ALMA 12-m array with the velocity resolution of 1\,$\mathrm{km}\,\mathrm{s^{-1}}$.}

\figsetgrpstart
\figsetgrpnum{figurenumber.27}
\figsetgrptitle{1kms-ch_co-sio_7kms}
\figsetplot{220405-1kms-co-sio-ch-37.eps}
\figsetgrpnote{Channel map zoomed in the FIR\,4 region at \vlsr\,= $+$7\,\kms\,for the CO\,($J$ = 2--1) and SiO\,($J$ = 5--4) line emissions obtained from the ALMA 12-m array with the velocity resolution of 1\,$\mathrm{km}\,\mathrm{s^{-1}}$ denoted by the black and red contours, respectively. The black contour levels are [10, 50, 100, 150, 200, 250] $\times 1\sigma$ ($1\sigma = 10$\,\mjy). The red contour levels are [5, 10, 15, 25, 35, 45] $\times 1\sigma$ ($1\sigma = 9.0$\,\mjy). The symbols show the positions of FraSCO sources: the green diamonds and blue crosses represent outflow driving sources and sources without outflow, respectively. The black and red ellipses at the bottom-left corner show the synthesized beam size of the CO and SiO images, respectively.}
\figsetgrpend

\figsetend

\bibliography{paper1}{}

\begin{thebibliography}{}
\expandafter\ifx\csname natexlab\endcsname\relax\def\natexlab#1{#1}\fi
\providecommand{\url}[1]{\href{#1}{#1}}
\providecommand{\dodoi}[1]{doi:~\href{http://doi.org/#1}{\nolinkurl{#1}}}
\providecommand{\doeprint}[1]{\href{http://ascl.net/#1}{\nolinkurl{http://ascl.net/#1}}}
\providecommand{\doarXiv}[1]{\href{https://arxiv.org/abs/#1}{\nolinkurl{https://arxiv.org/abs/#1}}}

\bibitem[{{Allen} {et~al.}(2007){Allen}, {Megeath}, {Gutermuth}, {Myers},
  {Wolk}, {Adams}, {Muzerolle}, {Young}, \& {Pipher}}]{allen2007}
{Allen}, L., {Megeath}, S.~T., {Gutermuth}, R., {et~al.} 2007, in Protostars
  and Planets V, ed. B.~{Reipurth}, D.~{Jewitt}, \& K.~{Keil}, 361.
\newblock \doarXiv{astro-ph/0603096}

\bibitem[{{Andr{\'e}} {et~al.}(2014){Andr{\'e}}, {Di Francesco},
  {Ward-Thompson}, {Inutsuka}, {Pudritz}, \& {Pineda}}]{andre2014}
{Andr{\'e}}, P., {Di Francesco}, J., {Ward-Thompson}, D., {et~al.} 2014, in
  Protostars and Planets VI, ed. H.~{Beuther}, R.~S. {Klessen}, C.~P.
  {Dullemond}, \& T.~{Henning}, 27,
  \dodoi{10.2458/azu_uapress_9780816531240-ch002}

\bibitem[{{Anglada} {et~al.}(1998){Anglada}, {Villuendas}, {Estalella},
  {Beltr{\'a}n}, {Rodr{\'\i}guez}, {Torrelles}, \& {Curiel}}]{anglada1998}
{Anglada}, G., {Villuendas}, E., {Estalella}, R., {et~al.} 1998, \aj, 116,
  2953, \dodoi{10.1086/300637}

\bibitem[{{Arce} {et~al.}(2007){Arce}, {Shepherd}, {Gueth}, {Lee}, {Bachiller},
  {Rosen}, \& {Beuther}}]{arce2007}
{Arce}, H.~G., {Shepherd}, D., {Gueth}, F., {et~al.} 2007, in Protostars and
  Planets V, ed. B.~{Reipurth}, D.~{Jewitt}, \& K.~{Keil}, 245.
\newblock \doarXiv{astro-ph/0603071}

\bibitem[{{Aso} {et~al.}(2000){Aso}, {Tatematsu}, {Sekimoto}, {Nakano},
  {Umemoto}, {Koyama}, \& {Yamamoto}}]{Aso2000}
{Aso}, Y., {Tatematsu}, K., {Sekimoto}, Y., {et~al.} 2000, \apjs, 131, 465,
  \dodoi{10.1086/317378}

\bibitem[{{Bachiller}(1996)}]{bachiller1996}
{Bachiller}, R. 1996, \araa, 34, 111, \dodoi{10.1146/annurev.astro.34.1.111}

\bibitem[{{Bally} {et~al.}(1987){Bally}, {Langer}, {Stark}, \&
  {Wilson}}]{bally1987}
{Bally}, J., {Langer}, W.~D., {Stark}, A.~A., \& {Wilson}, R.~W. 1987, \apjl,
  312, L45, \dodoi{10.1086/184817}

\bibitem[{{Bonnell} {et~al.}(2001){Bonnell}, {Bate}, {Clarke}, \&
  {Pringle}}]{bonnell2001}
{Bonnell}, I.~A., {Bate}, M.~R., {Clarke}, C.~J., \& {Pringle}, J.~E. 2001,
  \mnras, 323, 785, \dodoi{10.1046/j.1365-8711.2001.04270.x}

\bibitem[{{Bontemps} {et~al.}(1996){Bontemps}, {Andre}, {Terebey}, \&
  {Cabrit}}]{bontemps1996}
{Bontemps}, S., {Andre}, P., {Terebey}, S., \& {Cabrit}, S. 1996, \aap, 311,
  858

\bibitem[{{Buckle} {et~al.}(2012){Buckle}, {Davis}, {di Francesco}, {Graves},
  {Nutter}, {Richer}, {Roberts}, {Ward-Thompson}, {White}, {Brunt}, {Butner},
  {Cavanagh}, {Chrysostomou}, {Curtis}, {Duarte-Cabral}, {Etxaluze}, {Fich},
  {Friberg}, {Friesen}, {Fuller}, {Greaves}, {Hatchell}, {Hogerheijde},
  {Johnstone}, {Matthews}, {Matthews}, {Rawlings}, {Sadavoy}, {Simpson},
  {Tothill}, {Tsamis}, {Viti}, {Wouterloot}, \& {Yates}}]{buckle2012}
{Buckle}, J.~V., {Davis}, C.~J., {di Francesco}, J., {et~al.} 2012, \mnras,
  422, 521, \dodoi{10.1111/j.1365-2966.2012.20628.x}

\bibitem[{{Cao} {et~al.}(2022){Cao}, {Qiu}, {Zhang}, \& {Li}}]{cao2022}
{Cao}, Y., {Qiu}, K., {Zhang}, Q., \& {Li}, G.-X. 2022, \apj, 927, 106,
  \dodoi{10.3847/1538-4357/ac4696}

\bibitem[{{Caselli} {et~al.}(1997){Caselli}, {Hartquist}, \&
  {Havnes}}]{caselli1997}
{Caselli}, P., {Hartquist}, T.~W., \& {Havnes}, O. 1997, \aap, 322, 296

\bibitem[{{Cheng} {et~al.}(2018){Cheng}, {Tan}, {Liu}, {Kong}, {Lim},
  {Andersen}, \& {Da Rio}}]{cheng2018}
{Cheng}, Y., {Tan}, J.~C., {Liu}, M., {et~al.} 2018, \apj, 853, 160,
  \dodoi{10.3847/1538-4357/aaa3f1}

\bibitem[{{Chini} {et~al.}(1997){Chini}, {Reipurth}, {Ward-Thompson}, {Bally},
  {Nyman}, {Sievers}, \& {Billawala}}]{chini1997}
{Chini}, R., {Reipurth}, B., {Ward-Thompson}, D., {et~al.} 1997, \apjl, 474,
  L135, \dodoi{10.1086/310436}

\bibitem[{{Clarke} {et~al.}(2020){Clarke}, {Williams}, \& {Walch}}]{clarke2020}
{Clarke}, S.~D., {Williams}, G.~M., \& {Walch}, S. 2020, \mnras, 497, 4390,
  \dodoi{10.1093/mnras/staa2298}

\bibitem[{{Crimier} {et~al.}(2009){Crimier}, {Ceccarelli}, {Lefloch}, \&
  {Faure}}]{crimier2009}
{Crimier}, N., {Ceccarelli}, C., {Lefloch}, B., \& {Faure}, A. 2009, \aap, 506,
  1229, \dodoi{10.1051/0004-6361/200911651}

\bibitem[{{Davis} {et~al.}(2010){Davis}, {Gell}, {Khanzadyan}, {Smith}, \&
  {Jenness}}]{davis2009}
{Davis}, C.~J., {Gell}, R., {Khanzadyan}, T., {Smith}, M.~D., \& {Jenness}, T.
  2010, \aap, 511, A24, \dodoi{10.1051/0004-6361/200913561}

\bibitem[{{Evans}(1999)}]{evans1999}
{Evans}, Neal~J., I. 1999, \araa, 37, 311,
  \dodoi{10.1146/annurev.astro.37.1.311}

\bibitem[{{Evans} {et~al.}(2021){Evans}, {Fontani}, {Vastel}, {Ceccarelli},
  {Caselli}, {L{\'o}pez-Sepulcre}, {Neri}, {Alves}, {Chahine}, {Favre}, \&
  {Lattanzi}}]{evans2021}
{Evans}, L., {Fontani}, F., {Vastel}, C., {et~al.} 2021, arXiv e-prints,
  arXiv:2110.10427.
\newblock \doarXiv{2110.10427}

\bibitem[{{Favre} {et~al.}(2018){Favre}, {Ceccarelli}, {L{\'o}pez-Sepulcre},
  {Fontani}, {Neri}, {Manigand}, {Kama}, {Caselli}, {Jaber Al-Edhari},
  {Kahane}, {Alves}, {Balucani}, {Bianchi}, {Caux}, {Codella}, {Dulieu},
  {Pineda}, {Sims}, \& {Theul{\'e}}}]{favre2018}
{Favre}, C., {Ceccarelli}, C., {L{\'o}pez-Sepulcre}, A., {et~al.} 2018, \apj,
  859, 136, \dodoi{10.3847/1538-4357/aabfd4}

\bibitem[{{Feddersen} {et~al.}(2020){Feddersen}, {Arce}, {Kong}, {Suri},
  {S{\'a}nchez-Monge}, {Ossenkopf-Okada}, {Dunham}, {Nakamura}, {Shimajiri}, \&
  {Bally}}]{feddersen2020}
{Feddersen}, J.~R., {Arce}, H.~G., {Kong}, S., {et~al.} 2020, \apj, 896, 11,
  \dodoi{10.3847/1538-4357/ab86a9}

\bibitem[{{Fontani} {et~al.}(2017){Fontani}, {Ceccarelli}, {Favre}, {Caselli},
  {Neri}, {Sims}, {Kahane}, {Alves}, {Balucani}, {Bianchi}, {Caux}, {Jaber
  Al-Edhari}, {Lopez-Sepulcre}, {Pineda}, {Bachiller}, {Bizzocchi},
  {Bottinelli}, {Chacon-Tanarro}, {Choudhury}, {Codella}, {Coutens}, {Dulieu},
  {Feng}, {Rimola}, {Hily-Blant}, {Holdship}, {Jimenez-Serra}, {Laas},
  {Lefloch}, {Oya}, {Podio}, {Pon}, {Punanova}, {Quenard}, {Sakai}, {Spezzano},
  {Taquet}, {Testi}, {Theul{\'e}}, {Ugliengo}, {Vastel}, {Vasyunin}, {Viti},
  {Yamamoto}, \& {Wiesenfeld}}]{fontani2017}
{Fontani}, F., {Ceccarelli}, C., {Favre}, C., {et~al.} 2017, \aap, 605, A57,
  \dodoi{10.1051/0004-6361/201730527}

\bibitem[{{Furlan} {et~al.}(2016){Furlan}, {Fischer}, {Ali}, {Stutz}, {Stanke},
  {Tobin}, {Megeath}, {Osorio}, {Hartmann}, {Calvet}, {Poteet}, {Booker},
  {Manoj}, {Watson}, \& {Allen}}]{furlan2016}
{Furlan}, E., {Fischer}, W.~J., {Ali}, B., {et~al.} 2016, \apjs, 224, 5,
  \dodoi{10.3847/0067-0049/224/1/5}

\bibitem[{{Gonz{\'a}lez-Garc{\'\i}a} {et~al.}(2016){Gonz{\'a}lez-Garc{\'\i}a},
  {Manoj}, {Watson}, {Vavrek}, {Megeath}, {Stutz}, {Osorio}, {Wyrowski},
  {Fischer}, {Tobin}, {S{\'a}nchez-Portal}, {Diaz Rodriguez}, \&
  {Wilson}}]{gonzalez2016}
{Gonz{\'a}lez-Garc{\'\i}a}, B., {Manoj}, P., {Watson}, D.~M., {et~al.} 2016,
  \aap, 596, A26, \dodoi{10.1051/0004-6361/201527186}

\bibitem[{{Goodman} {et~al.}(2009){Goodman}, {Rosolowsky}, {Borkin}, {Foster},
  {Halle}, {Kauffmann}, \& {Pineda}}]{goodman2009}
{Goodman}, A.~A., {Rosolowsky}, E.~W., {Borkin}, M.~A., {et~al.} 2009, \nat,
  457, 63, \dodoi{10.1038/nature07609}

\bibitem[{{Gro{\ss}schedl} {et~al.}(2018){Gro{\ss}schedl}, {Alves}, {Meingast},
  {Ackerl}, {Ascenso}, {Bouy}, {Burkert}, {Forbrich}, {F{\"u}rnkranz},
  {Goodman}, {Hacar}, {Herbst-Kiss}, {Lada}, {Larreina}, {Leschinski},
  {Lombardi}, {Moitinho}, {Mortimer}, \& {Zari}}]{groschedl2018}
{Gro{\ss}schedl}, J.~E., {Alves}, J., {Meingast}, S., {et~al.} 2018, \aap, 619,
  A106, \dodoi{10.1051/0004-6361/201833901}

\bibitem[{{Gueth} {et~al.}(1998){Gueth}, {Guilloteau}, \&
  {Bachiller}}]{gueth1998}
{Gueth}, F., {Guilloteau}, S., \& {Bachiller}, R. 1998, \aap, 333, 287

\bibitem[{{Habel} {et~al.}(2021){Habel}, {Megeath}, {Booker}, {Fischer},
  {Kounkel}, {Poteet}, {Furlan}, {Stutz}, {Manoj}, {Tobin}, {Nagy}, {Pokhrel},
  \& {Watson}}]{habel2021}
{Habel}, N.~M., {Megeath}, S.~T., {Booker}, J.~J., {et~al.} 2021, arXiv
  e-prints, arXiv:2102.06717.
\newblock \doarXiv{2102.06717}

\bibitem[{{Hacar} {et~al.}(2018){Hacar}, {Tafalla}, {Forbrich}, {Alves},
  {Meingast}, {Grossschedl}, \& {Teixeira}}]{hacar2018}
{Hacar}, A., {Tafalla}, M., {Forbrich}, J., {et~al.} 2018, \aap, 610, A77,
  \dodoi{10.1051/0004-6361/201731894}

\bibitem[{{Hacar} {et~al.}(2013){Hacar}, {Tafalla}, {Kauffmann}, \&
  {Kov{\'a}cs}}]{hacar2013}
{Hacar}, A., {Tafalla}, M., {Kauffmann}, J., \& {Kov{\'a}cs}, A. 2013, \aap,
  554, A55, \dodoi{10.1051/0004-6361/201220090}

\bibitem[{{Hansen} {et~al.}(2012){Hansen}, {Klein}, {McKee}, \&
  {Fisher}}]{hansen2012}
{Hansen}, C.~E., {Klein}, R.~I., {McKee}, C.~F., \& {Fisher}, R.~T. 2012, \apj,
  747, 22, \dodoi{10.1088/0004-637X/747/1/22}

\bibitem[{{Hayashi}(1981)}]{hayashi1981}
{Hayashi}, C. 1981, Progress of Theoretical Physics Supplement, 70, 35,
  \dodoi{10.1143/PTPS.70.35}

\bibitem[{{Hill} {et~al.}(2011){Hill}, {Motte}, {Didelon}, {Bontemps},
  {Minier}, {Hennemann}, {Schneider}, {Andr{\'e}}, {Men'shchikov}, {Anderson},
  {Arzoumanian}, {Bernard}, {di Francesco}, {Elia}, {Giannini}, {Griffin},
  {K{\"o}nyves}, {Kirk}, {Marston}, {Martin}, {Molinari}, {Nguyen Luong},
  {Peretto}, {Pezzuto}, {Roussel}, {Sauvage}, {Sousbie}, {Testi},
  {Ward-Thompson}, {White}, {Wilson}, \& {Zavagno}}]{hill2011}
{Hill}, T., {Motte}, F., {Didelon}, P., {et~al.} 2011, \aap, 533, A94,
  \dodoi{10.1051/0004-6361/201117315}

\bibitem[{{Hirano} {et~al.}(2010){Hirano}, {Ho}, {Liu}, {Shang}, {Lee}, \&
  {Bourke}}]{hirano2010}
{Hirano}, N., {Ho}, P. P.~T., {Liu}, S.-Y., {et~al.} 2010, \apj, 717, 58,
  \dodoi{10.1088/0004-637X/717/1/58}

\bibitem[{{Ikeda} {et~al.}(2007){Ikeda}, {Sunada}, \& {Kitamura}}]{ikeda2007}
{Ikeda}, N., {Sunada}, K., \& {Kitamura}, Y. 2007, \apj, 665, 1194,
  \dodoi{10.1086/519484}

\bibitem[{{Ishii} {et~al.}(2019){Ishii}, {Nakamura}, {Shimajiri}, {Kawabe},
  {Tsukagoshi}, {Dobashi}, \& {Shimoikura}}]{ishii2019}
{Ishii}, S., {Nakamura}, F., {Shimajiri}, Y., {et~al.} 2019, \pasj, 71, S9,
  \dodoi{10.1093/pasj/psz065}

\bibitem[{{Jeans}(1902)}]{jeans1902}
{Jeans}, J.~H. 1902, Philosophical Transactions of the Royal Society of London
  Series A, 199, 1, \dodoi{10.1098/rsta.1902.0012}

\bibitem[{{Johnstone} \& {Bally}(1999)}]{johnstone1999}
{Johnstone}, D., \& {Bally}, J. 1999, \apjl, 510, L49, \dodoi{10.1086/311792}

\bibitem[{{Kainulainen} {et~al.}(2017){Kainulainen}, {Stutz}, {Stanke},
  {Abreu-Vicente}, {Beuther}, {Henning}, {Johnston}, \&
  {Megeath}}]{kainulainen2017}
{Kainulainen}, J., {Stutz}, A.~M., {Stanke}, T., {et~al.} 2017, \aap, 600,
  A141, \dodoi{10.1051/0004-6361/201628481}

\bibitem[{{Kang} {et~al.}(2021){Kang}, {Choi}, {Wyrowski}, {Kim}, {Bieging},
  {Kim}, {Park}, {Megeath}, {Choi}, {Kang}, {Yoo}, \& {Manoj}}]{kang2021}
{Kang}, M., {Choi}, M., {Wyrowski}, F., {et~al.} 2021, \apjs, 255, 2,
  \dodoi{10.3847/1538-4365/abfd35}

\bibitem[{{Kirk} {et~al.}(2013){Kirk}, {Myers}, {Bourke}, {Gutermuth},
  {Hedden}, \& {Wilson}}]{kirk2013}
{Kirk}, H., {Myers}, P.~C., {Bourke}, T.~L., {et~al.} 2013, \apj, 766, 115,
  \dodoi{10.1088/0004-637X/766/2/115}

\bibitem[{{Kirk} {et~al.}(2017){Kirk}, {Friesen}, {Pineda}, {Rosolowsky},
  {Offner}, {Matzner}, {Myers}, {Di Francesco}, {Caselli}, {Alves},
  {Chac{\'o}n-Tanarro}, {Chen}, {Chun-Yuan Chen}, {Keown}, {Punanova}, {Seo},
  {Shirley}, {Ginsburg}, {Hall}, {Singh}, {Arce}, {Goodman}, {Martin}, \&
  {Redaelli}}]{Kirk2017}
{Kirk}, H., {Friesen}, R.~K., {Pineda}, J.~E., {et~al.} 2017, \apj, 846, 144,
  \dodoi{10.3847/1538-4357/aa8631}

\bibitem[{{Kong} {et~al.}(2018){Kong}, {Arce}, {Feddersen}, {Carpenter},
  {Nakamura}, {Shimajiri}, {Isella}, {Ossenkopf-Okada}, {Sargent},
  {S{\'a}nchez-Monge}, {Suri}, {Kauffmann}, {Pillai}, {Pineda}, {Koda},
  {Bally}, {Lis}, {Padoan}, {Klessen}, {Mairs}, {Goodman}, {Goldsmith},
  {McGehee}, {Schilke}, {Teuben}, {Maureira}, {Hara}, {Ginsburg}, {Burkhart},
  {Smith}, {Schmiedeke}, {Pineda}, {Ishii}, {Sasaki}, {Kawabe}, {Urasawa},
  {Oyamada}, \& {Tanabe}}]{kong2018}
{Kong}, S., {Arce}, H.~G., {Feddersen}, J.~R., {et~al.} 2018, \apjs, 236, 25,
  \dodoi{10.3847/1538-4365/aabafc}

\bibitem[{{Krumholz} {et~al.}(2012){Krumholz}, {Klein}, \&
  {McKee}}]{krumholz2012}
{Krumholz}, M.~R., {Klein}, R.~I., \& {McKee}, C.~F. 2012, \apj, 754, 71,
  \dodoi{10.1088/0004-637X/754/1/71}

\bibitem[{{Lada} \& {Lada}(2003)}]{ladalada2003}
{Lada}, C.~J., \& {Lada}, E.~A. 2003, \araa, 41, 57,
  \dodoi{10.1146/annurev.astro.41.011802.094844}

\bibitem[{{Lada} {et~al.}(2008){Lada}, {Muench}, {Rathborne}, {Alves}, \&
  {Lombardi}}]{Lada2008}
{Lada}, C.~J., {Muench}, A.~A., {Rathborne}, J., {Alves}, J.~F., \& {Lombardi},
  M. 2008, \apj, 672, 410, \dodoi{10.1086/523837}

\bibitem[{{Li} {et~al.}(2013){Li}, {Kauffmann}, {Zhang}, \& {Chen}}]{Li2013}
{Li}, D., {Kauffmann}, J., {Zhang}, Q., \& {Chen}, W. 2013, \apjl, 768, L5,
  \dodoi{10.1088/2041-8205/768/1/L5}

\bibitem[{{Lis} {et~al.}(1998){Lis}, {Serabyn}, {Keene}, {Dowell}, {Benford},
  {Phillips}, {Hunter}, \& {Wang}}]{lis1998}
{Lis}, D.~C., {Serabyn}, E., {Keene}, J., {et~al.} 1998, \apj, 509, 299,
  \dodoi{10.1086/306500}

\bibitem[{{Liu} {et~al.}(2021){Liu}, {Tan}, {Marvil}, {Kong}, {Rosero},
  {Caselli}, \& {Cosentino}}]{liu2021}
{Liu}, M., {Tan}, J.~C., {Marvil}, J., {et~al.} 2021, \apj, 921, 96,
  \dodoi{10.3847/1538-4357/ac0829}

\bibitem[{{L{\'o}pez-Sepulcre} {et~al.}(2013){L{\'o}pez-Sepulcre}, {Taquet},
  {S{\'a}nchez-Monge}, {Ceccarelli}, {Dominik}, {Kama}, {Caux}, {Fontani},
  {Fuente}, {Ho}, {Neri}, \& {Shimajiri}}]{lopez2013}
{L{\'o}pez-Sepulcre}, A., {Taquet}, V., {S{\'a}nchez-Monge}, {\'A}., {et~al.}
  2013, \aap, 556, A62, \dodoi{10.1051/0004-6361/201220905}

\bibitem[{{Machida} {et~al.}(2007){Machida}, {Inutsuka}, \&
  {Matsumoto}}]{machida2007}
{Machida}, M.~N., {Inutsuka}, S.-i., \& {Matsumoto}, T. 2007, \apj, 670, 1198,
  \dodoi{10.1086/521779}

\bibitem[{{Machida} {et~al.}(2008){Machida}, {Inutsuka}, \&
  {Matsumoto}}]{machida2008}
---. 2008, \apj, 676, 1088, \dodoi{10.1086/528364}

\bibitem[{{Maddalena} \& {Thaddeus}(1985)}]{Maddalena1985}
{Maddalena}, R.~J., \& {Thaddeus}, P. 1985, \apj, 294, 231,
  \dodoi{10.1086/163291}

\bibitem[{{Mason} {et~al.}(2020){Mason}, {Dicker}, {Sadavoy}, {Stanchfield},
  {Mroczkowski}, {Romero}, {Friesen}, {Sarazin}, {Sievers}, {Stanke}, \&
  {Devlin}}]{Mason2020}
{Mason}, B., {Dicker}, S., {Sadavoy}, S., {et~al.} 2020, \apj, 893, 13,
  \dodoi{10.3847/1538-4357/ab734a}

\bibitem[{{Matsushita} {et~al.}(2021){Matsushita}, {Takahashi}, {Ishii},
  {Tomisaka}, {Ho}, {Carpenter}, \& {Machida}}]{Matsushita2021}
{Matsushita}, Y., {Takahashi}, S., {Ishii}, S., {et~al.} 2021, \apj, 916, 23,
  \dodoi{10.3847/1538-4357/ac069f}

\bibitem[{{Matsushita} {et~al.}(2019){Matsushita}, {Takahashi}, {Machida}, \&
  {Tomisaka}}]{matsushita2019}
{Matsushita}, Y., {Takahashi}, S., {Machida}, M.~N., \& {Tomisaka}, K. 2019,
  \apj, 871, 221, \dodoi{10.3847/1538-4357/aaf1b6}

\bibitem[{{McKee} \& {Tan}(2003)}]{mckee2003}
{McKee}, C.~F., \& {Tan}, J.~C. 2003, \apj, 585, 850, \dodoi{10.1086/346149}

\bibitem[{{McLaughlin} \& {Pudritz}(1996)}]{McLaughlin1996ApJ...469..194M}
{McLaughlin}, D.~E., \& {Pudritz}, R.~E. 1996, \apj, 469, 194,
  \dodoi{10.1086/177771}

\bibitem[{{Megeath} {et~al.}(2012){Megeath}, {Gutermuth}, {Muzerolle},
  {Kryukova}, {Flaherty}, {Hora}, {Allen}, {Hartmann}, {Myers}, {Pipher},
  {Stauffer}, {Young}, \& {Fazio}}]{megeath2012}
{Megeath}, S.~T., {Gutermuth}, R., {Muzerolle}, J., {et~al.} 2012, \aj, 144,
  192, \dodoi{10.1088/0004-6256/144/6/192}

\bibitem[{{Mikami} {et~al.}(1992){Mikami}, {Umemoto}, {Yamamoto}, \&
  {Saito}}]{mikami1992}
{Mikami}, H., {Umemoto}, T., {Yamamoto}, S., \& {Saito}, S. 1992, \apjl, 392,
  L87, \dodoi{10.1086/186432}

\bibitem[{{Motte} {et~al.}(2010){Motte}, {Zavagno}, {Bontemps}, {Schneider},
  {Hennemann}, {di Francesco}, {Andr{\'e}}, {Saraceno}, {Griffin}, {Marston},
  {Ward-Thompson}, {White}, {Minier}, {Men'shchikov}, {Hill}, {Abergel},
  {Anderson}, {Aussel}, {Balog}, {Baluteau}, {Bernard}, {Cox}, {Csengeri},
  {Deharveng}, {Didelon}, {di Giorgio}, {Hargrave}, {Huang}, {Kirk}, {Leeks},
  {Li}, {Martin}, {Molinari}, {Nguyen-Luong}, {Olofsson}, {Persi}, {Peretto},
  {Pezzuto}, {Roussel}, {Russeil}, {Sadavoy}, {Sauvage}, {Sibthorpe},
  {Spinoglio}, {Testi}, {Teyssier}, {Vavrek}, {Wilson}, \&
  {Woodcraft}}]{Motte2010A}
{Motte}, F., {Zavagno}, A., {Bontemps}, S., {et~al.} 2010, \aap, 518, L77,
  \dodoi{10.1051/0004-6361/201014690}

\bibitem[{{Myers}(2009)}]{myers2009}
{Myers}, P.~C. 2009, \apj, 700, 1609, \dodoi{10.1088/0004-637X/700/2/1609}

\bibitem[{{Nagahama} {et~al.}(1998){Nagahama}, {Mizuno}, {Ogawa}, \&
  {Fukui}}]{Nagahama1998}
{Nagahama}, T., {Mizuno}, A., {Ogawa}, H., \& {Fukui}, Y. 1998, \aj, 116, 336,
  \dodoi{10.1086/300392}

\bibitem[{{Nakamura} \& {Li}(2011)}]{nakamura2011}
{Nakamura}, F., \& {Li}, Z.-Y. 2011, \apj, 740, 36,
  \dodoi{10.1088/0004-637X/740/1/36}

\bibitem[{{Nakamura} {et~al.}(2019){Nakamura}, {Oyamada}, {Okumura}, {Ishii},
  {Shimajiri}, {Tanabe}, {Tsukagoshi}, {Kawabe}, {Momose}, {Urasawa}, {Nishi},
  {Lin}, {Lai}, {Dobashi}, {Shimoikura}, \& {Sugitani}}]{nakamura2019}
{Nakamura}, F., {Oyamada}, S., {Okumura}, S., {et~al.} 2019, \pasj, 71, S10,
  \dodoi{10.1093/pasj/psz001}

\bibitem[{{Nielbock} {et~al.}(2003){Nielbock}, {Chini}, \&
  {M{\"u}ller}}]{nielbock2003}
{Nielbock}, M., {Chini}, R., \& {M{\"u}ller}, S.~A.~H. 2003, \aap, 408, 245,
  \dodoi{10.1051/0004-6361:20030961}

\bibitem[{{Nutter} \& {Ward-Thompson}(2007)}]{nutter2007}
{Nutter}, D., \& {Ward-Thompson}, D. 2007, \mnras, 374, 1413,
  \dodoi{10.1111/j.1365-2966.2006.11246.x}

\bibitem[{{O'Dell} {et~al.}(2008){O'Dell}, {Muench}, {Smith}, \&
  {Zapata}}]{ODell2008}
{O'Dell}, C.~R., {Muench}, A., {Smith}, N., \& {Zapata}, L. 2008, in Handbook
  of Star Forming Regions, Volume I, ed. B.~{Reipurth}, Vol.~4, 544

\bibitem[{{Offner} \& {Chaban}(2017)}]{Offner2017}
{Offner}, S. S.~R., \& {Chaban}, J. 2017, \apj, 847, 104,
  \dodoi{10.3847/1538-4357/aa8996}

\bibitem[{{Ohashi} {et~al.}(2016){Ohashi}, {Sanhueza}, {Chen}, {Zhang},
  {Busquet}, {Nakamura}, {Palau}, \& {Tatematsu}}]{Ohashi2016ApJ...833..209O}
{Ohashi}, S., {Sanhueza}, P., {Chen}, H.-R.~V., {et~al.} 2016, \apj, 833, 209,
  \dodoi{10.3847/1538-4357/833/2/209}

\bibitem[{{Osorio} {et~al.}(2017){Osorio}, {D{\'\i}az-Rodr{\'\i}guez},
  {Anglada}, {Megeath}, {Rodr{\'\i}guez}, {Tobin}, {Stutz}, {Furlan},
  {Fischer}, {Manoj}, {G{\'o}mez}, {Gonz{\'a}lez-Garc{\'\i}a}, {Stanke},
  {Watson}, {Loinard}, {Vavrek}, \& {Carrasco-Gonz{\'a}lez}}]{osorio2017}
{Osorio}, M., {D{\'\i}az-Rodr{\'\i}guez}, A.~K., {Anglada}, G., {et~al.} 2017,
  \apj, 840, 36, \dodoi{10.3847/1538-4357/aa6975}

\bibitem[{{Ossenkopf} \& {Henning}(1994)}]{ossenkopf1994}
{Ossenkopf}, V., \& {Henning}, T. 1994, \aap, 291, 943

\bibitem[{{Palla} \& {Stahler}(1993)}]{palla1993}
{Palla}, F., \& {Stahler}, S.~W. 1993, \apj, 418, 414, \dodoi{10.1086/173402}

\bibitem[{{Pattle} {et~al.}(2015){Pattle}, {Ward-Thompson}, {Kirk}, {White},
  {Drabek-Maunder}, {Buckle}, {Beaulieu}, {Berry}, {Broekhoven-Fiene},
  {Currie}, {Fich}, {Hatchell}, {Kirk}, {Jenness}, {Johnstone}, {Mottram},
  {Nutter}, {Pineda}, {Quinn}, {Salji}, {Tisi}, {Walker-Smith}, {di Francesco},
  {Hogerheijde}, {Andr{\'e}}, {Bastien}, {Bresnahan}, {Butner}, {Chen},
  {Chrysostomou}, {Coude}, {Davis}, {Duarte-Cabral}, {Fiege}, {Friberg},
  {Friesen}, {Fuller}, {Graves}, {Greaves}, {Gregson}, {Griffin}, {Holland},
  {Joncas}, {Knee}, {K{\"o}nyves}, {Mairs}, {Marsh}, {Matthews},
  {Moriarty-Schieven}, {Rawlings}, {Richer}, {Robertson}, {Rosolowsky},
  {Rumble}, {Sadavoy}, {Spinoglio}, {Thomas}, {Tothill}, {Viti}, {Wouterloot},
  {Yates}, \& {Zhu}}]{Pattle2015}
{Pattle}, K., {Ward-Thompson}, D., {Kirk}, J.~M., {et~al.} 2015, \mnras, 450,
  1094, \dodoi{10.1093/mnras/stv376}

\bibitem[{{Pecaut} \& {Mamajek}(2013)}]{pecaut2013}
{Pecaut}, M.~J., \& {Mamajek}, E.~E. 2013, \apjs, 208, 9,
  \dodoi{10.1088/0067-0049/208/1/9}

\bibitem[{{Plunkett} {et~al.}(2015){Plunkett}, {Arce}, {Mardones}, {van
  Dokkum}, {Dunham}, {Fern{\'a}ndez-L{\'o}pez}, {Gallardo}, \&
  {Corder}}]{plunkett2015}
{Plunkett}, A.~L., {Arce}, H.~G., {Mardones}, D., {et~al.} 2015, \nat, 527, 70,
  \dodoi{10.1038/nature15702}

\bibitem[{{Reipurth} {et~al.}(1999){Reipurth}, {Rodr{\'\i}guez}, \&
  {Chini}}]{reipurth1999}
{Reipurth}, B., {Rodr{\'\i}guez}, L.~F., \& {Chini}, R. 1999, \aj, 118, 983,
  \dodoi{10.1086/300958}

\bibitem[{{Ren} {et~al.}(2021){Ren}, {Zhu}, {Shi}, {Yue}, {Li}, {Zhang},
  {Mardones}, {Wu}, {Jiao}, {Liu}, {Luo}, {Xie}, {Zhang}, \& {Xu}}]{ren2021}
{Ren}, Z., {Zhu}, L., {Shi}, H., {et~al.} 2021, \mnras, 505, 5183,
  \dodoi{10.1093/mnras/stab1509}

\bibitem[{{Rosolowsky} {et~al.}(2008){Rosolowsky}, {Pineda}, {Kauffmann}, \&
  {Goodman}}]{rosolowsky2008}
{Rosolowsky}, E.~W., {Pineda}, J.~E., {Kauffmann}, J., \& {Goodman}, A.~A.
  2008, \apj, 679, 1338, \dodoi{10.1086/587685}

\bibitem[{{Sadavoy} {et~al.}(2016){Sadavoy}, {Stutz}, {Schnee}, {Mason}, {Di
  Francesco}, \& {Friesen}}]{sadavoy2016}
{Sadavoy}, S.~I., {Stutz}, A.~M., {Schnee}, S., {et~al.} 2016, \aap, 588, A30,
  \dodoi{10.1051/0004-6361/201527364}

\bibitem[{{Sakamoto} {et~al.}(1994){Sakamoto}, {Hayashi}, {Hasegawa}, {Handa},
  \& {Oka}}]{sakamoto1994}
{Sakamoto}, S., {Hayashi}, M., {Hasegawa}, T., {Handa}, T., \& {Oka}, T. 1994,
  \apj, 425, 641, \dodoi{10.1086/174011}

\bibitem[{{Sanhueza} {et~al.}(2019){Sanhueza}, {Contreras}, {Wu}, {Jackson},
  {Guzm{\'a}n}, {Zhang}, {Li}, {Lu}, {Silva}, {Izumi}, {Liu}, {Miura},
  {Tatematsu}, {Sakai}, {Beuther}, {Garay}, {Ohashi}, {Saito}, {Nakamura},
  {Saigo}, {Veena}, {Nguyen-Luong}, \& {Tafoya}}]{Sanhueza2019}
{Sanhueza}, P., {Contreras}, Y., {Wu}, B., {et~al.} 2019, \apj, 886, 102,
  \dodoi{10.3847/1538-4357/ab45e9}

\bibitem[{{Schneider} {et~al.}(2010){Schneider}, {Csengeri}, {Bontemps},
  {Motte}, {Simon}, {Hennebelle}, {Federrath}, \& {Klessen}}]{schneider2010}
{Schneider}, N., {Csengeri}, T., {Bontemps}, S., {et~al.} 2010, \aap, 520, A49,
  \dodoi{10.1051/0004-6361/201014481}

\bibitem[{{Schneider} \& {Elmegreen}(1979)}]{schneider1979}
{Schneider}, S., \& {Elmegreen}, B.~G. 1979, \apjs, 41, 87,
  \dodoi{10.1086/190609}

\bibitem[{{Shimajiri} {et~al.}(2008){Shimajiri}, {Takahashi}, {Takakuwa},
  {Saito}, \& {Kawabe}}]{shimajiri2008}
{Shimajiri}, Y., {Takahashi}, S., {Takakuwa}, S., {Saito}, M., \& {Kawabe}, R.
  2008, \apj, 683, 255, \dodoi{10.1086/588629}

\bibitem[{{Smith} {et~al.}(2011){Smith}, {Glover}, {Bonnell}, {Clark}, \&
  {Klessen}}]{Smith2011}
{Smith}, R.~J., {Glover}, S. C.~O., {Bonnell}, I.~A., {Clark}, P.~C., \&
  {Klessen}, R.~S. 2011, \mnras, 411, 1354,
  \dodoi{10.1111/j.1365-2966.2010.17775.x}

\bibitem[{{Snell} {et~al.}(1980){Snell}, {Loren}, \& {Plambeck}}]{snell1980}
{Snell}, R.~L., {Loren}, R.~B., \& {Plambeck}, R.~L. 1980, \apjl, 239, L17,
  \dodoi{10.1086/183283}

\bibitem[{{Stanke} {et~al.}(2002){Stanke}, {McCaughrean}, \&
  {Zinnecker}}]{stanke2002}
{Stanke}, T., {McCaughrean}, M.~J., \& {Zinnecker}, H. 2002, \aap, 392, 239,
  \dodoi{10.1051/0004-6361:20020763}

\bibitem[{{Stutz} {et~al.}(2013){Stutz}, {Tobin}, {Stanke}, {Megeath},
  {Fischer}, {Robitaille}, {Henning}, {Ali}, {di Francesco}, {Furlan},
  {Hartmann}, {Osorio}, {Wilson}, {Allen}, {Krause}, \& {Manoj}}]{stutz2013}
{Stutz}, A.~M., {Tobin}, J.~J., {Stanke}, T., {et~al.} 2013, \apj, 767, 36,
  \dodoi{10.1088/0004-637X/767/1/36}

\bibitem[{{Takahashi} {et~al.}(2013){Takahashi}, {Ho}, {Teixeira}, {Zapata}, \&
  {Su}}]{takahashi2013}
{Takahashi}, S., {Ho}, P. T.~P., {Teixeira}, P.~S., {Zapata}, L.~A., \& {Su},
  Y.-N. 2013, \apj, 763, 57, \dodoi{10.1088/0004-637X/763/1/57}

\bibitem[{{Takahashi} {et~al.}(2008){Takahashi}, {Saito}, {Ohashi}, {Kusakabe},
  {Takakuwa}, {Shimajiri}, {Tamura}, \& {Kawabe}}]{takahashi2008}
{Takahashi}, S., {Saito}, M., {Ohashi}, N., {et~al.} 2008, \apj, 688, 344,
  \dodoi{10.1086/592212}

\bibitem[{{Tanabe} {et~al.}(2019){Tanabe}, {Nakamura}, {Tsukagoshi},
  {Shimajiri}, {Ishii}, {Kawabe}, {Feddersen}, {Kong}, {Arce}, {Bally},
  {Carpenter}, \& {Momose}}]{tanabe2019}
{Tanabe}, Y., {Nakamura}, F., {Tsukagoshi}, T., {et~al.} 2019, \pasj, 71, S8,
  \dodoi{10.1093/pasj/psz100}

\bibitem[{{Tanaka} {et~al.}(2013){Tanaka}, {Nakamura}, {Awazu}, {Shimajiri},
  {Sugitani}, {Onishi}, {Kawabe}, {Yoshida}, \& {Higuchi}}]{tanaka2013}
{Tanaka}, T., {Nakamura}, F., {Awazu}, Y., {et~al.} 2013, \apj, 778, 34,
  \dodoi{10.1088/0004-637X/778/1/34}

\bibitem[{{Tatematsu} {et~al.}(2016){Tatematsu}, {Ohashi}, {Sanhueza}, {Nguyen
  Luong}, {Umemoto}, \& {Mizuno}}]{tatematsu2016}
{Tatematsu}, K., {Ohashi}, S., {Sanhueza}, P., {et~al.} 2016, \pasj, 68, 24,
  \dodoi{10.1093/pasj/psw002}

\bibitem[{{Tatematsu} {et~al.}(1993){Tatematsu}, {Umemoto}, {Kameya}, {Hirano},
  {Hasegawa}, {Hayashi}, {Iwata}, {Kaifu}, {Mikami}, {Murata}, {Nakano},
  {Nakano}, {Ohashi}, {Sunada}, {Takaba}, \&
  {Yamamoto}}]{Tatematsu1993ApJ...404..643T}
{Tatematsu}, K., {Umemoto}, T., {Kameya}, O., {et~al.} 1993, \apj, 404, 643,
  \dodoi{10.1086/172318}

\bibitem[{{Teixeira} {et~al.}(2016){Teixeira}, {Takahashi}, {Zapata}, \&
  {Ho}}]{teixeira2016}
{Teixeira}, P.~S., {Takahashi}, S., {Zapata}, L.~A., \& {Ho}, P.~T.~P. 2016,
  \aap, 587, A47, \dodoi{10.1051/0004-6361/201526807}

\bibitem[{{THE CASA TEAM} {et~al.}(2022){THE CASA TEAM}, {Bean}, {Bhatnagar},
  {Castro}, {Donovan Meyer}, {Emonts}, {Garcia}, {Garwood}, {Golap}, {Gonzalez
  Villalba}, {Harris}, {Hayashi}, {Hoskins}, {Hsieh}, {Jagannathan},
  {Kawasaki}, {Keimpema}, {Kettenis}, {Lopez}, {Marvil}, {Masters},
  {McNichols}, {Mehringer}, {Miel}, {Moellenbrock}, {Montesino}, {Nakazato},
  {Ott}, {Petry}, {Pokorny}, {Raba}, {Rau}, {Schiebel}, {Schweighart},
  {Sekhar}, {Shimada}, {Small}, {Steeb}, {Sugimoto}, {Suoranta}, {Tsutsumi},
  {van Bemmel}, {Verkouter}, {Wells}, {Xiong}, {Szomoru}, {Griffith},
  {Glendenning}, \& {Kern}}]{CASA2022}
{THE CASA TEAM}, {Bean}, B., {Bhatnagar}, S., {et~al.} 2022, arXiv e-prints,
  arXiv:2210.02276.
\newblock \doarXiv{2210.02276}

\bibitem[{{Tobin} {et~al.}(2019){Tobin}, {Megeath}, {van{\textquoteright}t
  Hoff}, {D{\'\i}az-Rodr{\'\i}guez}, {Reynolds}, {Osorio}, {Anglada}, {Furlan},
  {Karnath}, {Offner}, {Sheehan}, {Sadavoy}, {Stutz}, {Fischer}, {Kama},
  {Persson}, {Di Francesco}, {Looney}, {Watson}, {Li}, {Stephens}, {Chandler},
  {Cox}, {Dunham}, {Kratter}, {Kounkel}, {Mazur}, {Murillo}, {Patel}, {Perez},
  {Segura-Cox}, {Sharma}, {Tychoniec}, \& {Wyrowski}}]{tobin2019}
{Tobin}, J.~J., {Megeath}, S.~T., {van{\textquoteright}t Hoff}, M., {et~al.}
  2019, \apj, 886, 6, \dodoi{10.3847/1538-4357/ab498f}

\bibitem[{{Tobin} {et~al.}(2020){Tobin}, {Sheehan}, {Reynolds}, {Megeath},
  {Osorio}, {Anglada}, {D{\'\i}az-Rodr{\'\i}guez}, {Furlan}, {Kratter},
  {Offner}, {Looney}, {Kama}, {Li}, {van't Hoff}, {Sadavoy}, \&
  {Karnath}}]{tobin2020b}
{Tobin}, J.~J., {Sheehan}, P.~D., {Reynolds}, N., {et~al.} 2020, \apj, 905,
  162, \dodoi{10.3847/1538-4357/abc5bf}

\bibitem[{{Trevi{\~n}o-Morales} {et~al.}(2019){Trevi{\~n}o-Morales}, {Fuente},
  {S{\'a}nchez-Monge}, {Kainulainen}, {Didelon}, {Suri}, {Schneider},
  {Ballesteros-Paredes}, {Lee}, {Hennebelle}, {Pilleri},
  {Gonz{\'a}lez-Garc{\'\i}a}, {Kramer}, {Garc{\'\i}a-Burillo}, {Luna},
  {Goicoechea}, {Tremblin}, \& {Geen}}]{trevino2019}
{Trevi{\~n}o-Morales}, S.~P., {Fuente}, A., {S{\'a}nchez-Monge}, {\'A}.,
  {et~al.} 2019, \aap, 629, A81, \dodoi{10.1051/0004-6361/201935260}

\bibitem[{{van Terwisga} {et~al.}(2019){van Terwisga}, {Hacar}, \& {van
  Dishoeck}}]{vanterwisga2019}
{van Terwisga}, S.~E., {Hacar}, A., \& {van Dishoeck}, E.~F. 2019, \aap, 628,
  A85, \dodoi{10.1051/0004-6361/201935378}

\bibitem[{{Wang} {et~al.}(2010){Wang}, {Li}, {Abel}, \& {Nakamura}}]{wang2010}
{Wang}, P., {Li}, Z.-Y., {Abel}, T., \& {Nakamura}, F. 2010, \apj, 709, 27,
  \dodoi{10.1088/0004-637X/709/1/27}

\bibitem[{{Williams} {et~al.}(2003){Williams}, {Plambeck}, \&
  {Heyer}}]{williams2003}
{Williams}, J.~P., {Plambeck}, R.~L., \& {Heyer}, M.~H. 2003, \apj, 591, 1025,
  \dodoi{10.1086/375396}

\bibitem[{{Wilson} {et~al.}(2005){Wilson}, {Dame}, {Masheder}, \&
  {Thaddeus}}]{wilson2005}
{Wilson}, B.~A., {Dame}, T.~M., {Masheder}, M.~R.~W., \& {Thaddeus}, P. 2005,
  \aap, 430, 523, \dodoi{10.1051/0004-6361:20035943}

\bibitem[{{Yu} {et~al.}(1997){Yu}, {Bally}, \& {Devine}}]{yu1997}
{Yu}, K.~C., {Bally}, J., \& {Devine}, D. 1997, \apjl, 485, L45,
  \dodoi{10.1086/310799}

\bibitem[{{Yuan} {et~al.}(2018){Yuan}, {Li}, {Wu}, {Ellingsen}, {Henkel},
  {Wang}, {Liu}, {Liu}, {Zavagno}, {Ren}, \& {Huang}}]{yuan2018}
{Yuan}, J., {Li}, J.-Z., {Wu}, Y., {et~al.} 2018, \apj, 852, 12,
  \dodoi{10.3847/1538-4357/aa9d40}

\bibitem[{{Zapata} {et~al.}(2006){Zapata}, {Ho}, {Rodr{\'\i}guez}, {O'Dell},
  {Zhang}, \& {Muench}}]{zapata2006}
{Zapata}, L.~A., {Ho}, P. T.~P., {Rodr{\'\i}guez}, L.~F., {et~al.} 2006, \apj,
  653, 398, \dodoi{10.1086/508319}

\bibitem[{{Zhang} {et~al.}(2020){Zhang}, {Ren}, {Wu}, {Li}, {Zhu}, {Zhang},
  {Mardones}, {Wang}, {Shi}, {Yue}, {Luo}, {Xie}, {Jiao}, {Liu}, {Xu}, \&
  {Wang}}]{zhang2020}
{Zhang}, C., {Ren}, Z., {Wu}, J., {et~al.} 2020, \mnras, 497, 793,
  \dodoi{10.1093/mnras/staa1958}

\bibitem[{{Zhang} {et~al.}(2002){Zhang}, {Hunter}, {Sridharan}, \&
  {Ho}}]{zhang2002}
{Zhang}, Q., {Hunter}, T.~R., {Sridharan}, T.~K., \& {Ho}, P. T.~P. 2002, \apj,
  566, 982, \dodoi{10.1086/338278}

\bibitem[{{Zhang} {et~al.}(2019){Zhang}, {Arce}, {Mardones}, {Cabrit},
  {Dunham}, {Garay}, {Noriega-Crespo}, {Offner}, {Raga}, \&
  {Corder}}]{zhang2019}
{Zhang}, Y., {Arce}, H.~G., {Mardones}, D., {et~al.} 2019, \apj, 883, 1,
  \dodoi{10.3847/1538-4357/ab3850}

\bibitem[{{Ziurys} {et~al.}(1989){Ziurys}, {Snell}, \& {Dickman}}]{ziurys1989}
{Ziurys}, L.~M., {Snell}, R.~L., \& {Dickman}, R.~L. 1989, \apj, 341, 857,
  \dodoi{10.1086/167544}

\end{thebibliography}
\bibliographystyle{aasjournal}

\end{document}